\documentclass[10pt]{book}
\usepackage{amsmath,epsfig,graphicx,amssymb}
\begin{document}
\setcounter{chapter}{1}

\begin{titlepage}

\begin{center}
{\Large {\bf Electronic structure of kinetic energy driven cuprate
superconductors}}
\end{center}

\vskip 5mm

\begin{center}
{\large Shiping Feng\footnote{Corresponding author}, Huaiming
Guo$^{*}$, Yu Lan, and Li Cheng} \\

\vskip 2mm
{\it Department of Physics, Beijing Normal University,
Beijing 100875, China}\\
\vskip 15mm

\begin{minipage}{5in}

In this paper, we review the low energy electronic structure of the
kinetic energy driven d-wave cuprate superconductors. We give a
general description of the charge-spin separation fermion-spin
theory, where the constrained electron is decoupled as the gauge
invariant dressed holon and spin. In particular, we show that under
the decoupling scheme, the charge-spin separation fermion-spin
representation is a natural representation of the constrained
electron defined in a restricted Hilbert space without double
electron occupancy. Based on the charge-spin separation fermion-spin
theory, we have developed the kinetic energy driven superconducting
mechanism, where the superconducting state is controlled by both
superconducting gap parameter and quasiparticle coherence. Within
this kinetic energy driven superconductivity, we have discussed the
low energy electronic structure of the single layer and bilayer
cuprate superconductors in both superconducting and normal states,
and qualitatively reproduced all main features of the angle-resolved
photoemission spectroscopy measurements on the single layer and
bilayer cuprate superconductors. We show that the superconducting
state in cuprate superconductors is the conventional
Bardeen-Cooper-Schrieffer like with the d-wave symmetry, so that the
basic Bardeen-Cooper-Schrieffer formalism with the d-wave gap
function is still valid in discussions of the low energy electronic
structure of cuprate superconductors, although the pairing mechanism
is driven by the kinetic energy by exchanging spin excitations. We
also show that the well pronounced peak-dip-hump structure of the
bilayer cuprate superconductors in the superconducting state and
double-peak structure in the normal state are mainly caused by the
bilayer splitting.
\end{minipage}
\end{center}

\end{titlepage}

\section{Introduction}

The parent compounds of cuprate superconductors are the Mott
insulators with an antiferromagnetic long-range order, then changing
the carrier concentration by ionic substitution or increasing the
oxygen content turns these compounds into the superconducting state
leaving the antiferromagnetic short-range correlation still intact
\cite{bednorz,kastner}. It has become clear in the past twenty years
that cuprate superconductors are among the most complex systems
studied in condensed matter physics
\cite{kastner,shen1,shen,campuzano}. The complications arise mainly
from (1) a layered crystal structure with one or more CuO$_{2}$
planes per unit cell separated by insulating layers which leads to a
quasi-two-dimensional electronic structure, and (2) extreme
sensitivity of the physical properties to the compositions
(stoichiometry) which control the carrier density in the CuO$_{2}$
plane. As a consequence, both experimental investigation
\cite{kastner,shen1,shen,campuzano} and theoretical understanding
\cite{anderson1,anderson2,anderson3,laughlin} are extremely
difficult. Moreover, the layered crystal structure leads to that
cuprate superconductors are highly anisotropic materials, then the
electron spectral function $A({\bf k},\omega)$ is dependent on the
in-plane momentum \cite{shen1,shen,campuzano}. After twenty years
extensive studies, it has been shown that many of the unusual
physical properties of cuprate superconductors have often been
attributed to particular characteristics of the low energy
excitations determined by the electronic structure
\cite{shen1,shen,campuzano}.

Experimentally, by virtue of systematic studies using the
angle-resolved photoemission spectroscopy (ARPES), the low energy
electronic structure of cuprate superconductors in both
superconducting and normal states has been well-established by now
\cite{shen1,shen,campuzano}. In particular, the information revealed
by ARPES experiments has shown that around the nodal and antinodal
points of the Brillouin zone contain the essentials of the whole low
energy quasiparticle excitation spectrum of cuprate superconductors
\cite{shen1,shen,campuzano}. In the normal-state, the charge
carriers doped into the parent compounds first enter into the ${\bf
k}=[\pi/2,\pi/2]$ (in units of inverse lattice constant) point of
the Brillouin zone \cite{shen1,kim,dessau,wells}. Moreover, the
electron spectral function $A({\bf k},\omega)$ has a flat band form
as a function of energy $\omega$ for momentum ${\bf k}$ in the
vicinity of the $[\pi,0]$ point just below the Fermi energy, which
leads to the unusual quasiparticle dispersion around the $[\pi,0]$
point with anomalously small changes of electron energy as a
function of momentum \cite{shen1,kim,dessau,wells}. However, the
lowest energy states are located at the $[\pi,0]$ point in the
superconducting state \cite{shen,campuzano,ding}, where the d-wave
superconducting gap function is maximal. Furthermore, the
improvements in the resolution of the ARPES experiments allowed for
an experimental verification of the particle-hole coherence in the
superconducting state and Bogoliubov-quasiparticle nature of the
sharp superconducting quasiparticle peak near the $[\pi,0]$ point
\cite{matsui,campuzano1}. It is striking that in spite of the
nonconventional superconducting mechanism and observed exotic
magnetic scattering \cite{yamada,dai,arai} in cuprate
superconductors, these ARPES experimental results
\cite{matsui,campuzano1} show that the superconducting coherence of
the quasiparticle peak is described by the simple
Bardeen-Cooper-Schrieffer (BCS) formalism \cite{bcs} with the d-wave
gap function. Although these common features are observed, there are
numerous anomalies for different families of cuprate
superconductors, which complicate the physical properties of the
electronic structure \cite{shen,campuzano}. Among these anomalies is
the dramatic change in the spectral lineshape around the $[\pi,0]$
point first observed on the bilayer cuprate superconductor
Bi$_{2}$Sr$_{2}$CaCu$_{2}$O$_{8+\delta}$ in the superconducting
state, where a sharp quasiparticle peak develops at the lowest
binding energy, followed by a dip and a hump, giving rise to the
so-called peak-dip-hump structure in the electron spectrum
\cite{dessau1,randeria,fedorov}. Later, this peak-dip-hump structure
was also found in YBa$_{2}$Cu$_{3}$O$_{7-\delta}$ \cite{lu} and in
Bi$_{2}$Sr$_{2}$Ca$_{2}$Cu$_{3}$O$_{10+\delta}$ \cite{sato}.
Moreover, although the sharp quasiparticle peaks are identified in
the superconducting state along the entire Fermi surface, the
peak-dip-hump structure is most strongly developed around the
$[\pi,0]$ point \cite{shen,dessau1,randeria,fedorov,lu,sato}. In
particular, the similar double-peak structure in the electron
spectrum around the $[\pi,0]$ point has been observed in the bilayer
cuprate superconductors in the normal-state \cite{fedorov,dlfeng2}.
These ARPES experimental measurements raise a question: whether the
behavior of the low energy excitations determined by the electronic
structure is universal or not.

One of the main concerns in the field of superconductivity in
cuprate superconductors is about the origin of the electron pairing.
From the experimental side, it has been well established that the
antiferromagnetic short-range correlation coexists with the
superconducting state in the whole superconducting regime
\cite{yamada,dai,arai}. In particular, an impurity-substitution
effect on the low energy dynamics of cuprate superconductors has
been investigated experimentally \cite{ding9}, which is a magnetic
analogue of the isotope effect used for the conventional
superconductors. It is shown \cite{ding9} that the impurity-induced
changes in the electron self-energy show a good correspondence to
those  of magnetic excitations. These experimental results provide a
clear link between the charge carrier pairing mechanism and magnetic
excitations, and also is an indication of the unconventional
superconducting mechanism that is responsible for a high
superconducting transition temperature. On the theoretical side,
there is an increasing theoretical evidence that purely electronic
models can indeed sustain a robust pairing, possibly leading to a
high superconducting transition temperature \cite{gros,dagotto}.
Recently, we \cite{feng1} have developed a charge-spin separation
fermion-spin theory, where the constrained electron operator is
decoupled as the gauge invariant dressed holon and spin, with the
dressed holon represents the charge degree of freedom together with
some effects of the spin configuration rearrangements due to the
presence of the doped hole itself, while the spin operator
represents the spin degree of freedom. Based on this charge-spin
separation fermion-spin theory, we \cite{feng2} have established a
kinetic energy driven superconducting mechanism, where the dressed
holon-spin interaction from the kinetic energy term induces the
dressed holon pairing state with d-wave symmetry by exchanging spin
excitations, then the electron Cooper pairs originating from the
dressed holon pairing state are due to the charge-spin
recombination, and their condensation reveals the d-wave
superconducting ground-state. In particular, this superconducting
state is the conventional BCS like with the d-wave symmetry, and is
controlled by both superconducting gap function and quasiparticle
coherence, then the maximal superconducting transition temperature
occurs around the optimal doping, and decreases in both underdoped
and overdoped regimes \cite{feng3}. Within this framework of the
kinetic energy driven superconductivity, we
\cite{feng4,feng5,guo1,guo2,lan1,lan2} have performed a systematic
calculation for the low energy electron spectral function of the
single layer and bilayer cuprate superconductors in both
superconducting and normal states, and qualitatively reproduced all
main features of the ARPES experimental measurements on the single
layer and bilayer cuprate superconductors \cite{shen,campuzano}. In
this paper, we give a brief review of our recent studies for the low
energy electronic structure of the kinetic energy driven d-wave
cuprate superconductors \cite{feng4,feng5,guo1,guo2,lan1,lan2}. We
show that in both superconducting and normal states, the spectral
weight increases with increasing doping, and decreases with
increasing temperatures \cite{feng4,feng5,guo1,guo2,lan1,lan2}.
Furthermore, the superconducting quasiparticles around the $[\pi,0]$
point disperse very weakly with momentum \cite{guo2,lan2}. In
corresponding to this weak dispersions in the superconducting state,
the quasiparticle dispersions in the normal state exhibit the flat
band around the $[\pi,0]$ point just below the Fermi energy
\cite{guo1,lan1}. Moreover, it is shown that the well pronounced
peak-dip-hump structure \cite{lan2} of the bilayer cuprate
superconductors in the superconducting state and double-peak
structure \cite{lan1} in the normal state are mainly caused by the
bilayer splitting. In particular, we show that one of the universal
features is that the d-wave superconducting state in cuprate
superconductors is the conventional BCS like, so that the basic BCS
formalism with the d-wave symmetry is still valid in discussions of
the low energy electronic structure of cuprate superconductors
\cite{feng4,feng5,guo2,lan2}, although the pairing mechanism is
driven by the kinetic energy by exchanging spin excitations, and
other exotic magnetic properties \cite{yamada,dai,arai} are beyond
the BCS formalism. Our these theoretical results
\cite{feng4,feng5,guo1,guo2,lan1,lan2} also show that the striking
behavior of the electronic structure in cuprate superconductors is
intriguingly related to the strong coupling between the electron
quasiparticles and collective magnetic excitations.

This paper is organized as follows. In section 2, the charge-spin
separation fermion-spin theory is presented \cite{feng1}. It is
shown that the charge-spin separation fermion-spin representation is
a natural representation of the constrained electron defined in a
restricted Hilbert space without double electron occupancy. Within
this charge-spin separation fermion-spin theory \cite{feng1} and
kinetic energy driven superconducting mechanism \cite{feng2,feng3},
the electronic structure of the single layer cuprate superconductors
in both superconducting and normal states
\cite{feng4,feng5,guo1,guo2} is presented in section 3. It is shown
that the superconducting coherence of the quasiparticle peak is
described by the simple BCS formalism. In section 4, the electronic
structure of the bilayer cuprate superconductors \cite{lan1,lan2} is
discussed by including the bilayer hopping and bilayer magnetic
exchange interaction. It is shown that the electron spectrum is
split into the bonding and antibonding components by the bilayer
splitting, then the superconducting peak is closely related to the
antibonding component, while the hump is mainly formed by the
bonding component. Finally, we give a summary and discussions in
section 5.

\section{Charge-spin separation fermion-spin theory}

In cuprate superconductors, the single common feature is the
presence of the two-dimensional CuO$_{2}$ plane \cite{kastner} as
mentioned in section 1, and it seems evident that the
nonconventional behaviors are dominated by this plane. Very soon
after the discovery of superconductivity in doped cuprates, Anderson
argued that the essential physics of the doped CuO$_{2}$ plane is
contained in the Hubbard model or its strong coupling limit, namely
the $t$-$J$ model on a square lattice \cite{anderson1}, which
includes an antiferromagnetic coupling between localized spins and a
kinetic energy term for the hole motion. Furthermore, Anderson
proposed a scenario of superconductivity in cuprate superconductors
based on the charge-spin separation within the two-dimensional
$t$-$J$ model \cite{anderson1}, where the internal degrees of
freedom of the electron are decoupled as the charge and spin degrees
of freedom, then the elementary excitations are collective modes for
the charge and spin degrees of freedom, and these collective modes
might be responsible for the nonconventional behaviors of cuprate
superconductors. Since then, many unusual physical properties of
cuprate superconductors are extensively studied along with this line
within the two-dimensional $t$-$J$ model.

The decoupling of the charge and spin degrees of freedom of electron
is undoubtedly correct in the one-dimensional interacting electron
systems \cite{haldane}, where the charge and spin degrees of freedom
are represented by boson operators that describe the excitations of
charge-density wave and spin-density wave, respectively. In
particular, the typical behavior of the Luttinger liquid, i.e., the
absence of the quasiparticle propagation and charge-spin separation,
has been demonstrated theoretically within the one-dimensional
$t$-$J$ model \cite{ogata}. Moreover, the collective modes for the
charge and spin degrees of freedom as the real elementary
excitations in the one-dimensional cuprates has been observed
directly by the ARPES experiment \cite{kim2}. Therefore both
theoretical and experimental studies indicate that the existence of
the real collective modes for the charge and spin degrees of freedom
is common in the one-dimensional interacting electron systems
\cite{maekawa}. However, the case in the two-dimensional is very
complex since there are many competing degrees of freedom
\cite{kastner,shen,campuzano}. Among the unusual physical properties
of cuprate superconductors in the normal state, a hallmark is the
charge transport \cite{cooper1,uchida1,uchida2,ando1}, where the
conductivity shows a non-Drude behavior at low energies, and is
carried by $\delta$ holes, with $\delta$ is the hole doping
concentration, while the resistivity exhibits a linear temperature
behavior over a wide range of temperatures. This is a strong
experimental evidence supporting the notion of the charge-spin
separation, since not even conventional electron-electron scattering
would show the striking linear rise of scattering rate above the
Debye frequency, and if there is no charge-spin separartion, the
phonons should affect these properties \cite{anderson2}. Moreover, a
compelling evidence for the charge-spin separartion in cuprate
superconductors has been found from the experimental test of the
Wiedemann-Franz law, where a clear departure from the universal
Wiedemann-Franz law for the typical Fermi liquid behavior is
observed \cite{hill}. In this case, a formal theory with the gauge
invariant collective modes for the charge and spin degrees of
freedom, i.e., the issue of whether these collective modes are real,
is centrally important for understanding of the physical properties
of cuprate superconductors \cite{laughlin}. In this section, we
review briefly the charge-spin separation fermion-spin theory
\cite{feng1}. It is shown that if the single occupancy local
constraint is treated properly, then the constrained electron in the
two-dimensional strong interaction systems can be decoupled
completely by introducing the dressed holon and spin, where the
dressed holon describes the charge degree of freedom together with
some effects of the spin configuration rearrangements due to the
presence of the doped hole itself, while the spin operator describes
the spin degree of freedom. Furthermore, these dressed holon and
spin are gauge invariant, i.e., they are real in the two-dimensional
strong interaction systems. In particular, we also show that the
charge-spin separation fermion-spin representation is a natural
representation for the constrained electron under the decoupling
scheme.

We start from the $t$-$J$ model defined on a square lattice as
\cite{anderson1},
\begin{eqnarray}
H=-t\sum_{i\hat{\eta}\sigma}C^{\dagger}_{i\sigma}
C_{i+\hat{\eta}\sigma}+t'\sum_{i\hat{\tau}\sigma}
C^{\dagger}_{i\sigma}C_{i+\hat{\tau}\sigma}+\mu_{0}\sum_{i\sigma}
C^{\dagger}_{i\sigma}C_{i\sigma}+J\sum_{i\hat{\eta}}{\bf S}_{i}
\cdot {\bf S}_{i+\hat{\eta}},
\end{eqnarray}
where $\hat{\eta}=\pm\hat{x},\pm\hat{y}$, $\hat{\tau}=\pm\hat{x}\pm
\hat{y}$, $C^{\dagger}_{i\sigma}$ ($C_{i\sigma}$) is the electron
creation (annihilation) operator, ${\bf S}_{i}=C^{\dagger}_{i}
{\vec\sigma}C_{i}/2$ is spin operator with ${\vec\sigma}=
(\sigma_{x},\sigma_{y},\sigma_{z})$ as Pauli matrices, and $\mu_{0}$
is the chemical potential. This $t$-$J$ model (1) is defined in a
restricted Hilbert space without double electron occupancy. In this
case, there are two ways to implement this requirement: either to
solve the $t$-$J$ model (1) combined with an important single
occupancy local constraint,
\begin{eqnarray}
\sum_{\sigma}C^{\dagger}_{i\sigma} C_{i\sigma}\leq 1,
\end{eqnarray}
or to introduce the constrained electron operators \cite{rice1},
replacing $C_{i\sigma}$ by $\hat{C}_{i\sigma}=C_{i\sigma}(1-
n_{i-\sigma})$ with $n_{i\sigma}=C^{\dagger}_{i\sigma}C_{i\sigma}$.
We will use both representations in this section.

The on-site single occupancy local constraint (2) is an reflection
of the strong on-site Mott-Hubbard Coulombic interaction. There is
much evidence that the interactions in cuprate superconductors are
dominated by this strong on-site Mott-Hubbard Coulombic interaction
\cite{anderson3}. Therefore we must deal with the on-site single
occupancy local constraint (2) before any analytical calculations
\cite{emery}. For a proper treating the electron single occupancy
local constraint (2), we follow the charge-spin separation scheme,
and decouple the electron operator in the $t$-$J$ model (1) as
\cite{feng1,feng6},
\begin{eqnarray}
C_{i\sigma}=h^{\dagger}_{i}a_{i\sigma},
\end{eqnarray}
supplemented by the local constraint $\sum_{\sigma}
a^{\dagger}_{i\sigma}a_{i\sigma}=1$, where the spinless fermion
operator $h_{i}$ keeps track of the charge degree of freedom, while
the boson operator $a_{i\sigma}$ keeps track of the spin degree of
freedom, then the Hamiltonian (1) can be rewritten as,
\begin{eqnarray}
H&=&t\sum_{i\hat{\eta}\sigma}h^{\dagger}_{i+\hat{\eta}}
h_{i}a^{\dagger}_{i\sigma}a_{i+\hat{\eta}\sigma}-
t'\sum_{i\hat{\tau}\sigma}h^{\dagger}_{i+\hat{\tau}}
h_{i}a^{\dagger}_{i\sigma}a_{i+\hat{\tau}\sigma}-\mu_{0}\sum_{i}
h^{\dagger}_{i}h_{i}  \nonumber \\
&+&J\sum_{i\hat{\eta}} (1-h^{\dagger}_{i}h_{i}) {\bf S}_{i}
\cdot{\bf S}_{i+\hat{\eta}}(1-h^{\dagger}_{i+\hat{\eta}}
h_{i+\hat{\eta}}),
\end{eqnarray}
with the spin operator ${\bf S}_{i}=a^{\dagger}_{i}{\vec\sigma}
a_{i}/2$. In this case, the electron single occupancy local
constraint (2) $\sum_{\sigma}C^{\dagger}_{i\sigma}C_{i\sigma}=1-
h^{\dagger}_{i}h_{i}\leq 1$ is exactly satisfied, with $n^{(h)}_{i}=
h^{\dagger}_{i}h_{i}$ is equal to 1 or 0. This decoupling scheme is
called as the CP$^{1}$ representation \cite{ioffe}, where the
elementary charge and spin excitations are called the holon and
spinon, respectively. We call such holon and spinon as the bare
holon and spinon, respectively, since an extra $U(1)$ gauge degree
of freedom related with the single occupancy local constraint
$\sum_{\sigma}a^{\dagger}_{i\sigma}a_{i\sigma}=1$ appears, i.e., the
CP$^{1}$ representation is invariant under a local $U(1)$ gauge
transformation,
\begin{eqnarray}
h_{i}\rightarrow h_{i}e^{i\theta_{i}}, ~~~ a_{i\sigma}\rightarrow
a_{i\sigma}e^{i\theta_{i}},
\end{eqnarray}
and then all physical quantities should be invariant with respect to
this transformation. However, the bare holon $h_{i}$ or bare spinon
$a_{i\sigma}$ itself is not gauge invariant, and they are strongly
coupled by the $U(1)$ gauge field fluctuations. In other words,
these bare holon and spinon are not real.

It has been shown \cite{feng1,feng6} that the CP$^{1}$ boson
$a_{i\sigma}$ together with the local constraint
$\sum_{\sigma}a^{\dagger}_{i\sigma}a_{i\sigma}=1$ can be mapped
exactly onto the spin representation defined with an additional
phase factor. This is because that the empty and doubly occupied
spin states have been ruled out due to the constraint
$\sum_{\sigma}a^{\dagger}_{i\sigma}a_{i\sigma}=1$, and only the
spin-up and spin-down singly occupied spin states are allowed.
Therefore the original four-dimensional representation space is
reduced to a two-dimensional space. Due to the symmetry of the
spin-up and spin-down states, $\mid{\rm occupied}\rangle_{\uparrow}
=\left (\begin{array}{cc} {1}\\{0}\end{array}\right)_{\uparrow}$
and $\mid{\rm empty}\rangle_{\uparrow}=\left (\begin{array}{cc}{0}\\
{1}\end{array}\right)_{\uparrow}$ are singly-occupied and empty
spin-up, while $\mid{\rm occupied}\rangle_{\downarrow}=\left(
\begin{array}{cc}{0}\\ {1}\end{array}\right)_{\downarrow}$ and
$\mid{\rm empty}\rangle_{\downarrow}=\left (\begin{array}{cc} {1}
\\ {0}\end{array}\right)_{\downarrow}$ are singly-occupied and empty
spin-down states, respectively, thus the constrained CP$^{1}$ boson
operators $a_{i\uparrow}$ and $a_{i\downarrow}$ can be represented
in this reduced two-dimensional space as \cite{feng1,feng6},
\begin{subequations}
\begin{eqnarray}
a_{\uparrow}&=&e^{i\Phi_{\uparrow}}\mid {\rm occupied}
\rangle_{\downarrow}~_{\uparrow}\langle {\rm occupied}\mid=
e^{i\Phi_{\uparrow}}\left (
\begin{array}{cc}
{0} & {0}\\{1} & {0}
\end{array}
\right)=e^{i\Phi_{\uparrow}}S^{-}, ~~~\\
a_{\downarrow}&=&e^{i\Phi_{\downarrow}}\mid {\rm occupied}
\rangle_{\uparrow}~_{\downarrow}\langle {\rm occupied}\mid=
e^{i\Phi_{\downarrow}}\left (
\begin{array}{cc}
{0} & {1}\\{0} & {0}
\end{array}
\right)= e^{i\Phi_{\downarrow}}S^{+},~~~
\end{eqnarray}
\end{subequations}
with $S^{-}$ is the $S^{z}$ lowering operator, while $S^{+}$ is the
$S^{z}$ raising operator, then the local constraint $\sum_{\sigma}
a^{\dagger}_{i\sigma}a_{i\sigma}=S^{+}_{i}S^{-}_{i}+S^{-}_{i}
S^{+}_{i}=1$ is exactly satisfied. In this case, the electron
decoupling form (3) combined with the electron single occupancy
local constraint can be expressed as,
\begin{eqnarray}
C_{i\uparrow}=h^{\dagger}_{i}e^{i\Phi_{i\uparrow}}S^{-}_{i},~~~~
C_{i\downarrow}=h^{\dagger}_{i}e^{i\Phi_{i\downarrow}}S^{+}_{i},
\end{eqnarray}
then the local $U(1)$ gauge transformation (5) can be rewritten as,
\begin{eqnarray}
h_{i}\rightarrow h_{i}e^{i\theta_{i}}, ~~~ \Phi_{i\sigma}\rightarrow
\Phi_{i\sigma}+\theta_{i}.
\end{eqnarray}
Furthermore, the phase factor $e^{i\Phi_{i\sigma}}$ is induced by
the spin configuration rearrangements due to the presence of the
doped hole itself, and therefore can be incorporated into the bare
holon operator $h^{\dagger}_{i}$. In this case, we can obtain a new
transformation from Eq. (7) as \cite{feng1},
\begin{eqnarray}
C_{i\uparrow}=h^{\dagger}_{i1}S^{-}_{i}=h^{\dagger}_{i\uparrow}
S^{-}_{i},~~~~C_{i\downarrow}=h^{\dagger}_{i2}S^{+}_{i}=
h^{\dagger}_{i\downarrow}S^{+}_{i},
\end{eqnarray}
with the {\it spinful fermion} operator $h_{i\sigma}=
e^{-i\Phi_{i\sigma}}h_{i}$ represents the charge degree of freedom
together with some effects of the spin configuration rearrangement
due to the presence of the doped hole itself (dressed holon), while
the spin operator $S_{i}$ represents the spin degree of freedom,
then electron single occupancy local constraint (2),
\begin{eqnarray}
\sum_{\sigma}C^{\dagger}_{i\sigma}C_{i\sigma}&=&S^{+}_{i}
h_{i\uparrow}h^{\dagger}_{i\uparrow}S^{-}_{i}+S^{-}_{i}
h_{i\downarrow}h^{\dagger}_{i\downarrow}S^{+}_{i}\nonumber \\
&=&h_{i} h^{\dagger}_{i}(S^{+}_{i}S^{-}_{i}+S^{-}_{i}S^{+}_{i})=1-
h^{\dagger}_{i}h_{i}\leq 1,
\end{eqnarray}
is always satisfied in analytical calculations, and the double
spinful fermion occupancy $h^{\dagger}_{i\sigma}
h^{\dagger}_{i-\sigma}=e^{i\Phi_{i\sigma}} h^{\dagger}_{i}
h^{\dagger}_{i}e^{i\Phi_{i-\sigma}}=0$ and $h_{i\sigma}h_{i-\sigma}
=e^{-i\Phi_{i\sigma}}h_{i}h_{i} e^{-i\Phi_{i-\sigma}}=0$ are ruled
out automatically. We call this electron decoupling form (9) as the
{\it charge-spin separation fermion-spin transformation}
\cite{feng1}. We emphasize that the dressed holon
$h_{i\sigma}=e^{-i\Phi_{i\sigma}} h_{i}$ is the spinless fermion
$h_{i}$ (bare holon) incorporated the spin cloud
$e^{-i\Phi_{i\sigma}}$ (magnetic flux) \cite{dagotto2}, and is a
magnetic dressing. Therefore the 'spin degree of freedom' $1$ ($2$)
or $\uparrow$ ($\downarrow$) in the dressed holon operator
$h_{i1}=h_{i\uparrow}$ ($h_{i2}=h_{i\downarrow}$) in Eq. (9) is
strongly dependent on the spin configuration $S^{-}_{i}$
($S^{+}_{i}$). In particular, these dressed holon and spin are
invariant under the local $U(1)$ gauge transformation (8), and
therefore all physical quantities from the dressed holon and spin
also are invariant with respect to the gauge transformation (8). In
this sense, the collective modes for these dressed holon and spin
are real and can be interpreted as the physical excitations for the
charge and spin degrees of freedom \cite{laughlin}. Although in
common sense $h_{i\sigma}$ is not an real spinful fermion operator,
it behaves like a spinful fermion. This is followed from that the
spinless fermion $h_{i}$ and spin operators $S^{+}_{i}$ and
$S^{-}_{i}$ obey the anticommutation relation and Pauli spin
algebra, respectively, it is then easy to show that the spinful
fermion $h_{i\sigma}$ also obey the same anticommutation relation as
the spinless fermion $h_{i}$. In this charge-spin separation
fermion-spin representation (9), the low-energy behavior of the
$t$-$J$ model in Eq. (4) can be expressed as \cite{feng1},
\begin{eqnarray}
H&=&t\sum_{i\hat{\eta}}(h^{\dagger}_{i+\hat{\eta}\uparrow}
h_{i\uparrow}S^{+}_{i}S^{-}_{i+\hat{\eta}}
+h^{\dagger}_{i+\hat{\eta}\downarrow}h_{i\downarrow}S^{-}_{i}
S^{+}_{i+\hat{\eta}})\nonumber\\
&-&t'\sum_{i\hat{\tau}}(h^{\dagger}_{i+\hat{\tau}\uparrow}
h_{i\uparrow}S^{+}_{i}S^{-}_{i+\hat{\tau}}+
h^{\dagger}_{i+\hat{\tau}\downarrow}h_{i\downarrow}S^{-}_{i}
S^{+}_{i+\hat{\tau}}) \nonumber \\
&-&\mu_{0}\sum_{i\sigma}h^{\dagger}_{i\sigma} h_{i\sigma}+J_{{\rm
eff}}\sum_{i\hat{\eta}}{\bf S}_{i}\cdot {\bf S}_{i+\hat{\eta}},
\end{eqnarray}
with $J_{{\rm eff}}=(1-\delta)^{2}J$, and $\delta=\langle
h^{\dagger}_{i\sigma}h_{i\sigma}\rangle=\langle h^{\dagger}_{i}
h_{i}\rangle$ is the hole doping concentration. As a consequence,
the kinetic energy term in the $t$-$J$ model has been expressed as
the dressed holon-spin interaction, which dominates the essential
physics of cuprate superconductors, while the magnetic energy term
is only to form an adequate spin configuration \cite{anderson2}.
This also reflects that even the kinetic energy term in the $t$-$J$
model has strong Coulombic contribution due to the restriction of no
doubly occupancy of a given site.

Now we show that the charge-spin separation fermion-spin
transformation (9) is a natural representation for the constrained
electron under the decoupling scheme. Since the $t$-$J$ model (1) is
defined in the restricted Hilbert space without double electron
occupancy, therefore there are two ways to implement this
requirement as mentioned above: either to solve the $t$-$J$ model
(1) combined with an important single occupancy local constraint (2)
or to introduce the constrained electron operators \cite{rice1},
replacing $C_{i\sigma}$ by $\hat{C}_{i\sigma}=C_{i\sigma}(1-
n_{i-\sigma})$. In the latter case, the $t$-$J$ model also can be
expressed in terms of these constrained electron operators as
\cite{rice1},
\begin{equation}
H=-t\sum_{i\hat{\eta}\sigma}\hat{C}^{\dagger}_{i\sigma}
\hat{C}_{i+\hat{\eta}\sigma}+t'\sum_{i\hat{\tau}\sigma}
\hat{C}^{\dagger}_{i\sigma}\hat{C}_{i+\hat{\tau}\sigma}+\mu_{0}
\sum_{i\sigma} \hat{C}^{\dagger}_{i\sigma}\hat{C}_{i\sigma}
+J\sum_{i\hat{\eta}}{\bf S}_{i} \cdot {\bf S}_{i+\hat{\eta}}.
\end{equation}
In the constrained electron operator, the operators
$C^{\dagger}_{i\sigma}$ and $C_{i\sigma}$ are to be thought of as
operating within the full Hilbert space (unprojected Hilbert space),
while the constrained electron operator
$\hat{C}^{\dagger}_{i\sigma}$ ($\hat{C}_{i\sigma}$) does not create
(destroy) any doubly occupied sites, and therefore represents
physical creation (annihilation) operator acting in the restricted
Hilbert space (projected Hilbert subspace) \cite{anderson3}. In
particular, these constrained electron operators
$\hat{C}_{i\uparrow}$ and $\hat{C}_{i\downarrow}$ can be expressed
in a different form as \cite{anderson3},
\begin{subequations}
\begin{eqnarray}
\hat{C}_{i\uparrow}=C_{i\uparrow}(1-n_{i\downarrow})=C_{i\downarrow}
C^{\dagger}_{i\downarrow}C_{i\uparrow}=C_{i\downarrow}S^{-}_{i},\\
\hat{C}_{i\downarrow}=C_{i\downarrow}(1-n_{i\uparrow})=C_{i\uparrow}
C^{\dagger}_{i\uparrow}C_{i\downarrow}=C_{i\uparrow}S^{+}_{i},
\end{eqnarray}
\end{subequations}
obviously, where the spin degree of freedom $\downarrow$
($\uparrow$) in the unprojected operator $C_{i\downarrow}$
($C_{i\uparrow}$) is not a free degree of freedom, and is strongly
dependent on the spin configuration $S^{-}_{i}$ ($S^{+}_{i}$).
Furthermore, this form of the constrained electron operators (13)
also show obviously that the spin operators $S^{+}_{i}$
($S^{-}_{i}$) represents the spin degree of freedom of the
constrained electron, while the unprojected operator $C_{i\sigma}$
represents the charge degree of freedom together with some effects
of the spin configuration rearrangement due to doping, which is
exact same as the charge-spin separation fermion-spin transformation
in Eq. (9) if the constrained electron is decoupled according to its
charge and spin degrees of freedom. To see this point clearly, the
constrained electron operators $\hat{C}_{i\uparrow}$ and
$\hat{C}_{i\downarrow}$ in Eq. (13) can be rewritten in terms of a
particle-hole transformation $C_{i\sigma} \rightarrow
h^{\dagger}_{i-\sigma}$ as,
\begin{eqnarray}
\hat{C}_{i\uparrow}=h^{\dagger}_{i\uparrow}S^{-}_{i},~~~~
\hat{C}_{i\downarrow}=h^{\dagger}_{i\downarrow}S^{+}_{i},
\end{eqnarray}
i.e., the creation (annihilation) of the spin-up hole is equivalent
to the annihilation (creation) of the spin-down '{\it electron}'.
This is why although the assumption of the charge-spin separation in
Eq. (9) which underlay that discussions are too radical
\cite{anderson3}, the essential physics of the constrained electron
is kept.

Although the charge-spin separation fermion-spin transformation (9)
is a natural representation for the constrained electron (13) under
the decoupling scheme, so long as $h^{\dagger}_{i}h_{i}=1$,
$\sum_{\sigma} C^{\dagger}_{i\sigma} C_{i\sigma}=0$, no matter what
the values of $S^{+}_{i}S^{-}_{i}$ and $S^{-}_{i}S^{+}_{i}$ are,
therefore it means that a '{\it spin}' even to an empty site has
been assigned. Obviously, this defect is originated from the
decoupling approximation. It has been shown \cite{feng6} that this
defect can be cured by introducing a projection operator $P_{i}$,
i.e., the electron operator $C_{i\sigma}$ with the single occupancy
local constraint (2) can be mapped exactly using the charge-spin
separation fermion-spin transformation (9) defined with an
additional projection operator $P_{i}$. However, this projection
operator is cumbersome to handle in the many cases, and it has been
dropped in the actual calculations
\cite{feng1,feng2,feng3,feng4,guo1,guo2,lan1,lan2,feng5,feng6}. It
has been shown \cite{feng1,feng6,plakida} that such treatment leads
to errors of the order $\delta$ in counting the number of spin
states, which is negligible for small doping. Moreover, the electron
single occupancy local constraint still is exactly obeyed even in
the mean-field approximation. In particular, the essential physics
of the gauge invariant dressed holon and spin is kept \cite{feng1}.
To show this point clearly, we can map electron operator
$C_{i\sigma}$ with the electron single occupancy local constraint
(2) onto the slave-fermion formulism \cite{feng1} as $C_{i\sigma}=
h^{\dagger}_{i}b_{i\sigma}$ with the local constraint
$h^{\dagger}_{i}h_{i}+\sum_{\sigma}b^{\dagger}_{i\sigma}
b_{i\sigma}=1$. We can solve the local constraint in the
slave-fermion formulism by rewriting the boson operators
$b_{i\sigma}$ in terms of the CP$^{1}$ boson operators $a_{i\sigma}$
as $b_{i\sigma}=a_{i\sigma}\sqrt{1-h^{\dagger}_{i} h_{i}}$
supplemented by the local constraint $\sum_{\sigma}
a^{\dagger}_{i\sigma}a_{i\sigma}=1$. As mentioned above, the
CP$^{1}$ boson operators $a_{i\uparrow}$ and $a_{i\downarrow}$ with
the constraint $\sum_{\sigma} a^{\dagger}_{i\sigma}a_{i\sigma}=1$
can be identified with the spin lowering and raising operators,
respectively, defined with the additional phase factor, therefore
the projection operator is approximately related to the dressed
holon number operator by $P_{i}\sim\sqrt{1-h^{\dagger}_{i\sigma}
h_{i\sigma}}=\sqrt{1-h^{\dagger}_{i}h_{i}}$, and its main role is to
remove the spurious spin when there is a holon at a given site $i$
\cite{feng1,feng6,plakida}. All these are also why the theoretical
results
\cite{feng1,feng2,feng3,feng4,guo1,guo2,lan1,lan2,feng5,feng6}
obtained from the $t$-$J$ model (11) based on the charge-spin
separation fermion-spin theory (9) are in qualitative agreement with
the numerical simulations and experimental observation on cuprate
superconductors.

\section{Electronic structure of the single layer cuprate
superconductors}

It has been known from the experimental measurements \cite{tanaka}
that the maximum superconducting transition temperature varies
strongly among different cuprate superconducting compounds. This
material dependence has been reduced to the crystal structure, since
one or more CuO$_{2}$ planes per unit cell separated by insulating
layers are found for different families of cuprate superconductors,
then the crystal structure determines the hybridizations of the Cu
orbital with those of other elements, resulting in different values
of hopping integrals within the CuO$_{2}$ plane and between the
CuO$_{2}$ planes in the unit cell for distinct compounds. This leads
to that some subtle structures in the electron spectrum for
different families of cuprate superconductors are observed
\cite{shen,campuzano}. In this section, we firstly discuss the
electronic structure of the single layer cuprate superconductors
within the kinetic energy driven superconductivity
\cite{feng3,feng4,guo1,guo2}.

\subsection{Electronic structure of the single layer cuprate
superconductors in the superconducting state}

As in the conventional superconductors \cite{bcs}, the
superconducting state in cuprate superconductors is also
characterized by the electron Cooper pairs \cite{tsuei}.
Furthermore, the ARPES measurements \cite{shen2} show that in the
real space the superconducting gap function and pairing force have a
range of one lattice spacing. In this case, the order parameter for
the electron Cooper pair can be expressed in the charge-spin
separation fermion-spin representation as,
\begin{eqnarray}
\Delta=\langle
C^{\dagger}_{i\uparrow}C^{\dagger}_{i+\hat{\eta}\downarrow}-
C^{\dagger}_{i\downarrow}C^{\dagger}_{i+\hat{\eta}\uparrow}\rangle
=\langle h_{i\uparrow}h_{i+\hat{\eta}\downarrow}S^{+}_{i}
S^{-}_{i+\hat{\eta}}-h_{i\downarrow}h_{i+\hat{\eta}\uparrow}
S^{-}_{i}S^{+}_{i+\hat{\eta}}\rangle .~~~
\end{eqnarray}
In the doped regime without the antiferromagnetic long-range order,
spins form a disordered spin liquid state, where the spin
correlation function $\langle
S^{+}_{i}S^{-}_{i+\hat{\eta}}\rangle=\langle S^{-}_{i}
S^{+}_{i+\hat{\eta}}\rangle$, then the order parameter for the
electron Cooper pair in Eq. (15) can be written as,
\begin{eqnarray}
\Delta=-\langle S^{+}_{i}S^{-}_{i+\hat{\eta}}\rangle \Delta_{h},
\end{eqnarray}
with the dressed holon pairing order parameter,
\begin{eqnarray}\label{E9}
\Delta_{h}=\langle h_{i+\hat{\eta}\downarrow}h_{i\uparrow}
-h_{i+\hat{\eta}\uparrow}h_{i\downarrow}\rangle,
\end{eqnarray}
which shows that the superconducting order parameter is related to
the dressed holon pairing amplitude, and is proportional to the
number of doped holes, and not to the number of electrons. However,
in the extreme low doped regime with the antiferromagnetic
long-range order, where the spin correlation function $\langle
S^{+}_{i}S^{-}_{i+\hat{\eta}}\rangle\neq\langle S^{-}_{i}
S^{+}_{i+\hat{\eta}}\rangle$, then the conduct is disrupted by the
antiferromagnetic long-range order, and therefore there is no mixing
of superconductivity and antiferromagnetic long-range order
\cite{bozovic}. Therefore in the following discussions, we only
focus on the doped regime without the antiferromagnetic long-range
order.

Within the mean-field approximation, the $t$-$J$ model in Eq. (11)
can be decoupled as \cite{feng2},
\begin{subequations}
\begin{eqnarray}
H_{{\rm MFA}}&=&H_{t}+H_{J}-2NZt\phi_{1}\chi_{1}
+2NZt'\phi_{2}\chi_{2}, \\
H_{t}&=&\chi_{1}t\sum_{i\hat{\eta}\sigma}
h^{\dagger}_{i+\hat{\eta}\sigma} h_{i\sigma}
-\chi_{2}t'\sum_{i\hat{\tau}\sigma} h^{\dagger}_{i+\hat{\tau}\sigma}
h_{i\sigma}-\mu_{0}\sum_{i\sigma}
h^{\dagger}_{i\sigma}h_{i\sigma}, ~~~~~\\
H_{J}&=& {1\over 2}J_{{\rm eff}}\epsilon\sum_{i\hat{\eta}}
(S^{+}_{i}S^{-}_{i+\hat{\eta}}+S^{-}_{i}S^{+}_{i+\hat{\eta}})
+J_{{\rm eff}}\sum_{i\hat{\eta}}S^{z}_{i}S^{z}_{i+\hat{\eta}}
\nonumber \\
&-&t'\phi_{2}\sum_{i\hat{\tau}}(S^{+}_{i}S^{-}_{i+\hat{\tau}}+
S^{-}_{i}S^{+}_{i+\hat{\tau}}),
\end{eqnarray}
\end{subequations}
with the anisotropic parameter $\epsilon=1+2t\phi_{1}/J_{{\rm eff}
}$, the dressed holon's particle-hole parameters $\phi_{1}=\langle
h^{\dagger}_{i\sigma} h_{i+\hat{\eta}\sigma}\rangle$ and
$\phi_{2}=\langle h^{\dagger}_{i\sigma}h_{i+\hat{\tau}\sigma}
\rangle$, the spin correlation functions $\chi_{1}=\langle S_{i}^{+}
S_{i+\hat{\eta}}^{-}\rangle$ and $\chi_{2}=\langle S_{i}^{+}
S_{i+\hat{\tau}}^{-}\rangle$, $Z$ is the number of the nearest
neighbor or second-nearest neighbor sites, and $N$ is the number of
sites. Before the discussions of the electronic structure, we now
define firstly the dressed holon normal and anomalous Green's
functions as,
\begin{subequations}
\begin{eqnarray}
g(i-j,t-t')&=&\langle\langle
h_{i\sigma}(t);h^{\dagger}_{j\sigma}(t')
\rangle\rangle ,\\
\Im (i-j,t-t')&=&\langle\langle h_{i\downarrow}(t);h_{j\uparrow}(t')
\rangle\rangle ,\\
\Im^{\dagger}(i-j,t-t')&=&\langle\langle h^{\dagger}_{i\uparrow}(t);
h^{\dagger}_{j\downarrow}(t')\rangle\rangle ,
\end{eqnarray}
\end{subequations}
respectively, and the spin Green's functions as,
\begin{subequations}
\begin{eqnarray}
D(i-j,t-t')&=&\langle\langle S^{+}_{i}(t);S^{-}_{j}(t')\rangle
\rangle,\\
D_{z}(i-j,t-t')&=&\langle\langle S^{z}_{i}(t);S^{z}_{j}(t')\rangle
\rangle ,
\end{eqnarray}
\end{subequations}
where $\langle \ldots \rangle$ is an average over the ensemble. In
the mean-field level, the spin system is an anisotropic away from
the half-filling \cite{feng2}, therefore we have defined the two
spin Green's function $D(i-j,t-t')$ and $D_{z}(i-j,t-t')$ to
describe the spin propagations. In the doped regime without the
antiferromagnetic long-range order, i.e., $\langle S^{z}_{i}\rangle
=0$, the mean-field theory of the $t$-$J$ model based on the
charge-spin separation fermion-spin theory has been developed
\cite{feng7} within the Kondo-Yamaji decoupling scheme \cite{kondo},
which is a stage one-step further than the Tyablikov's decoupling
scheme. Following their discussions \cite{feng7}, we can obtain the
the mean-field dressed holon normal Green's function as,
\begin{eqnarray}
g^{(0)}({\bf k},\omega)={1\over \omega-\xi_{{\bf k}}},
\end{eqnarray}
and mean-field spin Green's functions as,
\begin{subequations}
\begin{eqnarray}
D^{(0)}({\bf k},\omega)&=&{B_{{\bf k}}\over 2\omega_{{\bf k}}}\left
({1\over \omega-\omega_{{\bf k}}}-{1\over \omega+\omega_{{\bf k}}}
\right ),
\\
D^{(0)}_{z}({\bf k},\omega)&=& {B_{z}({\bf k})\over 2\omega_{z}
({\bf k})}\left ({1\over\omega-\omega_{z}({\bf k})}-{1\over \omega
+\omega_{z} ({\bf k})} \right ),
\end{eqnarray}
\end{subequations}
where $B_{{\bf k}}=2\lambda_{1}(A_{1}\gamma_{{\bf k}}-A_{2})-
\lambda_{2}(2\chi^{z}_{2} \gamma_{{\bf k}}'- \chi_{2})$, $B_{z}
({\bf k})=\epsilon\chi_{1}\lambda_{1}(\gamma_{{\bf k}}-1)-\chi_{2}
\lambda_{2}(\gamma_{{\bf k}}'-1)$, $\lambda_{1}=2ZJ_{eff}$,
$\lambda_{2}=4Z\phi_{2}t'$, $\gamma_{{\bf k}}=(1/Z)\sum_{\hat{\eta}}
e^{i{\bf k}\cdot\hat{\eta}}$, $\gamma_{{\bf k}}'=(1/Z)
\sum_{\hat{\tau}}e^{i{\bf k} \cdot\hat{\tau}}$, $A_{1}=\epsilon
\chi^{z}_{1}+\chi_{1}/2$, $A_{2}=\chi^{z}_{1}+\epsilon\chi_{1}/2$,
the spin correlation functions $\chi^{z}_{1}=\langle S_{i}^{z}
S_{i+\hat{\eta}}^{z}\rangle$ and $\chi^{z}_{2}=\langle S_{i}^{z}
S_{i+\hat{\tau}}^{z}\rangle$, the mean-field dressed holon
excitation spectrum,
\begin{eqnarray}
\xi_{{\bf k}}=\epsilon_{{\bf k}}-\mu_{0},
\end{eqnarray}
with $\epsilon_{{\bf k}}=Zt\chi_{1}\gamma_{{\bf k}}-Zt'\chi_{2}
\gamma_{{\bf k} }'$, and the mean-field spin excitation spectra,
\begin{subequations}
\begin{eqnarray}
\omega^{2}_{{\bf k}}&=& \lambda_{1}^{2}[(A_{4}-\alpha\epsilon
\chi^{z}_{1}\gamma_{{\bf k}}-{1\over 2Z}\alpha\epsilon\chi_{1})
(1-\epsilon\gamma_{{\bf k}})\nonumber \\
&+&{1\over 2}\epsilon(A_{3}-{1\over 2} \alpha\chi^{z}_{1}-\alpha
\chi_{1}\gamma_{{\bf k}})(\epsilon-\gamma_{{\bf k}})]
\nonumber \\
&+&\lambda_{2}^{2}[\alpha(\chi^{z}_{2}\gamma_{{\bf k}}'-{3\over
2Z}\chi_{2})\gamma_{{\bf k}}'+{1\over 2}(A_{5}-{1\over 2}
\alpha \chi^{z}_{2})]\nonumber \\
&+& \lambda_{1}\lambda_{2}[\alpha\chi^{z}_{1}(1-\epsilon
\gamma_{{\bf k}})\gamma_{{\bf k}}'+{1\over 2}\alpha(\chi_{1}
\gamma_{{\bf k}}'-C_{3})(\epsilon-\gamma_{{\bf k}})\nonumber\\
&+&\alpha \gamma_{{\bf k}}'(C^{z}_{3}-\epsilon \chi^{z}_{2}
\gamma_{{\bf k}})-{1\over 2}\alpha\epsilon(C_{3}- \chi_{2}
\gamma_{{\bf k}})], \\
\omega^{2}_{z}({\bf k})&=&\epsilon\lambda^{2}_{1}(\epsilon
A_{3}-{1\over Z}\alpha\chi_{1}-\alpha\chi_{1}\gamma_{{\bf k}})
(1-\gamma_{{\bf k}})+\lambda^{2}_{2}A_{5}(1-\gamma_{{\bf k}}')
\nonumber\\
&+&\lambda_{1}\lambda_{2}[\alpha\epsilon C_{3}(\gamma_{{\bf k}}+
\gamma_{{\bf k}}'-2)+\alpha\chi_{2} \gamma_{{\bf k}}(1 -
\gamma_{{\bf k}}')],
\end{eqnarray}
\end{subequations}
with $A_{3}=\alpha C_{1}+(1-\alpha)/(2Z)$, $A_{4}=\alpha C^{z}_{1}
+(1-\alpha)/(4Z)$, $A_{5}=\alpha C_{2}+(1-\alpha)/(2Z)$, and the
spin correlation functions
$C_{1}=(1/Z^{2})\sum_{\hat{\eta},\hat{\eta'}}\langle
S_{i+\hat{\eta}}^{+}S_{i+\hat{\eta'}}^{-}\rangle$,
$C^{z}_{1}=(1/Z^{2})\sum_{\hat{\eta},\hat{\eta'}}\langle
S_{i+\hat{\eta}}^{z}S_{i+\hat{\eta'}}^{z}\rangle$,
$C_{2}=(1/Z^{2})\sum_{\hat{\tau},\hat{\tau'}}\langle
S_{i+\hat{\tau}}^{+}S_{i+\hat{\tau'}}^{-}\rangle$, and $C_{3}=(1/Z)$
$\sum_{\hat{\tau}}\langle S_{i+\hat{\eta}}^{+}
S_{i+\hat{\tau}}^{-}\rangle$, $C^{z}_{3}=(1/Z)
\sum_{\hat{\tau}}\langle S_{i+\hat{\eta}}^{z}
S_{i+\hat{\tau}}^{z}\rangle$. In order to satisfy the sum rule of
the correlation function $\langle S^{+}_{i}S^{-}_{i}\rangle=1/2$ in
the case without the antiferromagnetic long-range order, the
important decoupling parameter $\alpha$ has been introduced in the
mean-field calculation \cite{feng7,kondo}, which can be regarded as
the vertex correction.

Within the charge-spin separation fermion-spin theory, we have
recently developed the kinetic energy driven mechanism
\cite{feng2,feng3}, where we have shown that the dressed holon-spin
interaction in the $t$-$J$ model (11) is quite strong, and can
induce the dressed holon pairing state (then the electron Cooper
pairing state) by exchanging spin excitations in the higher power of
the doping concentration $\delta$. Following our previous
discussions \cite{feng2,feng3}, the self-consistent equations that
satisfied by the full dressed holon normal and anomalous Green's
functions are obtained as,
\begin{subequations}
\begin{eqnarray}
g({\bf k},\omega)&=&g^{(0)}({\bf k},\omega)+g^{(0)}({\bf k},\omega)
[\Sigma^{(h)}_{1}({\bf k},\omega)g({\bf k},\omega)\nonumber \\
&-&\Sigma^{(h)}_{2}(-{\bf k},-\omega)\Im^{\dagger}({\bf k},\omega)],
~~~~~\\
\Im^{\dagger}({\bf k},\omega)&=&g^{(0)}(-{\bf k},-\omega)
[\Sigma^{(h)}_{1}(-{\bf k},-\omega)\Im^{\dagger}(-{\bf k},-\omega)
\nonumber \\
&+&\Sigma^{(h)}_{2}(-{\bf k},-\omega)g({\bf k},\omega)],~~~~~
\end{eqnarray}
\end{subequations}
respectively, where the corresponding dressed holon self-energy
functions are given by \cite{feng3},
\begin{subequations}
\begin{eqnarray}
\Sigma^{(h)}_{1}({\bf k},i\omega_{n})&=&{1\over N^{2}}\sum_{{\bf p,
p'}}\Lambda^{2}_{{\bf p}+{\bf p}'+{\bf k}}\nonumber \\
&\times&{1\over \beta} \sum_{ip_{m}}g({\bf p}+{\bf k},ip_{m}
+i\omega_{n})\Pi({\bf p},{\bf p}',ip_{m}),~~~~\\
\Sigma^{(h)}_{2}({\bf k},i\omega_{n})&=&{1\over N^{2}}\sum_{{\bf p,
p'}}\Lambda^{2}_{{\bf p}+{\bf p}'+{\bf k}}\nonumber \\
&\times&{1\over \beta} \sum_{ip_{m}}\Im (-{\bf p}-{\bf k},-ip_{m}
-i\omega_{n})\Pi({\bf p}, {\bf p}',ip_{m}),~~~~
\end{eqnarray}
\end{subequations}
with $\Lambda_{{\bf k}}=Zt\gamma_{{\bf k}}-Zt'\gamma_{{\bf k}}'$,
and the spin bubble,
\begin{eqnarray}
\Pi({\bf p},{\bf p}',ip_{m})={1\over\beta}\sum_{ip'_{m}}D^{(0)}
({\bf p}',ip_{m}')D^{(0)}({\bf p}+{\bf p}',ip'_{m}+ip_{m}).
\end{eqnarray}
In the above calculations of the dressed holon self-energies, the
spin part has been limited to the mean-field level, i.e., the full
spin Green's function $D({\bf k},\omega)$ in Eq. (27) has been
replaced by the mean-field spin Green's function $D^{(0)}({\bf k},
\omega)$ in Eq. (22a), since the theoretical results of the normal
state charge transport obtained at this level are consistent with
the experimental data of cuprate superconductors in the normal state
\cite{feng1,feng8}.

In the self-consistent equations (25), since both doping and
temperature dependence of the pairing force and dressed holon gap
function have been incorporated into the self-energy function
$\Sigma^{(h)}_{2}({\bf k},\omega)$, then the self-energy function
$\Sigma^{(h)}_{2}({\bf k},\omega)$ describes the effective dressed
holon pair gap function, while the self-energy function
$\Sigma^{(h)}_{1}({\bf k},\omega)$ renormalizes the mean-field
dressed holon spectrum $\xi_{{\bf k}}$ in Eq. (23), and therefore it
describes the dressed holon quasiparticle coherence. Furthermore,
the self-energy function $\Sigma^{(h)}_{2}({\bf k},\omega)$ is an
even function of $\omega$, while the other self-energy function
$\Sigma^{(h)}_{1}({\bf k},\omega)$ is not. For the convenience, the
self-energy function $\Sigma^{(h)}_{1}({\bf k},\omega)$ can be
broken up into its symmetric and antisymmetric parts as,
$\Sigma^{(h)}_{1}({\bf k}, \omega)=\Sigma^{(h)}_{1e}({\bf k},
\omega)+\omega\Sigma^{(h)}_{1o}({\bf k},\omega)$, then both
$\Sigma^{(h)}_{1e}({\bf k},\omega)$ and $\Sigma^{(h)}_{1o}({\bf k},
\omega)$ are even functions of $\omega$. In this case, the dressed
holon quasiparticle coherent weight can be defined as,
\begin{eqnarray}
{1\over Z_{hF}({\bf k},\omega)}=1-\Sigma^{(h)}_{1o}({\bf k}
,\omega),
\end{eqnarray}
then the dressed holon normal and anomalous Green's functions in
Eqs. (25) can be rewritten as \cite{feng2,feng3},
\begin{subequations}
\begin{eqnarray}
g({\bf k},\omega)&=&{\omega Z^{-1}_{hF}({\bf k},\omega)+\xi_{{\bf k}
}+\Sigma^{(h)}_{1e}({\bf k},\omega)\over [\omega Z^{-1}_{hF}({\bf k}
,\omega)]^{2}-[\xi_{{\bf k}}+\Sigma^{(h)}_{1e}({\bf k},\omega)]^{2}-
[\Sigma^{(h)}_{2}({\bf k},\omega)]^{2}} ,~~~~\\
\Im^{\dagger}({\bf k},\omega)&=&{-\Sigma^{(h)}_{2}({\bf k},\omega)
\over [\omega Z^{-1}_{hF}({\bf k},\omega)]^{2}-[\xi_{{\bf k}}+
\Sigma^{(h)}_{1e}({\bf k},\omega)]^{2}- [\Sigma^{(h)}_{2}({\bf k},
\omega)]^{2}}.~~~~
\end{eqnarray}
\end{subequations}
Recently, a universal high energy anomaly in the electronic
structure has been reported in cuprate superconductors by using the
ARPES \cite{graf}, where they found the dispersion anomalies marked
by two distinctive high energy scales. From Eqs. (29), we think that
the energy dependence of the effective dressed holon pair gap
parameter and quasiparticle coherent weight should be considered in
the discussions of this high energy anomaly in the electronic
structure. However, in this paper, we only focus on the low-energy
electronic structure of cuprate superconductors. In this case, the
effective dressed holon pair gap function and quasiparticle coherent
weight can be discussed in the static limit, i.e., $\bar{\Delta}_{h}
({\bf k})=\Sigma^{(h)}_{2}({\bf k},\omega)\mid_{\omega=0}$,
$Z^{-1}_{hF}({\bf k})=1-\Sigma^{(h)}_{1o} ({\bf k},\omega)
\mid_{\omega=0}$, and $\Sigma^{(h)}_{1e} ({\bf k})=
\Sigma^{(h)}_{1e}({\bf k},\omega)\mid_{\omega=0}$, then the dressed
holon normal and anomalous Green's functions in Eqs. (29) can be
expressed explicitly as,
\begin{subequations}
\begin{eqnarray}
g({\bf k},\omega)&=&Z_{hF}({\bf k})\left ({U^{2}_{h{\bf k}}\over
\omega- E_{h{\bf k}}}+{V^{2}_{h{\bf k}} \over \omega+E_{h{\bf k}}}
\right ),\\
\Im^{\dagger}({\bf k},\omega)&=&-Z_{hF}({\bf k}){\bar{\Delta}_{hZ}
({\bf k})\over 2E_{h{\bf k}}}\left ({1\over \omega-E_{h{\bf k}}}-
{1\over \omega+ E_{h{\bf k}}}\right ),
\end{eqnarray}
\end{subequations}
with the dressed holon quasiparticle coherence factors,
\begin{subequations}
\begin{eqnarray}
U^{2}_{h{\bf k}}={1\over 2}\left (1+{\bar{\xi_{{\bf k}}}\over
E_{h{\bf k}}}\right ),\\
V^{2}_{h{\bf k}}={1\over 2}\left (1-{\bar{\xi_{{\bf k}}}\over
E_{h{\bf k}}}\right ),
\end{eqnarray}
\end{subequations}
and
\begin{subequations}
\begin{eqnarray}
E_{h{\bf k}}&=&\sqrt{\bar{\xi^{2}_{{\bf k}}}+\mid\bar{\Delta}_{hZ}
({\bf k}) \mid^{2}}, \\
\bar{\xi_{{\bf k}}}&=&Z_{hF}\epsilon_{{\bf k}}-\mu\\
\bar{\Delta}_{hZ}({\bf k})&=&Z_{hF}({\bf k})\bar{\Delta}_{h}({\bf k}
),\\
\mu&=&Z_{hF}(\mu_{0}-\Sigma^{(h)}_{1e}),
\end{eqnarray}
\end{subequations}
are the dressed holon quasiparticle spectrum, the renormalized
dressed holon excitation spectrum, the renormalized dressed holon
pair gap function, and the renormalized chemical potential,
respectively. Obviously, the dressed holon pairing state is
described by the simple Bardeen-Cooper-Schrieffer (BCS) formalism
\cite{bcs}, although the pairing mechanism is driven by the kinetic
energy by exchanging spin excitations. In particular, this dressed
holon quasiparticle is the excitation of a single dressed holon
'adorned' with the attractive interaction between paired dressed
holons, while the dressed holon quasiparticle coherent weight
$Z_{hF}({\bf k})$ reduces the dressed holon (then electron)
quasiparticle bandwidth, and therefore the energy scale of the
electron quasiparticle band is controlled by the magnetic
interaction $J$. Although $Z_{hF}({\bf k} )$ and
$\Sigma^{(h)}_{1e}({\bf k})$ still are a function of ${\bf k} $, the
wave vector dependence may be unimportant \cite{eliashberg}. It has
been shown from the ARPES experiments \cite{shen,campuzano,ding}
that in the superconducting state of cuprate superconductors, the
lowest energy states are located at the $[\pi,0]$ point, which
indicates that the majority contribution for the electron spectrum
comes from the $[\pi,0]$ point. In this case, the wave vector ${\bf
k}$ in $Z_{hF}({\bf k})$ and $\Sigma^{(h)}_{1e}({\bf k})$ can be
chosen as $Z^{-1}_{hF}=1- \Sigma^{(h)}_{1o}({\bf k})\mid_{{\bf k}
=[\pi,0]}$ and $\Sigma^{(h)}_{1e}=\Sigma^{(h)}_{1e}({\bf k})
\mid_{{\bf k}= [\pi,0]}$. In particular, we \cite{feng3,feng4} have
shown within the kinetic energy driven superconducting mechanism
that the electron Cooper pairs have a dominated d-wave symmetry over
a wide range of the doping concentration, around the optimal doping.
In this case, the effective dressed holon pair gap function
$\bar{\Delta}_{hZ}({\bf k})$ can be expressed explicitly as the
d-wave form, $\bar{\Delta}_{hZ}({\bf k})=\bar{\Delta}_{hZ}
\gamma^{(d)}_{{\bf k}}$ with $\gamma^{(d)}_{{\bf k}}=({\rm cos}
k_{x}-{\rm cos}k_{y})/2$, then the quasiparticle coherent weight and
dressed holon effective gap parameter in Eqs. (26a) and (26b)
satisfy the following two equations,
\begin{subequations}
\begin{eqnarray}
1&=&{1\over N^{3}}\sum_{{\bf k,q,p}}\Lambda^{2}_{{\bf q}+{\bf k}}
\gamma^{(a)}_{{\bf k-p+q}} \gamma^{(a)}_{{\bf k}}{Z^{2}_{hF}\over
E_{h{\bf k}}}{B_{{\bf q}}B_{{\bf p}}\over\omega_{{\bf q}}
\omega_{{\bf p}}}\left({F^{(1)}_{1}({\bf k,q,p})\over(\omega_{{\bf p
}}-\omega_{{\bf q}})^{2}-E^{2}_{h{\bf k}}}\right . \nonumber\\
&-&\left . {F^{(2)}_{1}({\bf k,q,p}) \over (\omega_{{\bf p}}+
\omega_{{\bf q}})^{2}- E^{2}_{h{\bf k}}}\right ) , \\
{1\over Z_{hF}}&=& 1+{1\over N^{2}}\sum_{{\bf q,p}}
\Lambda^{2}_{{\bf p}+{\bf k}_{A}}Z_{hF}{B_{{\bf q}} B_{{\bf p}}\over
4\omega_{{\bf q}}\omega_{{\bf p}}}\left({F^{(1)}_{2}({\bf q,p})\over
(\omega_{{\bf p}}-\omega_{{\bf q}}-E_{h{\bf p-q+k}_{A}})^{2}} \right
.\nonumber \\
&+& {F^{(2)}_{2}({\bf q,p})\over(\omega_{{\bf p}}-\omega_{{\bf q}}+
E_{h{\bf p-q+k}_{A}})^{2}}+{F^{(3)}_{2}({\bf q,p})
\over(\omega_{{\bf p}}+\omega_{{\bf q}}-E_{h{\bf p-q+k}_{A}})^{2}}
\nonumber \\
&+&\left . {F^{(4)}_{2}({\bf q,p}) \over(\omega_{{\bf
p}}+\omega_{{\bf q}}+ E_{h{\bf p-q+k}_{A}})^{2}} \right ) ,~~~~~
\end{eqnarray}
\end{subequations}
respectively, where ${\bf k}_{A}=[\pi,0]$, and
\begin{subequations}
\begin{eqnarray}
F^{(1)}_{1}({\bf k,q,p})&=&(\omega_{{\bf p}}-\omega_{{\bf q}})
[n_{B}(\omega_{{\bf q}})-n_{B}(\omega_{{\bf p}})][1-2 n_{F}
(E_{h{\bf k}})] \nonumber \\
&+&E_{h{\bf k}}[n_{B}(\omega_{{\bf p}})n_{B} (-\omega_{{\bf q}})
+n_{B}(\omega_{{\bf q}})n_{B}(-\omega_{{\bf p}})], \\
F^{(2)}_{1}({\bf k,q,p})&=&(\omega_{{\bf p }}+\omega_{{\bf q}})
[n_{B}(-\omega_{{\bf p}})-n_{B}(\omega_{{\bf q}})][1-2 n_{F}
(E_{h{\bf k}})] \nonumber \\
&+&E_{h{\bf k}}[n_{B}(\omega_{{\bf p}}) n_{B} (\omega_{{\bf q}})+
n_{B}(-\omega_{{\bf p}})n_{B}(- \omega_{{\bf q}})],\\
F^{(1)}_{2}({\bf q,p})&=&n_{F}(E_{h{\bf p-q+k}_{A}})[n_{B}
(\omega_{{\bf q}})-n_{B}(\omega_{{\bf p}})]\nonumber \\
&-& n_{B}(\omega_{{\bf p}})n_{B}(-\omega_{{\bf q}}),~~~~~\\
F^{(2)}_{2}({\bf q,p})&=&n_{F}(E_{h{\bf p-q+k}_{A}})[n_{B}
(\omega_{{\bf p}})-n_{B} (\omega_{{\bf q}})]\nonumber \\
&-&n_{B}(\omega_{{\bf q}}) n_{B} (-\omega_{{\bf p}}),~~~~~~
\end{eqnarray}
\begin{eqnarray}
F^{(3)}_{2}({\bf q,p})&=& n_{F}(E_{h{\bf p-q+k}_{A}}) [n_{B}
(\omega_{{\bf q}})-n_{B}(-\omega_{{\bf p}})]\nonumber \\
&+&n_{B}(\omega_{{\bf p}})n_{B}(\omega_{{\bf q}}),~~~~~~\\
F^{(4)}_{2}({\bf q,p})&=&n_{F}(E_{h{\bf p-q+k}_{A}})
[n_{B}(-\omega_{{\bf q}})-n_{B} (\omega_{{\bf p}})]\nonumber\\
&+& n_{B}(-\omega_{{\bf p}})n_{B}(-\omega_{{\bf q}}). ~~~~~
\end{eqnarray}
\end{subequations}
These two equations (33a) and (33b) must be solved simultaneously
with other self-consistent equations \cite{feng2,feng3},
\begin{subequations}
\begin{eqnarray}
\phi_{1}&=&{1\over 2N}\sum_{{\bf k}}\gamma_{{\bf k}}Z_{hF}\left
(1-{\bar{\xi_{{\bf k}}}\over E_{h{\bf k}}}{\rm tanh} [{1\over 2}
\beta E_{h{\bf k}}]\right ),\\
\phi_{2}&=&{1\over 2N}\sum_{{\bf k}}\gamma_{{\bf k}}'Z_{hF}\left
(1-{\bar{\xi_{{\bf k}}}\over E_{h{\bf k}}}{\rm tanh}
[{1\over 2}\beta E_{h{\bf k}}]\right ),\\
\delta &=& {1\over 2N}\sum_{{\bf k}}Z_{hF}\left (1-{\bar{\xi_{{\bf
k}}} \over E_{h{\bf k}}}{\rm tanh}[{1\over 2}\beta E_{h{\bf k}}]
\right ),\\
\chi_{1}&=&{1\over N}\sum_{{\bf k}}\gamma_{{\bf k}} {B_{{\bf k}}
\over 2\omega_{{\bf k}}}{\rm coth} [{1\over 2}\beta\omega_{{\bf
k}}], \\
\chi_{2}&=&{1\over N}\sum_{{\bf k}}\gamma_{{\bf k}}'{B_{{\bf
k}}\over 2\omega_{{\bf k}}}{\rm coth} [{1\over 2}\beta\omega_{{\bf
k}}],\\
C_{1}&=&{1\over N}\sum_{{\bf k}}\gamma^{2}_{{\bf k}} {B_{{\bf
k}}\over 2\omega_{{\bf k}}}{\rm coth}
[{1\over 2}\beta\omega_{{\bf k}}],\\
C_{2}&=&{1\over N}\sum_{{\bf k}}\gamma'^{2}_{{\bf k}} {B_{{\bf
k}}\over 2\omega_{{\bf k}}}{\rm coth}  [{1\over 2}
\beta\omega_{{\bf k}}], \\
C_{3}&=&{1\over N}\sum_{{\bf k}}\gamma_{{\bf k}}\gamma_{{\bf k}}'
{B_{{\bf k}}\over 2\omega_{{\bf k}}}{\rm coth} [{1\over 2}
\beta\omega_{{\bf k}}],\\
{1\over 2} &=&{1\over N}\sum_{{\bf k}}{B_{{\bf k}} \over
2\omega_{{\bf k}}}{\rm coth} [{1\over 2}\beta\omega_{{\bf k}}],\\
\chi^{z}_{1}&=&{1\over N}\sum_{{\bf k}}\gamma_{{\bf k}} {B_{z}({\bf
k})\over 2\omega_{z}({\bf k})}{\rm coth}
[{1\over 2}\beta\omega_{z}({\bf k})],\\
\chi^{z}_{2}&=& {1\over N}\sum_{{\bf k}}\gamma_{{\bf k}}'
{B_{z}({\bf k})\over 2\omega_{z}({\bf k})}{\rm coth} [{1\over 2}
\beta\omega_{z}({\bf k})], \\
C^{z}_{1}&= &{1\over N}\sum_{{\bf k}}\gamma^{2}_{{\bf k}}
{B_{z}({\bf k})\over 2\omega_{z}({\bf k})}{\rm coth} [{1\over 2}
\beta\omega_{z}({\bf k})], \\
C^{z}_{3}&=&{1\over N}\sum_{{\bf k}}\gamma_{{\bf k}}\gamma_{{\bf
k}}'{B_{z}({\bf k})\over 2\omega_{z}({\bf k})}{\rm coth} [{1\over
2}\beta\omega_{z}({\bf k})],
\end{eqnarray}
\end{subequations}
then all the order parameters, decoupling parameter $\alpha$, and
chemical potential $\mu$ are determined by the self-consistent
calculation \cite{feng2,feng3,feng4}. With above discussions, the
dressed holon pair gap function can be obtained in terms of the
dressed holon anomalous Green's function (30b) as,
\begin{eqnarray}
\Delta_{h}({\bf k})=-{1\over \beta}\sum_{i\omega_{n}}
\Im^{\dagger}({\bf k},i\omega_{n})={1\over 2}Z_{hF}
{\bar{\Delta}_{hZ}({\bf k})\over E_{h{\bf k}}}{\rm tanh}[{1\over 2}
\beta E_{h{\bf k}}],
\end{eqnarray}
then the dressed holon pair order parameter in Eq. (17) can be
evaluated explicitly from this dressed holon pair gap function as,
\begin{eqnarray}
\Delta_{h}={2\over N}\sum_{{\bf k}}[\gamma^{(d)}_{{\bf k}}]^{2}
{Z_{hF}\bar{\Delta}^{(a)}_{hZ}\over E_{h{\bf k}}}{\rm tanh}[{1\over
2}\beta E_{h{\bf k}}].
\end{eqnarray}

In our previous discussions \cite{feng2,feng3,feng4}, we have shown
that the dressed holon pairing state originating from the kinetic
energy term by exchanging spin excitations also leads to form the
electron Cooper pairing state. In this case, the electron normal and
anomalous Green's functions $G(i-j,t-t') =\langle\langle
C_{i\sigma}(t);C^{\dagger}_{j\sigma}(t')\rangle \rangle$ and
$\Gamma^{\dagger}(i-j,t-t')=\langle\langle C^{\dagger}_{i\uparrow}
(t);C^{\dagger}_{j\downarrow}(t')\rangle \rangle$ are the
convolutions of the spin Green's function and dressed holon normal
and anomalous Green's functions, and can be obtained in terms of the
mean-field spin Green's function (22a) and dressed holon normal and
anomalous Green's functions (30a) and (30b) as,
\begin{subequations}
\begin{eqnarray}
G({\bf k},\omega)&=&{1\over N}\sum_{{\bf p}}Z_{hF}{B_{{\bf p}}\over
4\omega_{{\bf p}}}\left\{ {\rm coth}[{1\over 2}\beta\omega_{{\bf p}}
]\left ( {U^{2}_{h{\bf p+k}}\over\omega+E_{h{\bf p+k}}-
\omega_{{\bf p}}}\right . \right .\nonumber \\
&+&\left . {U^{2}_{h{\bf p+k}}\over\omega+E_{h{\bf p+k}}+
\omega_{{\bf p}}}+{V^{2}_{h{\bf p+k}}\over\omega-E_{h{\bf p+k}}+
\omega_{{\bf p}}}+{V^{2}_{h{\bf p+k}}\over\omega-E_{h{\bf p+k}}
-\omega_{{\bf p}}}\right )\nonumber\\
&+&{\rm tanh}[{1\over 2}\beta E_{h{\bf p+k} }]\left ({U^{2}_{h{\bf
p+k}}\over\omega+E_{h{\bf p+k}}+\omega_{{\bf p}}}-{U^{2}_{h{\bf p+
k}}\over\omega+E_{h{\bf p+k}}-\omega_{{\bf p}}}\right .\nonumber\\
&+&\left . \left . {V^{2}_{h{\bf p+k}}\over\omega-E_{h{\bf p+k}}-
\omega_{{\bf p}}}-{V^{2}_{h{\bf p+k}}\over\omega-E_{h{\bf p+k}}
+\omega_{{\bf p}}} \right ) \right \} ~~~,\\
\Gamma^{\dagger}({\bf k},\omega)&=&{1\over N}\sum_{{\bf p}}
Z_{hF}{\bar{\Delta}_{hZ}({\bf p+k})\over 2E_{h{\bf p+k}}}{B_{{\bf p
}}\over 4\omega_{{\bf p}}}\left \{{\rm coth}[{1\over
2}\beta\omega_{{\bf p}}]\left ({1\over\omega-E_{h{\bf p+k}}
-\omega_{{\bf p}}}\right .\right .\nonumber \\
&+&\left .{1\over\omega-E_{h{\bf p+k}}+\omega_{{\bf p}}}-{1\over
\omega +E_{h{\bf p+k}}+\omega_{{\bf p}}}-{1\over\omega +E_{h{\bf p+k
}}-\omega_{{\bf p}}}\right )\nonumber \\
&+&{\rm tanh}[{1\over 2}\beta E_{h{\bf p+k}}]\left ({1\over\omega -
E_{h{\bf p+k}}-\omega_{{\bf p}}}-{1\over \omega-E_{h{\bf p+k}}+
\omega_{{\bf p}}} \right .\nonumber \\
&-&\left. \left. {1\over \omega +E_{h{\bf p+k}}+\omega_{{\bf p}}}+
{1\over \omega +E_{h{\bf p+k}}-\omega_{{\bf p}}} \right )\right \},
\end{eqnarray}
\end{subequations}
respectively, these convolutions of the spin Green's function and
dressed holon normal and anomalous Green's functions reflect the
charge-spin recombination \cite{anderson2}, then the electron
spectral function and superconducting gap function are obtained from
the above electron normal and anomalous Green's functions as,
\begin{eqnarray}
A({\bf k},\omega)&=&-2{\rm Im}G({\bf k},\omega)\nonumber\\
&=&2\pi {1\over N}\sum_{{\bf p}}Z_{hF}{B_{{\bf p}} \over
4\omega_{{\bf p}}}\left \{{\rm coth}({1\over 2}\beta \omega_{{\bf
p}})[U^{2}_{h{\bf p+k}}\delta(\omega+E_{h{\bf p+k}}
-\omega_{{\bf p }})\right . \nonumber \\
&+&U^{2}_{h{\bf p+k}}\delta(\omega+E_{h{\bf p+k}}+\omega_{{\bf p}}
)+V^{2}_{h{\bf p+k}}\delta(\omega-E_{h{\bf p+k}}+\omega_{{\bf p}}
)\nonumber \\
&+&V^{2}_{h{\bf p+k}}\delta(\omega-E_{h{\bf p+k}}-\omega_{{\bf p}}
)]+{\rm tanh}({1\over 2}\beta E_{h{\bf p+k}}) \nonumber \\
&\times& [U^{2}_{h{\bf p+k}}\delta(\omega+E_{h{\bf p+k}}+
\omega_{{\bf p}})-U^{2}_{h{\bf p+k}}\delta(\omega+E_{h{\bf p+k}}-
\omega_{{\bf p}})\nonumber \\
&+& \left . V^{2}_{h{\bf p+k}}\delta(\omega-E_{h{\bf p+k}}-
\omega_{{\bf p}}) - V^{2}_{h{\bf p+k}}\delta(\omega-E_{h{\bf p+k}}
+\omega_{{\bf p}})]\right \},~~~~~~\\
\Delta({\bf k})&=&-{1\over\beta}\sum_{i\omega_{n}}
\Gamma^{\dagger}({\bf k},i\omega_{n})\nonumber\\
&=&{-1\over N}\sum_{{\bf p}}{Z_{hF} \bar{\Delta}_{hZ}({\bf p}-{\bf
k}) \over 2E_{h{\bf p-k}}}{\rm tanh} [{1\over 2}\beta E_{h{\bf p}
-{\bf k}}]{B_{{\bf p}}\over 2\omega_{{\bf p}}}{\rm coth} [{1\over 2}
\beta\omega_{{\bf p}}],~~~~~
\end{eqnarray}
which shows that the symmetry of the electron Cooper pair is
completely determined by the symmetry of the dressed holon pair.
With the above superconducting gap function (40), the
superconducting gap parameter in Eq. (15) is obtained as
$\Delta=-\chi_{1}\Delta_{h}$. Since both dressed holon (then
electron) pairing gap parameter and pairing interaction in cuprate
superconductors are doping dependent, then the experimental observed
doping dependence of the superconducting gap parameter should be an
effective superconducting gap parameter $\bar{\Delta}\sim-\chi_{1}
\bar{\Delta}_{h}$. On the other hand, the electron quasiparticle
coherent weight $Z_{F}$ can be obtained from the sum rule for the
electron spectral function $A({\bf k},\omega)$ in Eq. (39) as,
\begin{eqnarray}
Z_{F}=\int^{\infty}_{-\infty}{d\omega\over 2\pi}A({\bf k},\omega)
={1\over N}\sum_{{\bf p}}Z_{hF}{B_{{\bf p}}\over 2\omega_{{\bf p}}}
{\rm coth}\left ({\beta\omega_{{\bf p}}\over 2}\right)={1\over 2}
Z_{hF}.
\end{eqnarray}

\begin{figure}[t]
\begin{center}
\begin{minipage}[h]{85mm}
\epsfig{file=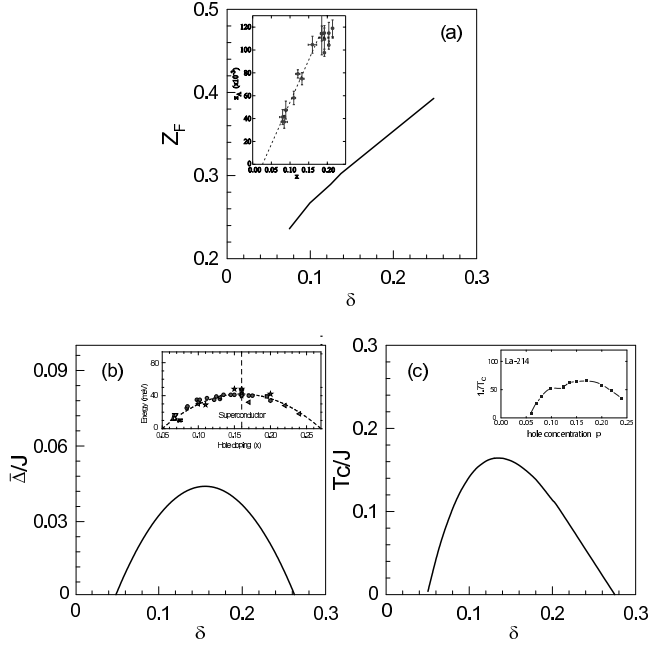, width=85mm}
\end{minipage}
\caption{(a) The electron quasiparticle coherent weight
$Z_{F}(T_{c})$ in the $[\pi,0]$ point, (b) the effective
superconducting gap parameter $\bar{\Delta}$ at $T=0.002J$, and (c)
the superconducting transition temperature $T_{c}$ as a function of
the doping concentration for $t/J=2.5$ and $t'/t=0.3$. Inset: the
corresponding experimental results of cuprate superconductors taken
from Refs. \protect\cite{ding}, \protect\cite{wen} and
\protect\cite{tallon}, respectively. }
\end{center}
\end{figure}

In Fig. 1, we plot (a) the electron quasiparticle coherent weight
$Z_{F}(T_{c})$, (b) the effective superconducting gap parameter
$\bar{\Delta}$ at temperature $T=0.002J$, and (c) the
superconducting transition temperature $T_{c}$ as a function of the
doping concentration $\delta$ for parameters $t/J=2.5$ and
$t'/t=0.3$. For comparison, the corresponding experimental results
(inset) of the quasiparticle coherent weight in the $[\pi,0]$ point
\cite{ding}, superconducting gap parameter \cite{wen}, and
superconducting transition temperature \cite{tallon} as a function
of the doping concentration are also shown in Fig. 1(a), 1(b), and
1(c), respectively. Our results show that the quasiparticle coherent
weight grows linearly with the doping concentration, i.e.,
$Z_{F}\propto\delta$, which together with the superconducting gap
parameter defined in Eq. (15) show that only $\delta$ number of
coherent doped carriers are recovered in the superconducting state,
consistent with the picture of a doped Mott insulator with $\delta$
holes \cite{anderson1}. In this case, the superconducting state of
cuprate superconductors is controlled by both superconducting gap
function and superconducting quasiparticle coherence
\cite{ding,feng3,feng4}. Since the dressed holons (then electrons)
interact by exchanging spin excitations and that this interaction is
attractive. This attractive interaction leads to form the dressed
holon pairs (then electron Cooper pairs). The perovskite parent
compound of doped cuprate superconductors is a Mott insulator, when
holes are doped into this insulator, there is a gain in the kinetic
energy per hole proportional to $t$ due to hopping, but at the same
time, the spin correlation is destroyed, costing an energy of
approximately $J$ per site, therefore the doped holes into the Mott
insulator can be considered as a competition between the kinetic
energy ($\delta t$) and magnetic energy ($J$), and the magnetic
energy decreases with increasing doping. In the underdoped and
optimally doped regimes, the magnetic energy is rather too large,
and the dressed holon (then electron) attractive interaction by
exchanging spin excitations is also rather strong to form the
dressed holon pairs (then electron Cooper pairs) for the most
dressed holons (then electrons), therefore the number of the dressed
holon pairs (then electron Cooper pairs) and superconducting
transition temperature \cite{uemura} are proportional to the hole
doping concentration. However, in the overdoped regime, the magnetic
energy is relatively small, and the dressed holon (then electron)
attractive interaction by exchanging spin excitations is also
relatively weak, in this case, not all dressed holons (then
electrons) can be bounden as dressed holon pairs (then electron
Cooper pairs) by the weak attractive interaction, and therefore the
number of the dressed holon pairs (then electron Cooper pairs) and
superconducting transition temperature \cite{tallon} decrease with
increasing doping. Using an reasonably estimative value of $J\sim
800$K to 1200K in cuprate superconductors \cite{kastner}, the
superconducting transition temperature in the optimal doping is
T$_{c}\approx 0.165J \approx 132{\rm K}\sim 198{\rm K}$, in
qualitative agreement with the experimental data \cite{tallon}.

\begin{figure}[t]
\begin{center}
\begin{minipage}[h]{75mm}
\epsfig{file=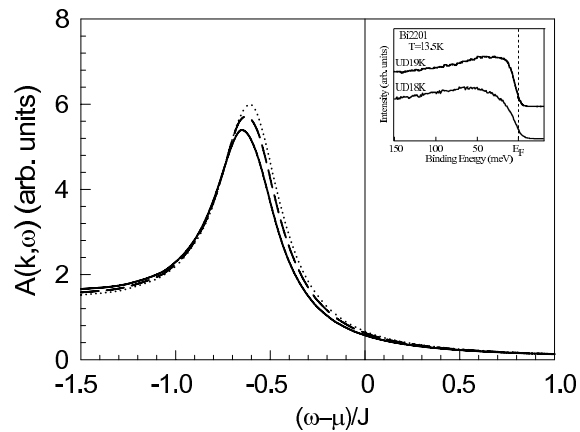, width=75mm}
\end{minipage}
\caption{ The electron spectral function $A({\bf k},\omega)$ at the
$[\pi,0]$ point in the superconducting state with $\delta=0.09$
(solid line), $\delta=0.12$ (dashed line), and $\delta=0.15$ (dotted
line) at $T=0.002J$ for $t/J=2.5$ and $t'/t=0.3$.  Inset: the
corresponding experimental result of the single layer cuprate
superconductor Bi$_{2}$Sr$_{2}$CuO$_{6+\delta}$ in the
superconducting state taken from Ref. \protect\cite{sato6}. }
\end{center}
\end{figure}

\begin{figure}[t]
\begin{center}
\begin{minipage}[h]{75mm}
\epsfig{file=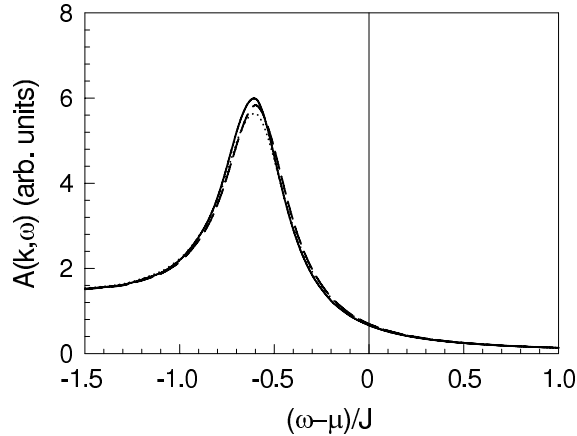, width=75mm}
\end{minipage}
\caption{The electron spectral function $A({\bf k},\omega)$ at the
$[\pi,0]$ point in the superconducting state in $\delta=0.15$ with
$T=0.002J$ (solid line), $T=0.10J$ (dashed line), and $T=0.15J$
(dotted line) for $t/J=2.5$ and $t'/t=0.3$.}
\end{center}
\end{figure}

Now we discuss the electronic structure of the single layer cuprate
superconductors in the superconducting state. We have performed a
calculation for the electron spectral function $A({\bf k},\omega)$
in Eq. (39), and the results of $A({\bf k},\omega)$ in the $[\pi,0]$
point with $\delta=0.09$ (solid line), $\delta=0.12$ (dashed line),
and $\delta=0.15$ (dotted line) at $T=0.002J$ for $t/J=2.5$ and
$t'/t=0.3$ are plotted in Fig. 2 in comparison with the
corresponding experimental result of the single layer cuprate
superconductor Bi$_{2}$Sr$_{2}$CuO$_{6+\delta}$ in the
superconducting state \cite{sato6} (inset). Obviously, (1) there is
a sharp superconducting quasiparticle peak near the electron Fermi
energy in the $[\pi,0]$ point, and the position of the
superconducting quasiparticle peak in $\delta=0.15$ is located at
$\omega_{{\rm peak}}\approx 0.6J\approx 0.042$eV$\sim 0.06$eV, which
is qualitatively consistent with $\omega_{{\rm peak}}\approx
0.035$eV observed \cite{sato6} in the underdoped single layer
cuprate superconductor Bi$_{2}$Sr$_{2}$CuO$_{6+\delta}$ in the
superconducting state; (2) The electron spectrum in the
superconducting state is doping dependent. With increasing the
doping concentration, the weight of the superconducting
quasiparticle peaks increases; (3) The position of the
superconducting quasiparticle peak moves to the Fermi energy with
increasing doping \cite{sato6}. Furthermore, we have discussed the
temperature dependence of the electron spectrum in the
superconducting state, and the results of $A({\bf k},\omega)$ in the
$[\pi,0]$ point with $\delta=0.15$ at $T=0.002J$ (solid line),
$T=0.10J$ (dashed line), and $T=0.15J$ (dotted line) for $t/J=2.5$
and $t'/t=0.3$ are plotted in Fig. 3. It is shown that the spectral
weight decreases as temperature is increased. These theoretical
results under the kinetic energy driven superconducting mechanism is
in qualitative agreement with the experimental data of the single
layer cuprate superconductors in the superconducting state
\cite{shen,sato6,zhou1}.

\begin{figure}[t]
\begin{center}
\begin{minipage}[h]{75mm}
\epsfig{file=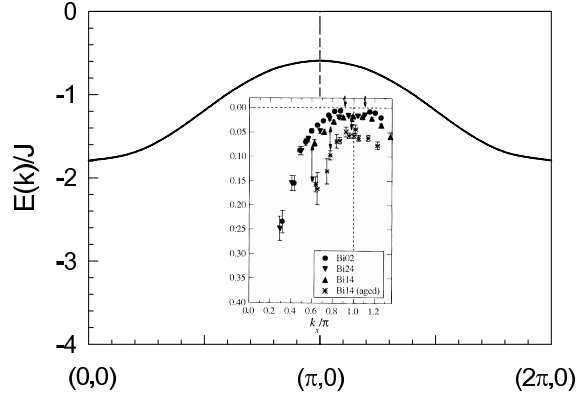, width=75mm}
\end{minipage}
\caption{The positions of the lowest energy superconducting
quasiparticle peaks in $A({\bf k},\omega)$ as a function of momentum
along the direction $[0,0]\rightarrow [\pi,0]\rightarrow [2\pi,0]$
with $\delta=0.15$ at $T=0.002J$ for $t/J=2.5$ and $t'/t=0.3$.
Inset: the corresponding experimental result of the single layer
cuprate superconductor Bi$_{2}$Sr$_{2}$CuO$_{6+\delta}$ in the
superconducting state taken from Ref. \protect\cite{takeuchi6}.}
\end{center}
\end{figure}

For the further understanding of the superconducting coherence of
the sharp quasiparticle peak around the $[\pi,0]$ point, we have
made a series of calculations for $A({\bf k},\omega)$ around the
$[\pi,0]$ point, and the results show that the sharp superconducting
quasiparticle peak persists in a very large momentum space region
around the $[\pi,0]$ point. To show this point clearly, we plot the
positions of the lowest energy superconducting quasiparticle peaks
in $A({\bf k}, \omega)$ as a function of momentum along the
direction $[0,0] \rightarrow [\pi,0]\rightarrow [2\pi,0]$ at
$\delta=0.15$ with $T=0.002J$ for $t/J=2.5$ and $t'/t=0.3$ in Fig. 4
in comparison with the corresponding experimental result of the
single layer cuprate superconductor Bi$_{2}$Sr$_{2}$CuO$_{6+\delta}$
in the superconducting state \cite{takeuchi6} (inset). Our results
show that the sharp low energy superconducting quasiparticle peaks
around the $[\pi,0]$ point disperse very weakly with momentum, which
also is qualitatively consistent with these obtained from ARPES
experimental measurements on the single layer cuprate
superconductors in the superconducting state
\cite{shen,zhou1,takeuchi6}.

A natural question is why the superconducting coherence of the
superconducting quasiparticle peak in cuprate superconductors can be
described qualitatively in the framework of the kinetic energy
driven superconductivity. The reason is that the superconducting
state in the kinetic energy driven superconductivity is the
conventional BCS like withe the d-wave symmetry \cite{feng4,guo2}.
This can be understood from the electron normal and anomalous
Green's functions in Eqs. (38a) and (38b). Since spins center around
the $[\pi,\pi]$ point in the mean-field level \cite{feng2,feng3},
then the main contributions for the spins comes from the $[\pi,\pi]$
point. In this case, the electron normal and anomalous Green's
functions in Eqs. (38a) and (38b) can be approximately reduced in
terms of $\omega_{{\bf p}= [\pi,\pi]}\sim 0$ and the self-consistent
equation (35i) as the simple BCS formalism with the d-wave gap
function,
\begin{subequations}
\begin{eqnarray}
G({\bf k},\omega)&\approx&Z_{F}\left ({U^{2}_{{\bf k}}\over\omega-
E_{{\bf k}}}+{V^{2}_{{\bf k}}\over\omega+E_{{\bf k}}}\right ),\\
\Gamma^{\dagger}({\bf k},\omega)&\approx& Z_{F}{\bar{\Delta}_{hZ}
({\bf k })\over 2E_{{\bf k}}}\left ({1\over \omega-E_{{\bf k}}}+
{1\over \omega+E_{{\bf k}}}\right ),
\end{eqnarray}
\end{subequations}
where the electron quasiparticle coherence factors,
\begin{subequations}
\begin{eqnarray}
U^{2}_{{\bf k} }\approx V^{2}_{h{\bf k+k}_{AF}},\\
V^{2}_{{\bf k}}\approx U^{2}_{h{\bf k+k}_{AF}},
\end{eqnarray}
\end{subequations}
and electron quasiparticle spectrum $E_{{\bf k}}\approx E_{h{\bf
k+k}_{AF}}$, with ${\bf k}_{AF}= [\pi,\pi]$, i.e., the hole-like
dressed holon quasiparticle coherence factors $V_{h{\bf k}}$ and
$U_{h{\bf k}}$ in Eqs. (31) and hole-like dressed holon
quasiparticle spectrum $E_{h{\bf k}}$ in Eq. (32a) have been
transferred into the electron quasiparticle coherence factors
$U_{{\bf k}}$ and $V_{{\bf k}}$ and electron quasiparticle spectrum
$E_{{\bf k}}$, respectively, by the convolutions of the spin Green's
function and dressed holon Green's functions, which means that the
dressed holon pairs condense with the d-wave symmetry in a wide
range of the doping concentration, then the electron Cooper pairs
originating from the dressed holon pairing state are due to the
charge-spin recombination, and their condensation automatically
gives the electron quasiparticle character. This electron
quasiparticle is the excitation of a single electron 'dressed' with
the attractive interaction between paired electrons. This is why the
basic BCS formalism with the d-wave gap function \cite{bcs} is still
valid in discussions of the doping dependence of the superconducting
gap parameter and superconducting transition temperature, and
superconducting coherence of the quasiparticle peak
\cite{matsui,campuzano1}, although the pairing mechanism is driven
by the kinetic energy by exchanging spin excitations, and other
exotic magnetic scattering \cite{yamada,dai,arai} is beyond the BCS
formalism.

As we have known that the quasiparticle is defined as in Fermi
liquid theory and gives a measure of how the quasiparticle is to
being a free electron \cite{landau}. The basis for BCS theory of the
conventional superconductivity is the formation first of a Fermi
liquid, i.e., a quantum coherent state. Then one looks for strong
interactions which crossover to an attractive pairing interaction in
some high angular momentum channel \cite{rice19}. In particular, the
quasiparticle coherent weight $Z_{F}=1$ in the simple BCS model for
the conventional superconductors \cite{bcs}. However, in the kinetic
energy driven superconducting mechanism, although the
superconducting coherence of the quasiparticle peak is described by
the simple BCS formalism with the d-wave gap function, the pairing
mechanism is driven by the kinetic energy by exchanging spin
excitations as mentioned above, which reflects that the strong
electron correlation does not suppress superconductivity, but rather
is to favor it because the main ingredient was identified into the
pairing mechanism not involving phonons as in the conventional
superconductors, but the internal spin degrees of freedom. We
\cite{guo3} have also calculated the superconducting quasiparticle
coherent weight $Z_{F}$ in Eq. (41) with different momenta, and the
results show that the overall spectral weight in cuprate
superconductors is heavily reduced ($Z_{F}\ll 1$) by the strong
electron correlation, in qualitative agreement with these obtained
from the variational Monte Carlo simulations \cite{lee}. All these
also are a consequence of the fact that the electron states of
cuprate superconductors are restricted in the Hilbert subspace
without double electron occupancy (projected Hilbert subspace),
where although the dressed holon quasiparticle coherent weight
$Z_{hF}$ in Eq. (28) obeys the usual perturbation theory identities
relating it to the dressed holon self-energy function, the value of
the true superconducting quasiparticle coherent weight $Z_{F}$ is
obtained in terms of the charge-spin recombination.

\subsection{Electronic structure of the single layer cuprate
superconductors in the normal state}

In correspondence with the above discussions of the electronic
structure of the single layer cuprate superconductors in the
superconducting state, we now turn to discuss the electronic
structure of the single layer cuprate superconductors in the normal
state. In the normal state, the dressed holon pairing order
parameter $\Delta_{h}=0$ (then superconducting gap parameter
$\Delta=0$), then the dressed holon normal Green's function in Eq.
(30a) is reduced as,
\begin{eqnarray}
g({\bf k},\omega)&=&{Z_{hF}\over \omega-\bar{\xi_{{\bf k}}}}.
\end{eqnarray}
It has been shown from the ARPES experiments
\cite{shen1,kim,dessau,wells} that in the normal-state, the lowest
energy states are located at the $[\pi/2,\pi/2]$ point, which
indicates that the majority contribution for the electron spectrum
comes from the $[\pi/2,\pi/2]$ point. In this case, the wave vector
${\bf k}$ in $Z_{hF}({\bf k})$ in Eq. (28) and $\Sigma^{(h)}_{1e}
({\bf k})$ can be chosen as $Z^{-1}_{hF}=1-\Sigma^{(h)}_{1o}({\bf k}
)\mid_{{\bf k}=[\pi/2,\pi/2]}$ and $\Sigma^{(h)}_{1e}=
\Sigma^{(h)}_{1e}({\bf k})\mid_{{\bf k}=[\pi/2,\pi/2]}$, then the
equation satisfied by the dressed holon quasiparticle coherent
weight $Z_{hF}$ in Eq. (33b) is reduced as,
\begin{eqnarray}
{1\over Z_{hF}} &=& 1 + {1\over N^{2}}\sum_{{\bf q,p}}
\Lambda^{2}({\bf p}+{\bf k}_{N})Z_{hF}{B_{{\bf q}}B_{{\bf p}}\over
4\omega_{{\bf q}}\omega_{{\bf p}}}\nonumber \\
&\times& \left({F_{1}({\bf q},{\bf p})\over (\omega_{{\bf p}}-
\omega_{{\bf q}}-\bar{\xi}_{{\bf p}-{\bf q}+{\bf k}_{N}})^{2}}
+{F_{2}({\bf q},{\bf p})\over (\omega_{{\bf p}}-\omega_{{\bf q}}
-\bar{\xi}_{{\bf p}-{\bf q}+{\bf k}_{N}})^{2}}\right .\nonumber\\
&+&\left . {F_{3}({\bf q},{\bf p})\over (\omega_{{\bf p}}+
\omega_{{\bf q}}-\bar{\xi}_{{\bf p}-{\bf q}+{\bf k}_{N}})^{2}} +
{F_{4}({\bf q},{\bf p})\over (\omega_{{\bf p}}+ \omega_{{\bf q}}
+\bar{\xi}_{{\bf p}-{\bf q}+{\bf k}_{N}})^{2}}\right ),
\end{eqnarray}
where ${\bf k}_{N}=[\pi/2,\pi/2]$, and
\begin{subequations}
\begin{eqnarray}
F_{1}({\bf q},{\bf p})&=&n_{F}(\bar{\xi}_{{\bf p}-{\bf q}+{\bf k
}_{N}})[n_{B}(\omega_{{\bf q}})-n_{B}(\omega_{{\bf p}})]
-n_{B}(\omega_{{\bf p}})n_{B}(-\omega_{{\bf q}}),~~~~~~ \\
F_{2} ({\bf q},{\bf p})&=&n_{F}(\bar{\xi}_{{\bf p}-{\bf q}+{\bf k
}_{N}})[n_{B}(\omega_{{\bf p}})-n_{B}(\omega_{{\bf q}})]
-n_{B}(\omega_{{\bf q}})n_{B}(-\omega_{{\bf p}}),~~~~~\\
F_{3}({\bf q},{\bf p})&=& n_{F}(\bar{\xi}_{{\bf p}-{\bf q}+{\bf k
}_{N}})[n_{B}(\omega_{{\bf q}})-n_{B}(-\omega_{{\bf p}})]
+n_{B}(\omega_{{\bf p}})n_{B}(\omega_{{\bf q}}),~~~~~~\\
F_{4}({\bf q},{\bf p})&=&n_{F}(\bar{\xi}_{{\bf p}-{\bf q}+{\bf k
}_{N}})n_{B}(-\omega_{{\bf q}})-n_{B}(\omega_{{\bf p}})]+
n_{B}(-\omega_{{\bf p}})n_{B}(-\omega_{{\bf q}}).~~~~~~~~~~
\end{eqnarray}
\end{subequations}
As in the superconducting state, this self-consistent equation must
be solved simultaneously with other self-consistent equations in Eqs
(35), then the electron normal Green's function in Eq. (38a) and the
electron spectral function in Eq. (39) in the normal state are
reduced as,
\begin{subequations}
\begin{eqnarray}
G({\bf k},\omega)&=&{1\over N}\sum_{{\bf p}}Z_{hF}{B_{{\bf p}}\over
2\omega_{{\bf p}}}\left ({L_{1}({\bf k},{\bf p})\over\omega+
\bar{\xi}_{{\bf p}+{\bf k}}-\omega_{{\bf p}}}+{L_{2}({\bf k},{\bf p}
)\over\omega+\bar{\xi}_{{\bf p}+{\bf k}}+\omega_{{\bf p}}}\right),
~~~~\\
A({\bf k},\omega)&=&2\pi {1\over N}\sum_{{\bf p}}Z_{hF}{B_{{\bf p}}
\over 2\omega_{{\bf p}}}[L_{1}({\bf k},{\bf p})\delta(\omega+
\bar{\xi}_{{\bf p}+{\bf k}}-\omega_{{\bf p}})\nonumber \\
&+& L_{2}({\bf k},{\bf p})\delta(\omega+\bar{\xi}_{{\bf p}+{\bf k}}
+\omega_{{\bf p}})],~~~
\end{eqnarray}
\end{subequations}
where $L_{1}({\bf k},{\bf p})=n_{F}(\bar{\xi}_{{\bf p}+{\bf k}})+
n_{B}(\omega_{{\bf p}})$ and $L_{2}({\bf k},{\bf p})=1-n_{F}
(\bar{\xi}_{{\bf p}+{\bf k}})+n_{B}(\omega_{{\bf p}})$.

\begin{figure}[t]
\begin{center}
\begin{minipage}[h]{90mm}
\epsfig{file=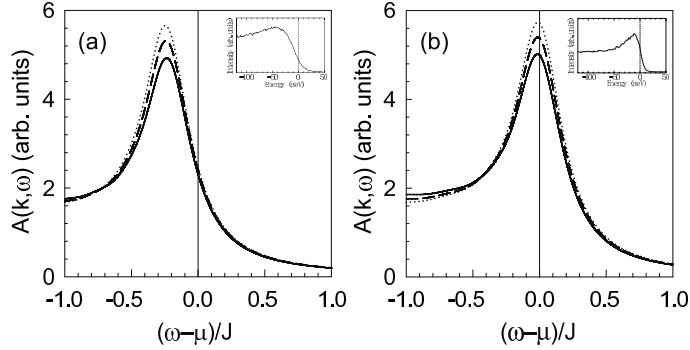, width=90mm}
\end{minipage}
\caption{The electron spectral function $A({\bf k},\omega)$ at (a)
the $[\pi,0]$ point (b) the $[\pi/2,\pi/2]$ point in the normal
state with $T=0.1J$ in $\delta=0.09$ (solid line), $\delta=0.12$
(dashed line), and $\delta=0.15$ (dotted line) for $t/J=2.5$ and
$t'/t=0.15$. Inset: the corresponding experimental result of the
single layer cuprate superconductor Bi$_{2}$Sr$_{2}$CuO$_{6+\delta}$
at the $[\pi,0]$ point and $[\pi/2,\pi/2]$ point, respectively, in
the normal state taken from Ref. \protect\cite{kondo6}.}
\end{center}
\end{figure}

We have performed a calculation for the electron spectral function
in Eq. (47b) in the normal state, and the results at (a) the
$[\pi,0]$ point and (b) the $[\pi/2,\pi/2]$ point with $T=0.1J$ for
$t/J=2.5$ and $t'/t=0.15$ at $\delta=0.09$ (solid line),
$\delta=0.12$ (dashed line), and $\delta=0.15$ (dotted line) are
plotted in Fig. 5 in comparison with the corresponding experimental
result \cite{kondo6} of the single layer cuprate superconductor
Bi$_{2}$Sr$_{2}$CuO$_{6+\delta}$ in the normal state (inset). It is
shown that (1) although both positions of the quasiparticle peaks at
the $[\pi,0]$ and $[\pi/2,\pi/2]$ points are below the Fermi energy,
the position of the quasiparticle peak at the $[\pi/2,\pi/2]$ point
is more close to the Fermi energy, which indicates that the lowest
energy states are located at the $[\pi/2,\pi/2]$ point. In other
words, the low energy spectral weight with the majority contribution
to the low-energy properties of cuprate superconductors in the
normal state comes from the $[\pi/2,\pi/2]$ point; (2) The electron
spectrum in the normal state is doping dependence as in the
superconducting state. The quasiparticle peaks at the $[\pi,0]$ and
$[\pi/2,\pi/2]$ points become sharper, while the spectral weight of
these peaks increases in intensity with increasing doping.
Furthermore, we have also discussed the temperature dependence of
the electron spectrum in the normal state, and the results show that
the spectral weight is suppressed with increasing temperatures. Our
these results are qualitatively consistent with the ARPES
experimental data of the single layer cuprate superconductors in the
norma state \cite{shen1,kim,dessau,wells,zhou1,kondo6}.

\begin{figure}[t]
\begin{center}
\begin{minipage}[h]{75mm}
\epsfig{file=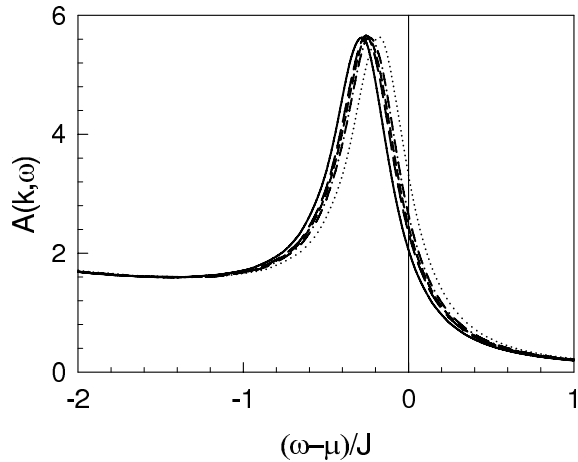, width=75mm}
\end{minipage}
\caption{The electron spectral function $A({\bf k},\omega)$ in the
normal state with $T=0.1J$ in $\delta=0.15$ at the $[0.9\pi,0]$
(solid line), $[0.95\pi,0]$ (long dashed line), $[\pi,0]$ (short
dashed line), $[\pi,0.05\pi]$ (dash-dotted line), and $[\pi,0.1\pi]$
(dotted line) points for $t/J=2.5$ and $t'/t=0.15$.}
\end{center}
\end{figure}

\begin{figure}[t]
\begin{center}
\begin{minipage}[h]{90mm}
\epsfig{file=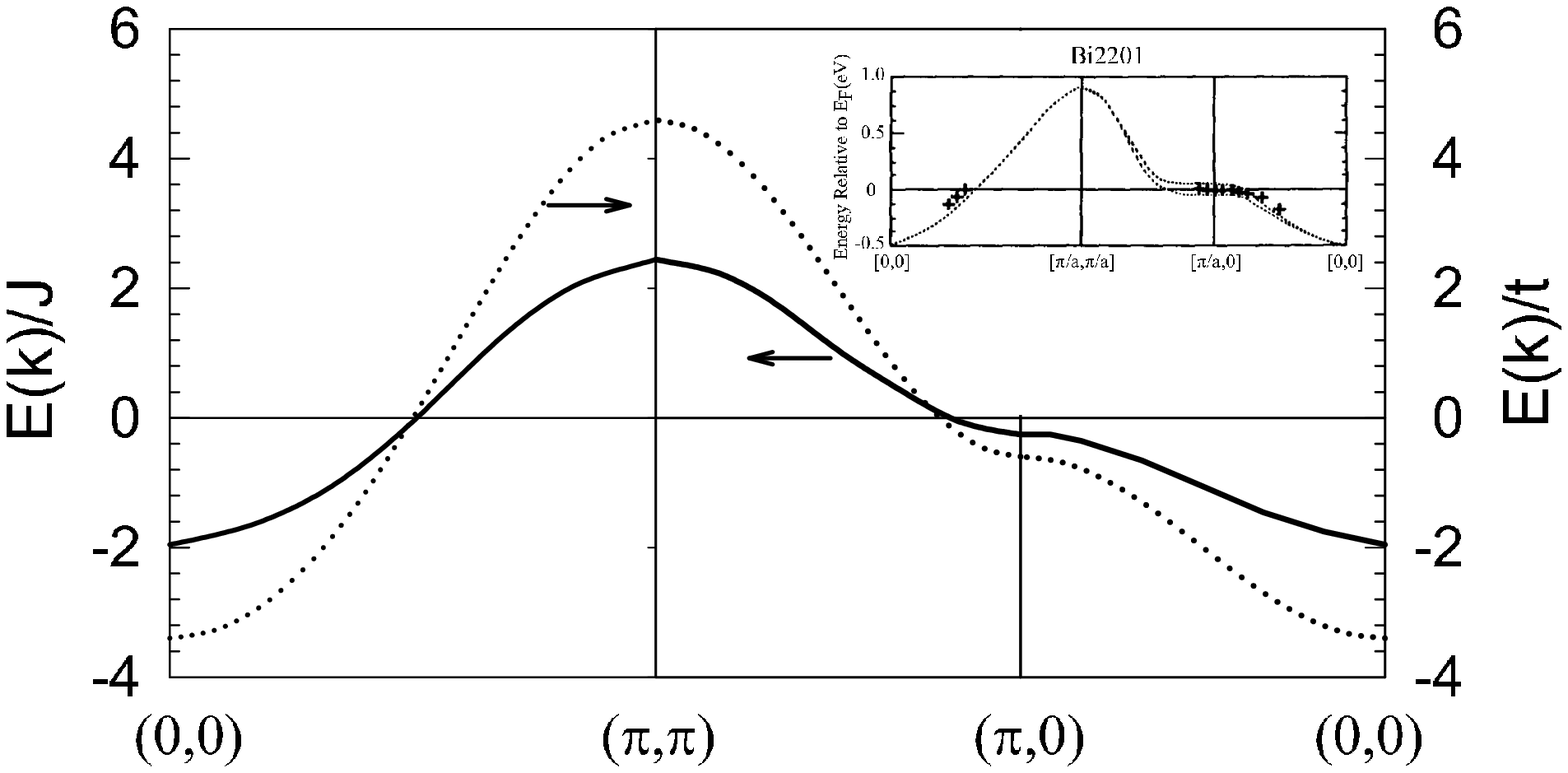, width=90mm}
\end{minipage}
\caption{The position of the lowest energy quasiparticle peaks in
$A({\bf k},\omega)$ in the normal state as a function of momentum
for $t/J=2.5$ and $t'/t=0.15$ with $T=0.1J$ at $\delta=0.15$. The
dotted line is corresponding result of the bare electron dispersion
of the $t$-$t'$ model. Inset: the corresponding experimental result
of the single layer cuprate superconductor
Bi$_{2}$Sr$_{2}$CuO$_{6+\delta}$ in the normal state taken from Ref.
\protect\cite{shen1}.}
\end{center}
\end{figure}

To elucidate the anomalous form of the electron spectrum $A({\bf
k},\omega)$ in the normal state as a function of energy $\omega$ for
${\bf k}$ in the vicinity of the $[\pi,0]$ point, we have discussed
the electron spectral function around the $[\pi,0]$ point, and the
results with $T=0.1J$ in $\delta=0.15$ for $t/J=2.5$ and $t'/t=0.15$
at the $[0.9\pi,0]$ (solid line), $[0.95\pi,0]$ (long dashed line),
$[\pi,0]$ (short dashed line), $[\pi,0.05\pi]$ (dash-dotted line),
and $[\pi,0.1\pi]$ (dotted line) points are plotted in Fig. 6.
Obviously, the positions of these peaks of the electron spectral
function $A({\bf k},\omega)$ around the $[\pi,0]$ point in the
normal state are almost not changeable, which leads to the unusual
quasiparticle dispersion around the $[\pi,0]$ point. In particular,
the lowest energy peaks in the normal state are well defined at all
momenta. To show this broad feature in the electron spectrum around
the $[\pi,0]$ point clearly, we plot the positions of the lowest
energy quasiparticle peaks in the electron spectral function as a
function of momentum along the high symmetry directions with
$T=0.1J$ at $\delta=0.15$ for $t/J=2.5$ and $t'/t=0.15$ in Fig. 7.
For comparison, the corresponding result of the bare electron
dispersion of the $t$-$t'$ model (dotted line), and experimental
result (inset) of the electron dispersion from the single layer
cuprate superconductor Bi$_{2}$Sr$_{2}$CuO$_{6+\delta}$ in the
normal state \cite{shen1} are also shown in Fig. 7. In accordance
with the anomalous property of the electron spectrum in Fig. 6, the
electron quasiparticles around the $[\pi,0]$ point disperse very
weakly with momentum, and then the unusual flat band appears, while
the Fermi energy is only slightly above this flat band, in
qualitative agreement with these obtained from ARPES experimental
measurements on the single layer cuprate superconductors in the
normal state \cite{shen1,kim,dessau,wells,zhou1,kondo6}.

Since the full electron Green's function (then the electron spectral
function) in the normal state is obtained beyond the mean-field
approximation by considering the fluctuation due to the spin pair
bubble, therefore the nature of the electron spectrum in the normal
state is closely related to the strong coupling between the dressed
holon (then electron quasiparticles) and collective magnetic
excitations. This can be understood from a comparison between the
bare electron dispersion of the $t$-$t'$ model and renormalized
electron quasiparticle dispersion of the $t$-$J$ model in Fig. 7.
Our results show that the single-particle hopping in the $t$-$J$
model is strongly renormalized by the magnetic interaction. As a
consequence, the quasiparticle bandwidth is reduced to the order of
(a few) $J$, and therefore the energy scale of the quasiparticle
band is controlled by the magnetic interaction. This renormalization
due to the strong interaction is then responsible for the unusual
electron quasiparticle spectrum and production of the flat band.
Moreover, these results also show that the electron quasiparticle
excitations originating from the dressed holons and spins are due to
the charge-spin recombination, this reflects the composite nature of
the electron quasiparticle excitations, and then the unconventional
normal state properties in cuprate superconductors are attributed to
the presence of the dressed holon, spin, and electron quasiparticle
excitations.

\section{Electronic structure of the bilayer cuprate
superconductors}

Recently, the improvements in the resolution of the ARPES
experiments \cite{shen,campuzano,dlfeng1,kordyuk,chuang} allowed to
resolve additional features in the electron spectral function
$A({\bf k},\omega)$. Among these new achievements is the observation
of the bilayer splitting in the bilayer cuprate superconductors in
both normal and superconducting states
\cite{shen,dlfeng1,kordyuk,chuang}. In this case, whether the
electronic structure of cuprate superconductors can be influenced by
the interaction between CuO$_{2}$ planes has been an interesting
issue. The study of the electronic structure is complicated by the
bilayer splitting, that is, the bilayer splitting of the CuO$_{2}$
planes derived the electronic structure in the bonding and
antibonding bands due to the present of CuO$_{2}$ bilayer blocks in
the unit cell \cite{dlfeng1,kordyuk,chuang}. The magnitude of the
bilayer splitting is the doping independent, and increases upon
approaching the $[\pi,0]$ point, where the bilayer splitting
exhibits the largest value. As a result of the maximal bilayer
splitting at the $[\pi,0]$ point, there are two main flat bands
around the $[\pi,0]$ point in the normal state
\cite{dlfeng1,kordyuk,chuang}. In corresponding to this double-peak
structure in the normal state, the peak-dip-hump structure around
the $[\pi,0]$ point in the superconducting state is observed
\cite{shen,dessau1,randeria,fedorov,lu,sato} as mentioned in section
1. Therefore an important issue is whether the behavior of the low
energy excitations determined by the electronic structure of cuprate
superconductors is universal or not. The earlier works
\cite{campuzano2,shen} gave the main impetus for a phenomenological
description of the single-particle excitations in terms of an
interaction between quasiparticles and collective modes, which is of
fundamental relevance to the nature of superconductivity and the
pairing mechanism in cuprate superconductors. However, the different
interpretive scenario has been proposed \cite{kordyuk,shen}, where
it has been argued that the main features of the peak-dip-hump
structure is caused by the bilayer splitting
\cite{kordyuk,dlfeng1,chuang,borisenko}, with the peak and hump
corresponding to the antibonding and bonding bands, respectively.
Furthermore, some ARPES experimental data measured above and below
the superconducting transition temperature show that this
peak-dip-hump structure is totally unrelated to superconductivity
\cite{dlfeng1}. In particular, the recent ARPES experimental results
reported by several groups support this scenario, and most
convincingly suggested that the peak-dip-hump structure originates
from the bilayer splitting at any doping levels \cite{dlfeng2}. In
this section we show explicitly if the bilayer interaction is
included in the framework of the kinetic energy driven
superconductivity, one can reproduce some main features in both
normal and superconducting states observed experimentally on the
bilayer cuprate superconductors \cite{shen,dlfeng1,kordyuk,chuang}.

\subsection{Electronic structure of the bilayer cuprate
superconductors in the superconducting state}

For discussions of the electronic structure of the bilayer cuprate
superconductors, the $t$-$J$ model in Eq. (1) can be generalized by
including the bilayer hopping and bilayer magnetic exchange
interaction as,
\begin{eqnarray}
H&=&-t\sum_{i\hat{\eta}a\sigma}C^{\dagger}_{ia\sigma}
C_{i+\hat{\eta}a\sigma}+t'\sum_{i\hat{\tau}a\sigma}
C^{\dagger}_{ia\sigma}C_{i+\hat{\tau}a\sigma}
-\sum_{i\sigma}t_{\perp}(i)(C^{\dagger}_{i1\sigma}C_{i2\sigma}+H.c.)
\nonumber\\
&+&\mu_{0}\sum_{ia\sigma}C^{\dagger}_{ia\sigma}C_{ia\sigma}
+J\sum_{i\hat{\eta}a}{\bf S}_{ia} \cdot {\bf
S}_{i+\hat{\eta}a}+J_{\perp}\sum_{i}{\bf S}_{i1} \cdot {\bf S}_{i2},
\end{eqnarray}
where $a=1,2$ is plane index, and the interlayer hopping
\cite{chakarvarty},
\begin{eqnarray}
t_{\perp}({\bf k})={t_{\perp}\over 4}(\cos k_{x} -\cos k_{y})^{2},
\end{eqnarray}
describes coherent hopping between the CuO$_{2}$ planes. This
functional form of the interlayer hopping in Eq. (49) is firstly
predicted on the basis of the local density approximation
calculations \cite{chakarvarty}, and later the experimental observed
bilayer spilitting agrees well with it
\cite{shen,dlfeng1,kordyuk,chuang}. In the charge-spin separation
fermion-spin representation (9), this bilayer $t$-$J$ model (48) can
be expressed as,
\begin{eqnarray}
H&=&t\sum_{i\hat{\eta}a}(h^{\dagger}_{i+\hat{\eta}a\uparrow}
h_{ia\uparrow}S^{+}_{ia}S^{-}_{i+\hat{\eta}a}+
h^{\dagger}_{i+\hat{\eta}a\downarrow}h_{ia\downarrow}S^{-}_{ia}
S^{+}_{i+\hat{\eta}a})\nonumber\\
&-&t'\sum_{i\hat{\tau}a} (h^{\dagger}_{i+\hat{\tau}a\uparrow}
h_{ia\uparrow}S^{+}_{ia}S^{-}_{i+\hat{\tau}a}+
h^{\dagger}_{i+\hat{\tau}a\downarrow}h_{ia\downarrow}
S^{-}_{ia}S^{+}_{i+\hat{\tau}a})\nonumber \\
&+&\sum_{i}t_{\perp}(i)(h^{\dagger}_{i2\uparrow}h_{i1\uparrow}
S^{+}_{i1}S^{-}_{i2}+h^{\dagger}_{i1\uparrow}h_{i2\uparrow}
S^{+}_{i2}S^{-}_{i1}\nonumber\\
&+&h^{\dagger}_{i2\downarrow}h_{i1\downarrow} S^{-}_{i1}
S^{+}_{i2}+h^{\dagger}_{i1\downarrow}h_{i2\downarrow} S^{-}_{i2}
S^{+}_{i1})-\mu_{0}\sum_{ia\sigma}h^{\dagger}_{ia\sigma}
h_{ia\sigma}\nonumber\\
&+&{J_{\rm eff}}\sum_{i\hat{\eta}a}{\bf S}_{ia}\cdot {\bf S}_{i+
\hat{\eta}a} +{J_{\rm eff\perp}}\sum_{i}{\bf S}_{i1}\cdot {\bf S}
_{i2},
\end{eqnarray}
where $J_{\rm eff\perp}=J_{\perp}(1-\delta)^{2}$. Since there are
two coupled CuO$_{2}$ planes in the unit cell, therefore the
superconducting order parameter for the electron Cooper pair is a
matrix $\Delta=\Delta_{L}+\sigma_{x} \Delta_{T}$, with the
longitudinal and transverse superconducting order parameters can be
expressed as \cite{lan2},
\begin{subequations}
\begin{eqnarray}
\Delta_{L}&=&\langle C^{\dagger}_{ia\uparrow}
C^{\dagger}_{i+\hat{\eta}a\downarrow}-C^{\dagger}_{ia\downarrow}
C^{\dagger}_{i+\hat{\eta}a\uparrow}\rangle\nonumber \\
&=&\langle h_{ia\uparrow} h_{i+\hat{\eta}a\downarrow}S^{+}_{ia}
S^{-}_{i+\hat{\eta}a}-h_{ia\downarrow}h_{i+\hat{\eta}a\uparrow}
S^{-}_{ia}S^{+}_{i+\hat{\eta}a}\rangle=-\chi_{1}\Delta_{hL},
~~~~~\\
\Delta_{T}&=&\langle C^{\dagger}_{i1\uparrow}
C^{\dagger}_{i2\downarrow}- C^{\dagger}_{i1\downarrow}
C^{\dagger}_{i2\uparrow}\rangle\nonumber \\
&=&\langle h_{i1\uparrow} h_{i2\downarrow}S^{+}_{i1}S^{-}_{i2}-
h_{i1\downarrow}h_{i2\uparrow}S^{-}_{i1}S^{+}_{i2}\rangle
=-\chi_{\perp}\Delta_{hT},~~~~~
\end{eqnarray}
\end{subequations}
respectively, where the spin correlation functions $\chi_{1}=\langle
S_{ia}^{+}S_{i+\hat{\eta}a}^{-}\rangle$ and $\chi_{\perp}=\langle
S^{+}_{i1}S^{-}_{i2}\rangle$, and the longitudinal and transverse
dressed holon pairing order parameters,
\begin{subequations}
\begin{eqnarray}
\Delta_{hL}&=&\langle h_{i+\hat{\eta}a\downarrow}h_{ia\uparrow}-
h_{i+\hat{\eta}a\uparrow} h_{ia\downarrow}\rangle \\
\Delta_{hT}&=&\langle h_{i2\downarrow} h_{i1\uparrow}-h_{i2\uparrow}
h_{i1\downarrow}\rangle.
\end{eqnarray}
\end{subequations}

In the following discussions, our main goal is to generalize the
analytical calculation from the single layer case \cite{guo2} in
section 3 to the bilayer system \cite{lan2}. As in the case for the
superconducting order parameter, the full dressed holon normal and
anomalous Green's functions also can be expressed as matrices
$g({\bf k},\omega)= g_{L}({\bf k},\omega)+\sigma_{x}g_{T}({\bf k},
\omega)$ and $\Im^{\dagger}({\bf k},\omega)=\Im^{\dagger}_{L}({\bf k
},\omega) +\sigma_{x}\Im^{\dagger}_{L}({\bf k},\omega)$,
respectively. We now can follow the discussions for the single layer
case in section 3 \cite{guo2,feng4}, and evaluate explicitly these
corresponding longitudinal and transverse parts of the full dressed
holon normal and anomalous Green's functions as,
\begin{subequations}
\begin{eqnarray}
g_{L}({\bf k},\omega)&=&{1\over 2}\sum_{\nu=1,2}Z^{(\nu)}_{hFA}\left
({U^{2}_{h\nu{\bf k}}\over\omega-E_{h\nu{\bf k}}}+ {V^{2}_{h\nu{\bf
k}}\over\omega+E_{h\nu{\bf k}}}\right),~~~~~~\\
g_{T}({\bf k},\omega)&=&{1\over 2}\sum_{\nu=1,2}(-1)^{\nu+1}
Z^{(\nu)}_{hFA}\left ({U^{2}_{h\nu{\bf k}}\over\omega- E_{h\nu{\bf
k}}}+{V^{2}_{h\nu{\bf k}}\over\omega+E_{h\nu{\bf k}}} \right),
\end{eqnarray}
\begin{eqnarray}
\Im^{\dagger}_{L}({\bf k},\omega)&=&-{1\over 2}\sum_{\nu=1,2}
Z^{(\nu)}_{hFA}{\bar{\Delta}_{hz}^{(\nu)}({\bf k})\over 2E_{h\nu{\bf
k}}}\left ({1\over \omega-E_{h\nu{\bf k}}}-{1\over \omega+
E_{h\nu{\bf k}}}\right ),~~~~~~\\
\Im^{\dagger}_{T}({\bf k},\omega)&=&-{1\over 2}\sum_{\nu=1,2}
(-1)^{\nu+1}Z^{(\nu)}_{hFA}{\bar{\Delta}_{hz}^{(\nu)}({\bf k}) \over
2E_{h\nu{\bf k}}}\left ({1\over\omega-E_{h\nu{\bf k}}}-{1\over
\omega +E_{h\nu{\bf k}}}\right ),~~~~~~~~~~~
\end{eqnarray}
\end{subequations}
with the dressed holon quasiparticle coherence factors,
\begin{subequations}
\begin{eqnarray}
U^{2}_{h\nu{\bf k}}={1\over 2}\left (1+{\bar{\xi}_{\nu{\bf k}}\over
E_{h\nu{\bf k}}}\right ),
\\
V^{2}_{h\nu{\bf k}}={1\over 2}\left (1-{\bar{\xi}_{\nu{\bf k}}\over
E_{h\nu{\bf k}}}\right),
\end{eqnarray}
\end{subequations}
and
\begin{subequations}
\begin{eqnarray}
E_{h\nu{\bf k}}&=&\sqrt{[{\bar\xi}_{\nu{\bf k}}]^{2}+\mid
\bar{\Delta}_{hz}^{(\nu)}({\bf k})\mid^{2}}, \\
{\bar\xi}_{\nu{\bf k}}&=&Z^{(\nu)}_{hFA}\epsilon_{\nu{\bf k}}-\mu,\\
\bar{\Delta}_{hz}^{(\nu)}({\bf k})&=&Z^{(\nu)}_{hFA}
[\bar{\Delta}_{hL}({\bf k})+(-1)^{\nu+1}\bar{\Delta}_{hT}({\bf k})],
\\
\mu&=&Z^{(\nu)}_{hFA}(\mu_{0}-\Sigma^{(h)}_{1Le}),
\end{eqnarray}
\end{subequations}
are the dressed holon quasiparticle spectrum, the renormalized
dressed holon excitation spectrum, the renormalized dressed holon
pair gap function, and renormalized chemical potential,
respectively, where $\epsilon_{\nu{\bf k}}=Zt\chi_{1}
\gamma_{\bf{k}}-Zt'\chi_{2}\gamma{'}_{\bf{k}}+ (-1)^{\nu+1}
\chi_{\perp} t_{\perp}({\bf{k}})$, the spin correlation function
$\chi_{2}=\langle S_{ia}^{+}S_{i+\hat{\tau}a}^{-}\rangle$, the
longitudinal and transverse effective dressed holon pair gap
functions, $\bar{\Delta}_{hL}({\bf k})=\Sigma^{(h)}_{2L}({\bf k},
\omega)\mid_{\omega=0}=\bar{\Delta}_{hL}\gamma^{(d)}_{\bf k}$,
$\bar{\Delta}_{hT}({\bf k})=\Sigma^{(h)}_{2T}({\bf k},\omega)
\mid_{\omega=0}=\bar{\Delta}_{hT}$, and the dressed holon
quasiparticle coherent weights,
\begin{subequations}
\begin{eqnarray}
{1\over Z^{(1)}_{hFA}}&=&{1\over Z_{hF1}}-{1\over Z_{hF2}},\\
{1\over Z^{(2)}_{hFA}}&=&{1\over Z_{hF1}}+{1\over Z_{hF2}},
\end{eqnarray}
\end{subequations}
with the longitudinal and transverse dressed holon quasiparticle
coherent weights,
\begin{subequations}
\begin{eqnarray}
{1\over Z_{hF1}}&=&1-\Sigma^{(ho)}_{1L}({\bf k}_{A},\omega)
\mid_{\omega=0}, \\
{1\over Z_{hF2}}&=& \Sigma^{(ho)}_{1T}({\bf k}_{A},\omega)
\mid_{\omega=0},
\end{eqnarray}
\end{subequations}
where $\Sigma^{(ho)}_{1L}({\bf k},\omega)$ and
$\Sigma^{(ho)}_{1T}({\bf k},\omega)$ are the corresponding
antisymmetric parts of the longitudinal and transverse dressed holon
self-energy functions $\Sigma^{(h)}_{1L}({\bf k},\omega)$ and
$\Sigma^{(h)}_{1T}({\bf k},\omega)$, while the longitudinal and
transverse parts of the dressed holon self-energy functions
$\Sigma^{(h)}_{1}({\bf k},\omega)$ and $\Sigma^{(h)}_{2}({\bf k},
\omega)$ are given by \cite{lan2},
\begin{subequations}
\begin{eqnarray}
\Sigma^{(h)}_{1L}({\bf k},i\omega_{n})&=&{1\over N^{2}}\sum_{\bf p,
q}[ R^{(1)}_{\bf{p+q+k}}{1\over\beta}\sum_{ip_{m}}g_{L} ({\bf p+k},
ip_{m}+i\omega_{n})\Pi_{LL}({\bf p},{\bf q},ip_{m})\nonumber \\
&+&R^{(2)}_{\bf{p+q+k}}{1\over\beta}\sum_{ip_{m}}g_{T}({\bf p+k},
ip_{m}+i\omega_{n})\Pi_{TL}({\bf p},{\bf q},ip_{m})],~~~\\
\Sigma^{(h)}_{1T}({\bf k},i\omega_{n})&=&{1\over N^{2}}\sum_{\bf p,
q}[R^{(1)}_{\bf{p+q+k}}{1\over\beta}\sum_{ip_{m}} g_{T}({\bf p+k},
ip_{m}+i\omega_{n})\Pi_{TT}({\bf p},{\bf q},ip_{m})\nonumber \\
&+&R^{(2)}_{\bf{p+q+k}}{1\over\beta}\sum_{ip_{m}}g_{L}({\bf p+k},
ip_{m}+i\omega_{n})\Pi_{LT}({\bf p},{\bf q},ip_{m})],~~~\\
\Sigma^{(h)}_{2L}({\bf k},i\omega_{n})&=&{1\over N^{2}}\sum_{\bf p,
q}[R^{(1)}_{\bf{p+q+k}}{1\over \beta}\sum_{ip_{m}} \Im^{\dag}_{L}
({\bf p+k},ip_{m}+i\omega_{n})\Pi_{LL}({\bf p},{\bf q},ip_{m})
\nonumber \\
&+&R^{(2)}_{\bf{p+q+k}}{1\over\beta}\sum_{ip_{m}}\Im^{\dag}_{T}
({\bf p+k},ip_{m}+i\omega_{n})\Pi_{TL}({\bf p},{\bf q},ip_{m})],~~~
\\
\Sigma^{(h)}_{2T}({\bf k},i\omega_{n})&=&{1\over N^{2}}\sum_{\bf p,
q} [R^{(1)}_{\bf{p+q+k}}{1\over \beta}\sum_{ip_{m}} \Im^{\dag}_{T}
({\bf p+k},ip_{m}+i\omega_{n})\Pi_{TT}({\bf p},{\bf q},ip_{m})
\nonumber \\
&+&R^{(2)}_{\bf{p+q+k}}{1\over \beta}\sum_{ip_{m}} \Im^{\dag}_{L}
({\bf p+k},ip_{m}+i\omega_{n})\Pi_{LT}({\bf p},{\bf q},ip_{m})],~~~
\end{eqnarray}
\end{subequations}
where $R^{(1)}_{\bf k}=[Z(t\gamma_{\bf k}-t'\gamma'_{\bf k})]^{2}+
t_{\perp}^{2}({\bf k})$, $R^{(2)}_{\bf k}=2Z(t\gamma_{\bf k}- t'
\gamma'_{\bf k})t_{\perp}({\bf k})$, and the spin bubbles,
\begin{eqnarray}
\Pi_{\eta,\eta'}({\bf p},{\bf q},ip_{m})={1\over\beta}\sum_{iq_{m}}
D^{(0)}_{\eta}({\bf q},iq_{m})D^{(0)}_{\eta'}({\bf q+p},iq_{m}+
ip_{m}),
\end{eqnarray}
with $\eta=L,T$ and $\eta'=L,T$, and the mean-field spin Green's
function $D^{(0)}({\bf k},\omega)=D^{(0)}_{L}({\bf k},\omega) +
\sigma_{x}D^{(0)}_{T}({\bf k},\omega)$, with the corresponding
longitudinal and transverse parts have been obtained as \cite{lan2},
\begin{subequations}
\begin{eqnarray}
D^{(0)}_{L}({\bf k},\omega)&=&{1\over 2}\sum_{\nu=1,2} {B_{\nu{\bf
k}}\over \omega^{2}-\omega^{2}_{\nu{\bf k}}}, \\
D^{(0)}_{T}({\bf k},\omega)&=&{1\over 2}
\sum_{\nu=1,2}(-1)^{\nu+1}{B_{\nu{\bf k}}\over \omega^{2}-
\omega^{2}_{\nu{\bf k}}},
\end{eqnarray}
\end{subequations}
where $B_{\nu{\bf k}}=\lambda(A_{1}\gamma_{\bf k}-A_{2})-\lambda{'}
(2\chi_{2}^{z}\gamma {'}_{\bf k}-\chi_{2})-J_{\rm{eff}\perp}
[\chi_{\perp}+2\chi_{\perp}^{z}(-1)^{\nu}][\epsilon_{\perp}({\bf k}
)+(-1)^{\nu}]$, $A_{1}=2\epsilon_{\parallel}\chi_{1}^{z}+\chi_{1}$,
$A_{2}=\epsilon_{\parallel}\chi_{1}+2\chi_{1}^{z}$, $\lambda=
2ZJ_{\rm eff}$, $\lambda{'}=4Z\phi_{2}t'$, $\epsilon_{\parallel}=
1+2t\phi_{1}/J_{\rm eff}$, $\epsilon_{\perp}({\bf{k}})=1+
4\phi_{\perp}t_{\perp}({\bf{k}})/J_{\rm eff\perp}$, the spin
correlation functions $\chi_{1}^{z}=\langle S_{ia}^{z}
S_{i+\hat{\eta}a}^{z}\rangle$, $\chi_{2}^{z}=\langle S_{ia}^{z}
S_{i+\hat{\tau}a}^{z}\rangle$,  $\chi^{z}_{\perp}=\langle S_{i1}^{z}
S_{i2}^{z}\rangle$, the dressed holon particle-hole order parameters
$\phi_{1}=\langle h^{\dagger}_{ia\sigma} h_{i+\hat{\eta}a\sigma}
\rangle$, $\phi_{2}=\langle h^{\dagger}_{ia\sigma}
h_{i+\hat{\tau}a\sigma}\rangle$, $\phi_{\perp}=\langle
h^{\dagger}_{i1\sigma}h_{i2\sigma}\rangle$, and the mean-field spin
excitation spectrum,
\begin{eqnarray}
\omega^{2}_{\nu{\bf k}}&=&\lambda^{2}\left [\left (A_{4}-\alpha
\epsilon_{\parallel}\chi_{1}^{z}\gamma_{\bf k}-{1\over 2Z}\alpha
\epsilon_{\parallel}\chi_{1}\right )(1-\epsilon_{\parallel}
\gamma_{\bf k})\right .\nonumber \\
&+&\left . {1\over 2}\epsilon_{\parallel}\left (A_{3}-{2\over
Z}\alpha\chi_{1}^{z} -\alpha\chi_{1}\gamma_{\bf k}\right )
(\epsilon_{\parallel}-\gamma_{\bf k})\right ]
\nonumber\\
&+&\lambda{'}^{2}\left [\alpha\left (\chi_{2}^{z}\gamma{'}_{\bf k}
-{Z-1\over 2Z}\chi_{2}\right )\gamma{'}_{\bf k}+{1\over 2}\left
(A_{5}-{2\over Z}\alpha \chi_{2}^{z}\right )\right ]\nonumber\\
&+&\lambda\lambda{'}\alpha \left [\chi_{1}^{z}(1-
\epsilon_{\parallel}\gamma_{\bf k})\gamma{'}_{\bf k}+{1\over 2}
(\chi_{1}\gamma{'}_{\bf k}-C_{2})(\epsilon_{\parallel}-\gamma_{\bf
k})+\gamma{'}_{\bf k}(C_{2}^{z} -\epsilon_{\parallel} \chi_{2}^{z}
\gamma_{\bf k})\right .\nonumber\\
&-&\left . {1 \over 2} \epsilon_{\parallel}(C_{2}-\chi_{2}
\gamma_{\bf k})\right ]+\lambda J_{\rm eff \perp}\alpha\left \{
{1\over 2}\epsilon_{\perp}({\bf k})(\epsilon_{\parallel}-\gamma_{\bf
k})[C_{\perp}+\chi_{1}(-1)^{\nu}]\right .\nonumber\\
&+&(1-\epsilon_{\parallel}\gamma_{\bf k})
[C_{\perp}^{z}+\chi_{1}^{z}\epsilon_{\perp}({\bf k})(-1)^{\nu}]
\nonumber\\
&+&\left . [\epsilon_{\perp}({\bf k})+(-1)^{\nu}]\left [{1\over 2}
\epsilon_{\parallel}(C_{\perp}-\chi_{\perp}\gamma_{\bf k})+
(C_{\perp}^{z}-\epsilon_{\parallel}\chi_{\perp}^{z}\gamma_{\bf k})
(-1)^{\nu}\right ]\right \}\nonumber\\
&+&\lambda{'}J_{\rm eff \perp}\alpha\left \{\gamma{'}_{\bf k}
[C{'}_{\perp}^{z}+\chi_{2}^{z}\epsilon_{\perp}({\bf k})(-1)^{\nu}]
-{1 \over 2}\epsilon_{\perp}({\bf k})[C'_{\perp}+\chi_{2}(-1)^{\nu}]
\right . \nonumber\\
&+&\left . \left [{1 \over 2}(\chi_{\perp}\gamma{'}_{\bf k}-
C'_{\perp})+\chi_{\perp}^{z}\gamma{'}_{\bf k}(-1)^{\nu}\right ]
[\epsilon_{\perp}({\bf k})+(-1)^{\nu}]\right \} \nonumber \\
&+&{1\over 4} J_{\rm{eff}\perp}^{2}[\epsilon_{\perp} ({\bf k})
+(-1)^{\nu}]^{2},
\end{eqnarray}
where $A_{3}=\alpha C_{1}+(1-\alpha)/2Z$, $A_{4}=\alpha C_{1}^{z}+
(1-\alpha)/4Z$, $A_{5}=\alpha C_{3}+(1-\alpha)/2Z$, and the spin
correlation functions $C_{1}=(1/Z^{2})\sum_{\hat{\eta}\hat{\eta'}}
\langle S_{i+\hat{\eta}a}^{+}S_{i+\hat{\eta'}a}^{-}\rangle$, $C_{2}
=(1/Z^{2})\sum_{\hat{\eta}\hat{\tau}}\langle S_{i+\hat{\eta}a}^{+}
S_{i+\hat{\tau}a}^{-}\rangle$, $C_{3}=(1/Z^{2})
\sum_{\hat{\tau}\hat{\tau'}}\langle S_{i+\hat{\tau}a}^{+}
S_{i+\hat{\tau'}a}^{-}\rangle$, $C_{1}^{z}= (1/Z^{2})
\sum_{\hat{\eta}\hat{\eta'}}$ $\langle S_{i+\hat{\eta}a}^{z}
S_{i+\hat{\eta'}a}^{z}\rangle$, $C_{2}^{z}=(1/Z^{2})
\sum_{\hat{\eta}\hat{\tau}}\langle S_{i+\hat{\eta}a}^{z}
S_{i+\hat{\tau}a}^{z}\rangle$, $C_{\perp}=(1/Z)\sum_{\hat{\eta}}
\langle S_{i1}^{+} S_{i+\hat{\eta}2}^{-}\rangle$, $C{'}_{\perp}=
(1/Z)\sum_{\hat{\tau}}\langle S_{i1}^{+} S_{i+\hat{\tau}2}^{-}
\rangle$, $C_{\perp}^{z}=(1/Z) \sum_{\hat{\eta}}\langle S_{i1}^{z}
S_{i+ \hat{\eta}2}^{z}\rangle$, $C{'}_{\perp}^{z}=(1/Z)
\sum_{\hat{\tau}}$ $\langle S_{i1}^{z} S_{i+\hat{\tau}2}^{z}
\rangle$.

As in the single layer case \cite{guo2,feng4} discussed in section
3, we now can calculate the electron normal and anomalous Green's
functions $G(i-j,t-t')=\langle\langle C_{i\sigma}(t);
C^{\dagger}_{j\sigma}(t')\rangle \rangle=G_{L}(i-j,t-t')+\sigma_{x}
G_{T}(i-j,t-t')$ and $\Gamma^{\dagger}(i-j,t-t')=\langle\langle
C^{\dagger}_{i\uparrow} (t);C^{\dagger}_{j\downarrow}(t')\rangle
\rangle= \Gamma^{\dagger}_{L}(i-j,t-t')+\sigma_{x}
\Gamma^{\dagger}_{T} (i-j,t-t')$ in terms of the full dressed holon
normal and anomalous Green's functions in Eqs. (53) and mean-field
spin Green's function in Eqs. (60), and can be evaluated explicitly
as,
\begin{subequations}
\begin{eqnarray}
G_{L}({\bf k},\omega)&=&{1\over 8N}\sum_{\bf p}\sum_{\mu\nu}
Z_{hFA}^{(\mu)}{B_{\nu{\bf p}}\over\omega_{\nu{\bf p}}}\left [
L^{(1)}_{\mu\nu}({\bf k,p}) \left ({U^{2}_{h\mu{\bf p-k}}\over
\omega+E_{h\mu{\bf p-k}}-\omega_{\nu{\bf p}}}\right .\right .
\nonumber \\
&+&\left . {V^{2}_{h\mu{\bf p-k}}\over\omega-E_{h\mu{\bf p-k}}+
\omega_{\nu{\bf p}}}\right )+L^{(2)}_{\mu\nu}({\bf k,p})\left (
{U^{2}_{h\mu{\bf p-k}}\over \omega+E_{h\mu{\bf p-k}}+
\omega_{\nu{\bf p}}}\right . \nonumber \\
&+&\left .\left. {V^{2}_{h\mu{\bf p-k}}\over\omega-E_{h\mu{\bf p-k}}
-\omega_{\nu{\bf p}}}\right )\right ] ,
\end{eqnarray}
\begin{eqnarray}
G_{T}({\bf k},\omega)&=&{1 \over 8N}\sum_{\bf p}\sum_{\mu\nu}
(-1)^{\mu+\nu}Z_{hFA}^{(\mu)}{B_{\nu{\bf p}}\over\omega_{\nu{\bf p}}
}\left [ L^{(1)}_{\mu\nu}({\bf k,p}) \left ({U^{2}_{h\mu{\bf p-k}}
\over\omega+E_{h\mu{\bf p-k}}-\omega_{\nu{\bf p}}}\right.\right.
\nonumber \\
&+&\left. {V^{2}_{h\mu{\bf p-k}}\over \omega-E_{h\mu{\bf
p-k}}+\omega_{\nu{\bf p}}}\right )+L^{(2)}_{\mu\nu}({\bf k,p})\left
({U^{2}_{h\mu{\bf p-k}}\over\omega+E_{h\mu{\bf p-k}}+\omega_{\nu{\bf
p}}}\right .\nonumber \\
&+&\left .\left . {V^{2}_{h\mu{\bf p-k}}\over\omega-E_{h\mu{\bf p-k}
}-\omega_{\nu{\bf p}}}\right )\right ] ,\\
\Gamma^{\dagger}_{L}({\bf k},\omega)&=&{1\over 8N}\sum_{\bf p}
\sum_{\mu\nu}Z^{(\mu)}_{hFA}{\bar{\Delta}_{hz}^{(\mu)}({\bf p-k})
\over 2E_{h\mu{\bf p-k}}}{B_{\nu{\bf p}}\over\omega_{\nu{\bf p}}}
\nonumber \\
&\times& \left [ L^{(1)}_{\mu\nu}({\bf k,p}) \left ({1\over\omega-
E_{h\mu{\bf p-k}}+\omega_{\nu{\bf p}}}-{1\over\omega+E_{h\mu{\bf
p-k}}-\omega_{\nu{\bf p}}}\right)\right .\nonumber \\
&+&\left . L^{(2)}_{\mu\nu}({\bf k,p})\left ({1\over\omega-
E_{h\mu{\bf p-k}}-\omega_{\nu{\bf p}}}-{1\over\omega+E_{h\mu{\bf
p-k}}+\omega_{\nu{\bf p}}}\right )\right ],~~~\\
\Gamma^{\dagger}_{T}({\bf k},\omega)&=&{1\over 8N}\sum_{\bf p}
\sum_{\mu\nu}(-1)^{\mu+\nu}Z^{(\mu)}_{hFA}{\bar{\Delta}_{hz}^{(\mu)}
({\bf p-k})\over 2E_{h\mu{\bf p-k}}}{B_{\nu{\bf p}}\over
\omega_{\nu{\bf p}}}\nonumber \\
&\times& \left [ L^{(1)}_{\mu\nu}({\bf k,p})\left ({1 \over \omega-
E_{h\mu{\bf p-k}}+\omega_{\nu{\bf p}}}-{1\over\omega+E_{h\mu{\bf p-
k}}-\omega_{\nu{\bf p}}} \right) \right . \nonumber \\
&+& \left . L^{(2)}_{\mu\nu}({\bf k,p})\left ({1\over\omega-
E_{h\mu{\bf p-k}} -\omega_{\nu{\bf p}}}- {1\over\omega+E_{h\mu{\bf
p-k}}+ \omega_{\nu{\bf p}}}\right )\right ],~~~
\end{eqnarray}
\end{subequations}
where $L^{(1)}_{\mu\nu}({\bf k,p})=[{\rm coth}(\beta\omega_{\nu{\bf
p}}/2)-{\rm tanh}(\beta E_{h\mu{\bf p-k}}/2)]/2$ and
$L^{(2)}_{\mu\nu}({\bf k,p})=[{\rm coth}(\beta\omega_{\nu{\bf p}
}/2)$ $+{\rm tanh}(\beta E_{h\mu{\bf p-k}}/2)]/2$, and the dressed
holon effective gap parameters and quasiparticle coherent weights
satisfy the following four equations,
\begin{subequations}
\begin{eqnarray}
\bar{\Delta}_{hL}&=&-{4\over 32N^{3}}\sum_{{\bf k,q,p}}
\sum_{\nu,\nu',\nu''}\gamma^{(d)}_{\bf k-p+q}C_{\nu\nu''}({\bf k+q})
{Z^{(\nu'')}_{hFA}B_{\nu'{\bf p}} B_{\nu{\bf q}}\over
\omega_{\nu'{\bf p}}\omega_{\nu{\bf q}}}\bar{\Delta}^{(\nu'')}_{hz}
({\bf k})\nonumber \\
&\times&\left ({F^{(1)}_{\nu\nu'\nu''}({\bf q,p})+
F^{(2)}_{\nu\nu'\nu''}({\bf k,q,p})\over [\omega_{\nu'{\bf p}}
-\omega_{\nu{\bf q}}]^2-E^{2}_{h\nu''{\bf k}}} \right .\nonumber \\
&+&\left . {F^{(3)}_{\nu\nu'\nu''}({\bf q,p})+
F^{(4)}_{\nu\nu'\nu''}({\bf k, q, p})\over [\omega_{\nu'{\bf p}}+
\omega_{\nu{\bf q}}]^{2}-E^{2}_{h\nu''{\bf k}}} \right ),~~~~~~~\\
\bar{\Delta}_{hT}&=&-{1\over 32N^{3}}\sum_{{\bf k,q,p}}
\sum_{\nu,\nu',\nu''}(-1)^{\nu+\nu'+\nu''+1}C_{\nu\nu''}({\bf k+q})
{Z^{(\nu'')}_{hFA}B_{\nu'{\bf p}} B_{\nu{\bf q}}\over
\omega_{\nu'{\bf p}}\omega_{\nu{\bf q}}}\bar{\Delta}^{(\nu'')}_{hz}
({\bf k})\nonumber \\
&\times&\left ({F^{(1)}_{\nu\nu'\nu''}({\bf q,p})+
F^{(2)}_{\nu\nu'\nu''}({\bf k,q,p})\over [\omega_{\nu'{\bf p}}
-\omega_{\nu{\bf q}}]^2-E^{2}_{h\nu''{\bf k}}}\right . \nonumber \\
&+&\left . {F^{(3)}_{\nu\nu'\nu''}({\bf q,p})+
F^{(4)}_{\nu\nu'\nu''}({\bf k, q, p})\over [\omega_{\nu'{\bf p}}+
\omega_{\nu{\bf q}}]^{2} -E^{2}_{h\nu''{\bf k}}} \right ),~~~~~~~~
\end{eqnarray}
\begin{eqnarray}
{1\over Z_{hFA}^{(1)}}&=&1+{1\over 32N^{2}}\sum_{{\bf q,p}}
\sum_{\nu,\nu',\nu''}[1+(-1)^{\nu+\nu'+\nu''+1}]C_{\nu\nu''}({\bf p+
k}_{A})\nonumber\\
&\times& {Z^{(\nu'')}_{hFA}B_{\nu'{\bf p}}B_{\nu{\bf q}}\over
\omega_{\nu'{\bf p}}\omega_{\nu{\bf q}}} \left (
{H^{(1)}_{\nu\nu'\nu''}({\bf q,p})\over [\omega_{\nu'{\bf p}}-
\omega_{\nu{\bf q}}+E_{h\nu''{\bf p-q+ k}_{A}}]^{2}}\right .
\nonumber\\
&+&{H^{(2)}_{\nu\nu'\nu''}({\bf q,p})\over [\omega_{\nu'{\bf p}}
-\omega_{\nu{\bf q}}-E_{h\nu''{\bf p-q+ k}_{A}}]^{2}}+
{H^{(3)}_{\nu\nu'\nu''}({\bf q,p})\over [\omega_{\nu'{\bf p}}+
\omega_{\nu{\bf q}}+E_{h\nu''{\bf p-q+k}_{A}}]^{2}}\nonumber \\
&+&\left. {H^{(4)}_{\nu\nu'\nu''}({\bf q,p})\over [\omega_{\nu' {\bf
p}}+\omega_{\nu{\bf q}}-E_{h\nu''{\bf p-q+k}_{A}}]^{2}}\right),\\
{1\over Z_{hFA}^{(2)}}&=&1+{1\over 32N^{2}}\sum_{{\bf q,p}}
\sum_{\nu,\nu',\nu''}[1-(-1)^{\nu+\nu'+\nu''+1}]C_{\nu\nu''}({\bf p+
k}_{A})\nonumber \\
&\times& {Z^{(\nu'')}_{hFA} B_{\nu'{\bf p}}B_{\nu{\bf q}}\over
\omega_{\nu'{\bf p}}\omega_{\nu{\bf q}}} \left (
{H^{(1)}_{\nu\nu'\nu''}({\bf q,p})\over [\omega_{\nu'{\bf p} }-
\omega_{\nu{\bf q}}+E_{h\nu''{\bf p-q+ k}_{A}}]^{2}}\right .
\nonumber \\
&+& {H^{(2)}_{\nu\nu'\nu''}({\bf q,p})\over [\omega_{\nu'{\bf p}}
-\omega_{\nu{\bf q}}-E_{h\nu''{\bf p-q+ k}_{A}}]^{2}}+
{H^{(3)}_{\nu\nu'\nu''}({\bf q,p})\over[\omega_{\nu'{\bf p}}+
\omega_{\nu{\bf q}}+E_{h\nu''{\bf p-q+k}_{A}}]^{2}} \nonumber \\
&+&\left. {H^{(4)}_{\nu\nu'\nu''}({\bf q,p})\over [\omega_{\nu' {\bf
p}}+\omega_{\nu{\bf q}}-E_{h\nu''{\bf p-q+k}_{A}}]^{2}}\right),
\end{eqnarray}
\end{subequations}
where $C_{\nu\nu''}({\bf k})=[Z(t\gamma_{\bf k}-t'\gamma{'}_{\bf k})
+(-1)^{\nu+\nu''}t_{\perp}({\bf k})]^{2}$, and
\begin{subequations}
\begin{eqnarray}
F^{(1)}_{\nu\nu'\nu''}({\bf q, p})&=&n_{B} (\omega_{\nu{\bf q}})+
n_{B}(\omega_{\nu'{\bf p}})+2n_{B} (\omega_{\nu{\bf q}})
n_{B}(\omega_{\nu'{\bf p}}), \\
F^{(2)}_{\nu\nu'\nu''}({\bf k,q,p})&=&{\omega_{\nu{\bf q}}-
\omega_{\nu'{\bf p}}\over E_{h\nu''{\bf k}}}{\rm tanh}[{1\over 2}
\beta E_{h\nu''{\bf k}}][n_{B}(\omega_{\nu{\bf q}})
-n_{B}(\omega_{\nu'{\bf p}})],\\
F^{(3)}_{\nu\nu'\nu''}({\bf q,p})&=&[1+n_{B}(\omega_{\nu{\bf q}})]
[1+n_{B}(\omega_{\nu'{\bf p}})]
+n_{B}(\omega_{\nu{\bf q}})n_{B}(\omega_{\nu'{\bf p}}), \\
F^{(4)}_{\nu\nu'\nu''}({\bf k,q,p})&=&{\omega_{\nu'{\bf p}}+
\omega_{\nu{\bf q}}\over -E_{h\nu''{\bf k}}}{\rm tanh}[{1\over 2}
\beta E_{h\nu''{\bf k}}][1+n_{B}(\omega_{\nu{\bf q}})
+n_{B}(\omega_{\nu'{\bf p}})],~~~~~~~\\
H^{(1)}_{\nu\nu'\nu''}({\bf q,p})&=&n_{F}(E_{h\nu''{\bf p-q+k}_{A}})
[n_{B}(\omega_{\nu'{\bf p}})-n_{B}(\omega_{\nu{\bf q}})]\nonumber \\
&+&n_{B}(\omega_{\nu{\bf q}})[1+n_{B}(\omega_{\nu'{\bf p}})], \\
H^{(2)}_{\nu\nu'\nu''}({\bf q,p})&=&n_{F}(E_{h\nu''{\bf p-q+k}_{A}})
[n_{B}(\omega_{\nu{\bf q}})-n_{B}(\omega_{\nu'{\bf p}})]\nonumber \\
&+&n_{B}(\omega_{\nu'{\bf p}})[1+n_{B}(\omega_{\nu{\bf q}})],\\
H^{(3)}_{\nu\nu'\nu''}({\bf q, p})&=&[1-n_{F}(E_{h\nu''{\bf p-q+
k}_{A}})][1+n_{B}(\omega_{\nu{\bf q}})+n_{B}(\omega_{\nu'{\bf p}})]
\nonumber \\
&+&n_{B}(\omega_{\nu{\bf q}})n_{B}(\omega_{\nu'{\bf p}}),\\
H^{(4)}_{\nu\nu'\nu''}({\bf q,p})&=&n_{F}(E_{h\nu''{\bf p-q+k}_{A}})
[1+n_{B}(\omega_{\nu{\bf q}})+ n_{B}(\omega_{\nu'{\bf
p}})]\nonumber\\
&+&n_{B}(\omega_{\nu{\bf q}}) n_{B} (\omega_{\nu'{\bf p}}).~~~
\end{eqnarray}
\end{subequations}
As in the single layer case \cite{guo2,feng4} discussed in section
3, these four equations must be solved self-consistently in
combination with other equations,
\begin{subequations}
\begin{eqnarray}
\delta&=&{1\over 4N}\sum_{\nu,{\bf k}}Z^{(\nu)}_{hFA}\left (1-
{\bar{\xi}_{\nu{\bf k}}\over E_{h\nu{\bf k}}}{\rm tanh}[{1\over 2}
\beta E_{h\nu{\bf k}}]\right ), \\
\phi_{1}&=&{1\over 4N}\sum_{\nu,{\bf k}}\gamma_{\bf k}
Z^{(\nu)}_{hFA}\left (1-{\bar{\xi}_{\nu{\bf k}}\over E_{h\nu{\bf k}
}}{\rm tanh}[{1\over 2}\beta E_{h\nu{\bf k}}]\right ), \\
\phi_{2}&=&{1\over 4N}\sum_{\nu,{\bf k}} \gamma{'}_{\bf k}
Z^{(\nu)}_{hFA} \left (1-{\bar{\xi}_{\nu{\bf k}}\over E_{h\nu{\bf k}
}}{\rm tanh}[{1\over 2}\beta E_{h\nu{\bf k}}]\right ), \\
\phi_{\perp}&=&{1\over 4N}\sum_{\nu,{\bf k}} (-1)^{\nu+1}
Z^{(\nu)}_{hFA}\left (1-{\bar{\xi}_{\nu{\bf k}}\over E_{h\nu{\bf k}
}}{\rm tanh}[{1\over 2}\beta E_{h\nu{\bf k}}]\right ),\\
1\over 2&=&{1\over 4N}\sum_{\nu,{\bf k}}{B_{\nu{\bf k}}\over
\omega_{\nu{\bf k}}}{\rm coth}[{1\over 2}\beta \omega_{\nu{\bf k}}],
\\
\chi_{1}&=&{1\over 4N}\sum_{\nu,{\bf k}}\gamma_{\bf k}{B_{\nu{\bf k}
}\over\omega_{\nu{\bf k}}}{\rm coth}
[{1\over 2}\beta\omega_{\nu{\bf k}}], \\
\chi_{2}&=&{1\over 4N}\sum_{\nu,{\bf k}} \gamma{'}_{\bf k}
{B_{\nu{\bf k}}\over\omega_{\nu{\bf k}}}{\rm coth}
[{1\over 2}\beta\omega_{\nu{\bf k}}], \\
C_{1}&=&{1\over 4N}\sum_{\nu,{\bf k}}\gamma_{\bf k}^{2}{B_{\nu{\bf k
}}\over\omega_{\nu{\bf k}}}{\rm coth}
[{1\over 2}\beta\omega_{\nu{\bf k}}], \\
C_{2}&=&{1\over 4N}\sum_{\nu,{\bf k}} \gamma_{\bf k}\gamma{'}_{\bf
k}{B_{\nu{\bf k}}\over\omega_{\nu{\bf k}}}{\rm coth}
[{1\over 2}\beta\omega_{\nu{\bf k}}], \\
C_{3}&=&{1\over 4N}\sum_{\nu,{\bf k}} \gamma{'}_{\bf
k}^{2}{B_{\nu{\bf k}}\over\omega_{\nu{\bf k}}}{\rm coth}
[{1\over 2}\beta\omega_{\nu{\bf k}}], \\
\chi_{1}^{z}&=&{1\over 4N}\sum_{\nu,{\bf k}} \gamma_{\bf k}
{B_{z\nu{\bf k}}\over\omega_{z\nu{\bf k}}}{\rm coth}
[{1\over 2}\beta\omega_{z\nu{\bf k}}], \\
\chi_{2}^{z}&=&{1\over 4N}\sum_{\nu,{\bf k}} \gamma{'}_{\bf k}
{B_{z\nu{\bf k}}\over\omega_{z\nu{\bf k}}}{\rm coth}
[{1\over 2}\beta\omega_{z\nu{\bf k}}], \\
C_{1}^{z}&=&{1\over 4N}\sum_{\nu,{\bf k}} \gamma_{\bf k}^{2}
{B_{z\nu{\bf k}}\over\omega_{z\nu{\bf k}}}{\rm coth}
[{1\over 2}\beta\omega_{z\nu{\bf k}}], \\
C_{2}^{z}&=&{1\over 4N}\sum_{\nu,{\bf k}} \gamma_{\bf k}
\gamma{'}_{\bf k} {B_{z\nu{\bf k}}\over\omega_{z\nu{\bf k}}}{\rm
coth} [{1\over 2}\beta\omega_{z\nu{\bf k}}],\\
\chi_{\perp}&=&{1\over 4N}\sum_{\nu,{\bf k}}(-1)^{\nu+1} {B_{\nu{\bf
k}}\over\omega_{\nu{\bf k}}}{\rm coth}[{1\over
2}\beta\omega_{\nu{\bf k}}], \\
C_{\perp}&=&{1\over 4N}\sum_{\nu,{\bf k}}(-1)^{\nu+1} \gamma_{\bf
k}{B_{\nu{\bf k}}\over\omega_{\nu{\bf k}}}
{\rm coth}[{1\over 2}\beta\omega_{\nu{\bf k}}], \\
C{'}_{\perp}&=&{1\over 4N}\sum_{\nu,{\bf k}}(-1)^{\nu+1}
\gamma{'}_{\bf k}{B_{\nu{\bf k}}\over\omega_{\nu{\bf k}}} {\rm
coth}[{1\over 2}\beta\omega_{\nu{\bf k}}],
\end{eqnarray}
\begin{eqnarray}
\chi^{z}_{\perp}&=&{1\over 4N}\sum_{\nu,{\bf k}}(-1)^{\nu+1}
{B_{z\nu{\bf k}}\over\omega_{z\nu{\bf k}}}{\rm coth}[{1\over
2}\beta\omega_{z\nu{\bf k}}], \\
C_{\perp}^{z}&=&{1\over 4N}\sum_{\nu,{\bf k}}(-1)^{\nu+1}
\gamma_{\bf k}{B_{z\nu{\bf k}}\over\omega_{z\nu{\bf k}}}{\rm
coth}[{1\over 2}\beta\omega_{z\nu{\bf k}}],\\
C{'}_{\perp}^{z}&=&{1\over 4N}\sum_{\nu,{\bf k}}(-1)^{\nu+1}
\gamma{'}_{\bf k}{B_{z\nu{\bf k}}\over\omega_{z\nu{\bf k}}}{\rm
coth}[{1\over 2}\beta\omega_{z\nu{\bf k}}],
\end{eqnarray}
\end{subequations}
then all order parameters, decoupling parameter $\alpha$, and
chemical potential $\mu$ are determined by the self-consistent
calculation \cite{lan2}.

From the electron normal and anomalous Green's functions in Eqs.
(62), we obtain the longitudinal and transverse parts of the
electron spectral function and superconducting gap function as,
\begin{subequations}
\begin{eqnarray}
A_{L}({\bf k},\omega)&=&\pi{1\over 4N}\sum_{\bf p}\sum_{\mu\nu}
Z_{hFA}^{(\mu)}{B_{\nu{\bf p}}\over\omega_{\nu{\bf p}}}\{
L^{(1)}_{\mu\nu}({\bf k,p})[U^{2}_{h\mu{\bf p-k}}
\delta(\omega+E_{h\mu{\bf p-k}}-\omega_{\nu{\bf p}}) \nonumber\\
&+&V^{2}_{h\mu{\bf p-k}}\delta(\omega-E_{h\mu{\bf p-k}}+
\omega_{\nu{\bf p}})]\nonumber \\
&+&L^{(2)}_{\mu\nu}({\bf k,p})[U^{2}_{h\mu{\bf
p-k}}\delta(\omega+E_{h\mu{\bf p-k}}+ \omega_{\nu{\bf p}})
\nonumber \\
&+&V^{2}_{h\mu{\bf p-k}}\delta(\omega-E_{h\mu{\bf p-k}}-
\omega_{\nu{\bf p}})]\}, \\
A_{T}({\bf k},\omega)&=&\pi{1\over 4N}\sum_{\bf p}\sum_{\mu\nu}
(-1)^{\mu+\nu}Z_{hFA}^{(\mu)}{B_{\nu{\bf p}}\over\omega_{\nu{\bf p}}
}\nonumber\\
&\times&\{ L^{(1)}_{\mu\nu}({\bf k,p})[U^{2}_{h\mu{\bf p-k}}
\delta(\omega+E_{h\mu{\bf p-k}}-\omega_{\nu{\bf p}})\nonumber\\
&+&V^{2}_{h\mu{\bf p-k}}\delta(\omega-E_{h\mu{\bf p-k}}+
\omega_{\nu{\bf p}})]\nonumber \\
&+&L^{(2)}_{\mu\nu}({\bf k,p})[U^{2}_{h\mu{\bf
p-k}}\delta(\omega+E_{h\mu{\bf p-k}}+ \omega_{\nu{\bf p}})\nonumber
\\
&+&V^{2}_{h\mu{\bf p-k}}\delta(\omega-E_{h\mu{\bf p-k}}-
\omega_{\nu{\bf p}})]\},\\
\Delta_{L}(\bf k)&=&-{1\over 16N}\sum_{{\bf p},\mu,\nu}
Z^{(\mu)}_{hFA} {\bar{\Delta}_{hz}^{(\mu)}({\bf p-k}) \over
E_{h\mu{\bf p-k}}}\nonumber\\
&\times&{B_{\nu{\bf p}}\over\omega_{\nu{\bf p}}}{\rm tanh} [{1\over
2}\beta E_{h\mu{\bf p-k}}]{\rm coth}
[{1\over 2}\beta\omega_{\nu{\bf p}}], \\
\Delta_{T}(\bf k)&=&-{1\over 16N}\sum_{{\bf p},\mu,\nu}
(-1)^{\mu+\nu}Z^{(\mu)}_{hFA}{\bar{\Delta}_{hz}^{(\mu)}({\bf p-k})
\over E_{h\mu{\bf p-k}}}\nonumber\\
&\times&{B_{\nu{\bf p}}\over\omega_{\nu{\bf p}}} {\rm tanh}[{1\over
2}\beta E_{h\mu{\bf p-k}}]{\rm coth} [{1\over 2}
\beta\omega_{\nu{\bf p}}].
\end{eqnarray}
\end{subequations}
With the help of these longitudinal and transverse parts of the
superconducting gap functions in Eqs. (66c) and (66d), the
corresponding longitudinal and transverse superconducting gap
parameters are obtained as $\Delta_{L}=- \chi_{1}\Delta_{hL}$ and
$\Delta_{T}=-\chi_{\perp}\Delta_{hT}$, respectively. In the bilayer
coupling case, the more appropriate classification is in terms of
the spectral function and superconducting gap function within the
basis of the antibonding and bonding components
\cite{dlfeng2,dlfeng1,kordyuk,chuang,borisenko}. In this case, the
electron spectral function and superconducting gap parameter can be
transformed from the plane representation to the antibonding-bonding
representation as,
\begin{subequations}
\begin{eqnarray}
A^{(a)}({\bf k},\omega)&=&{1\over 2}[A_{L}({\bf k},\omega)-A_{T}
({\bf k},\omega)],\\
A^{(b)}({\bf k},\omega)&=&{1\over 2}[A_{L}({\bf k},\omega)+ A_{T}
({\bf k},\omega)],\\
\Delta^{(a)}&=&\Delta_{L}-\Delta_{T}, \\
\Delta^{(b)}&=&\Delta_{L}+\Delta_{T}.
\end{eqnarray}
\end{subequations}
respectively, then the antibonding and bonding parts have odd and
even symmetries, respectively.

\begin{figure}[t]
\begin{center}
\begin{minipage}[h]{75mm}
\epsfig{file=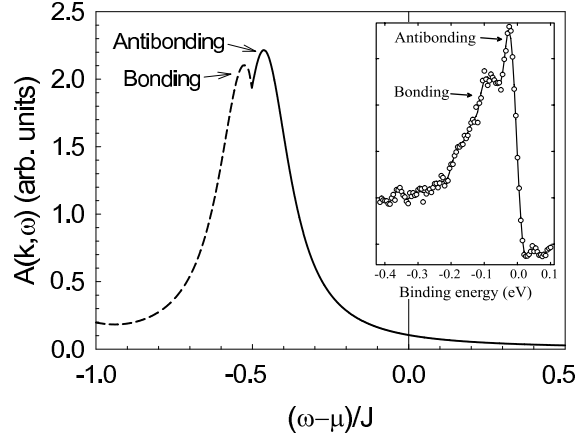, width=75mm}
\end{minipage}
\caption{The antibonding (solid line) and bonding (dashed line)
electron spectral functions of the bilayer cuprate superconductors
at the $[\pi,0]$ point in the superconducting state for $t/J=2.5$,
$t'/t=0.3$, and $t_{\perp}/t=0.35$ with $T=0.002J$ at $\delta=0.15$.
Inset: the corresponding ARPES experimental result
\protect\cite{borisenko} of the bilayer cuprate superconductor
Bi$_{2}$Sr$_{2}$CaCu$_{2}$O$_{8+\delta}$ in the superconducting
state.}
\end{center}
\end{figure}

In Fig. 8, we firstly plot the antibonding (solid line) and bonding
(dashed line) electron spectral functions in the $[\pi,0]$ point for
$t/J=2.5$, $t'/t=0.3$, and $t_{\perp}/t=0.35$ with $T=0.002J$ at
$\delta=0.15$. For comparison, the corresponding ARPES experimental
result \cite{borisenko} of the bilayer cuprate superconductor
Bi$_{2}$Sr$_{2}$CaCu$_{2}$O$_{8+\delta}$ in the superconducting
state is also shown in Fig. 8 (inset). In comparison with the single
layer case \cite{guo2} in section 3, the electron spectrum of the
bilayer system has been split into the bonding and antibonging
components, with the bonding and antibonding superconducting
quasiparticle peaks in the $[\pi,0]$ point are located at the
different positions. In this sense, the differentiation between the
bonding and antibonding components of the electron spectral function
is essential. The antibonding spectrum consists of a low energy
antibonding peak, corresponding to the superconducting peak, and the
bonding spectrum has a higher energy bonding peak, corresponding to
the hump, while the spectral dip is in between them, then the total
contributions for the electron spectrum from both antibonding and
bonding components give rise to the peak-dip-hump structure, in
qualitative agreement with the experimental observation on the
bilayer cuprate superconductors in the superconducting state
\cite{shen,dessau,randeria,fedorov,lu,sato}.

\begin{figure}[t]
\begin{center}
\begin{minipage}[h]{75mm}
\epsfig{file=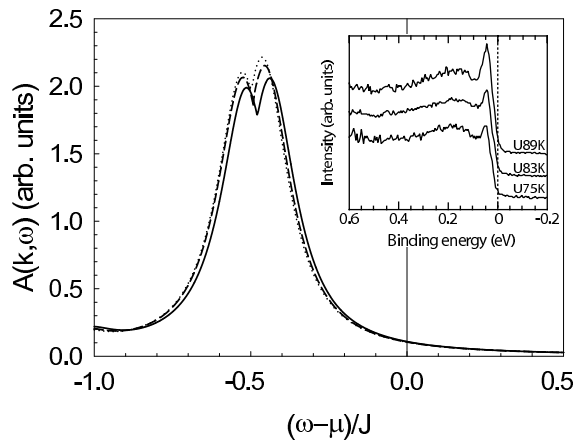, width=75mm}
\end{minipage}
\caption{The electron spectral functions of the bilayer cuprate
superconductors at $[\pi,0]$ point in the superconducting state for
$t/J=2.5$, $t'/t=0.3$, and $t_{\perp}/t=0.35$ with $T=0.002J$ at
$\delta=0.09$ (solid line), $\delta=0.12$ (dashed line), and
$\delta=0.15$ (dotted line). Inset: the corresponding ARPES
experimental results \protect\cite{campuzano2} of the bilayer
cuprate superconductor Bi$_{2}$Sr$_{2}$CaCu$_{2}$O$_{8+\delta}$ in
the superconducting state.}
\end{center}
\end{figure}

We now turn to discuss the doping evolution of the electron spectrum
of bilayer cuprate superconductors in the superconducting state. We
have calculated the electron spectrum at different doping
concentrations, and the result of the electron spectral functions at
the $[\pi,0]$ point for $t/J=2.5$, $t'/t=0.3$, and
$t_{\perp}/t=0.35$ with $T=0.002J$ at $\delta=0.09$ (solid line),
$\delta=0.12$ (dashed line), and $\delta=0.15$ (dotted line) are
plotted in Fig. 9 in comparison with the corresponding ARPES
experimental results \cite{campuzano2} of the bilayer cuprate
superconductor Bi$_{2}$Sr$_{2}$CaCu$_{2}$O$_{8+\delta}$ in the
superconducting state (inset). Obviously, the doping evolution of
the spectral weight of the bilayer superconductor
Bi$_{2}$Sr$_{2}$CaCu$_{2}$O$_{8+\delta}$ in the superconducting
state is reproduced. With increasing the doping concentration, both
superconducting peak and hump become sharper, and then the spectral
weights increase in intensity. Furthermore, we have also calculated
the electron spectrum with different temperatures, and the results
show that the spectral weights of both superconducting peak and hump
are suppressed with increasing temperatures. These results are also
qualitatively consistent with the ARPES experimental results on the
bilayer cuprate superconductors in the superconducting state
\cite{shen,fedorov,campuzano2}.

\begin{figure}[t]
\begin{center}
\begin{minipage}[h]{75mm}
\epsfig{file=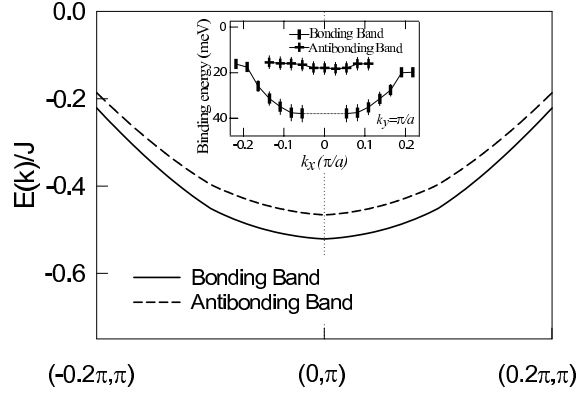, width=75mm}
\end{minipage}
\caption{The positions of the antibonding peaks and bonding humps in
the electron spectrum of the bilayer cuprate superconductors in the
superconducting state as a function of momentum along the direction
$[-0.2\pi,\pi]\rightarrow [0,\pi]\rightarrow [0.2\pi,\pi]$ with
$T=0.002J$ at $\delta=0.15$ for $t/J=2.5$, $t'/t=0.3$, and
$t_{\perp}/t=0.35$. Inset: the corresponding experimental data
\protect\cite{dlfeng1} of the bilayer cuprate superconductor
Bi$_{2}$Sr$_{2}$CaCu$_{2}$O$_{8+\delta}$ in the superconducting
state.}
\end{center}
\end{figure}

To better perceive the superconducting coherence of the antibonding
quasiparticle peak and bonding quasiparticle hump around the
$[\pi,0]$ point, we have made a series of calculations for the
electron spectral function at different momenta, and the result
shows that the sharp superconducting peak from the electron
antibonding spectral function and hump from the bonding spectral
function persist in a very large momentum space region around the
$[\pi,0]$ point. To show this point clearly, we plot the positions
of the antibonding peak and bonding hump in the electron spectrum as
a function of momentum along the direction $[-0.2\pi,\pi]\rightarrow
[0,\pi]\rightarrow [0.2\pi,\pi]$ with $T=0.002J$ at $\delta=0.15$
for $t/J=2.5$, $t'/t=0.3$, and $t_{\perp}/t=0.35$ in Fig. 10 in
comparison with the corresponding experimental data \cite{dlfeng1}
of the bilayer cuprate superconductor
Bi$_{2}$Sr$_{2}$CaCu$_{2}$O$_{8+\delta}$ in the superconducting
state (inset). It is shown that there are two branches in the
quasiparticle dispersion, with upper branch corresponding to the
antibonding quasiparticle dispersion, and lower branch corresponding
to the bonding quasiparticle dispersion. Furthermore, the bilayer
splitting reaches its maximum at the $[\pi,0]$ point, in qualitative
agreement with the ARPES experimental measurements on the bilayer
cuprate superconductors in the superconducting state
\cite{shen,dessau,randeria,fedorov,lu,sato}.

\begin{figure}[t]
\begin{center}
\begin{minipage}[h]{75mm}
\epsfig{file=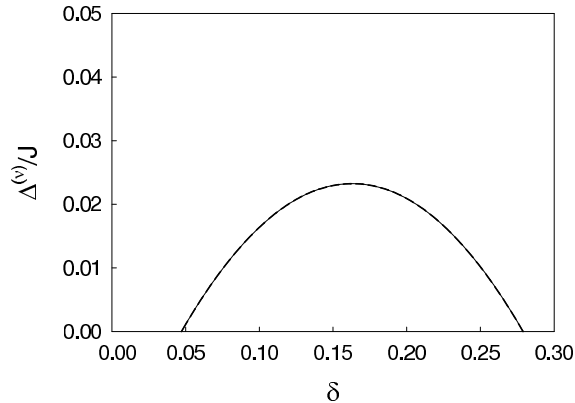, width=75mm}
\end{minipage}
\caption{The antibonding (solid line) and bonding (dashed line)
superconducting gap parameters of the bilayer cuprate
superconductors as a function of the doping concentration with
$T=0.002J$ for $t/J=2.5$, $t'/t=0.3$, and $t_{\perp}/t=0.35$.}
\end{center}
\end{figure}

In the above calculations for the bilayer systems, we find that
although the antibonding superconducting peak and bonding hump have
different dispersions, the transverse part of the superconducting
gap parameter $\Delta_{T}\approx 0$. To show this point clearly, we
plot the antibonding and bonding gap parameters in Eqs. (67c) and
(67d) as a function of the doping concentration with $T=0.002J$ for
$t/J=2.5$, $t'/t=0.3$, and $t_{\perp}/t=0.35$ in Fig. 11. As seen
from Fig. 11, both antibonding and bonding gap parameters have the
same d-wave superconducting gap magnitude in a given doping
concentration, i.e., $\Delta^{a}\approx\Delta^{b}$. This result
shows that although there is a single electron interlayer coherent
hopping (49) in the bilayer cuprate superconductors in the
superconducting state, the electron interlayer pairing interaction
vanishes. This reflects that in the kinetic energy driven
superconducting mechanism, the weak dressed holon-spin interaction
due to the interlayer coherent hopping (49) from the kinetic energy
term of the bilayer $t$-$J$ model does not induce the dressed holon
interlayer pairing state by exchanging spin excitations in the
higher power of the doping concentration. This is different from the
dressed holon-spin interaction due to the intralayer hopping from
the kinetic energy term of the bilayer $t$-$J$ model, it can induce
superconductivity by exchanging spin excitations in the higher power
of the doping concentration \cite{feng2}. This result is also
consistent with the ARPES experimental results of the bilayer
cuprate superconductor Bi(Pb)$_{2}$Sr$_{2}$CaCu$_{2}$O$_{8+\delta}$
in superconducting state \cite{dlfeng1,borisenko}, where the
superconducting gap separately for the bonding and antibonding bands
has been measured, and it is found that both d-wave superconducting
gaps from the antibonding and bonding components are identical
within the experimental uncertainties.

Now we give some physical interpretation to the above obtained
results. We find that there are two main reasons why the electronic
structure of the bilayer cuprate superconductors in the
superconducting state can be described qualitatively in the
framework of the kinetic energy driven superconductivity by
considering the bilayer interaction: Firstly, the bilayer
interaction causes the bilayer slitting, this leads to that the full
electron normal (anomalous) Green's function is divided into the
longitudinal and transverse parts, respectively, then the bonding
and antibonding electron spectral functions (superconducting gap
functions) are obtained from these longitudinal and transverse parts
of the electron normal (anomalous) Green's function, respectively.
Although the transverse part of the superconducting gap parameter
$\Delta_{T}\approx 0$, the antibonding peak around the $[\pi,0]$
point is always at lower binding energy than the bonding peak (hump)
due to the bilayer slitting. In this sense, the peak-dip-hump
structure in the bilayer cuprate superconductors in the
superconducting state is mainly caused by the bilayer splitting.
Secondly, the superconducting state in the bilayer cuprate
superconductors is the conventional BCS like with the d-wave
symmetry as in the single layer case \cite{guo2,feng4} discussed in
section 3. This can be understood from the electron normal and
anomalous Green's functions in Eqs. (62). Since the spins center
around the $[\pi,\pi]$ point in the mean-field level
\cite{guo2,feng2,lan2}, then the main contributions for the spins
comes from the $[\pi,\pi]$ point. In this case, the longitudinal and
transverse parts of the electron normal and anomalous Green's
functions in Eqs. (62) can be approximately reduced in terms of
$\omega_{\nu{\bf p}=[\pi,\pi]}\sim 0$ and one of the self-consistent
equations in Eq. (65e) as \cite{lan2},
\begin{subequations}
\begin{eqnarray}
G_{L}({\bf k},\omega)&\approx&{1\over 2}\sum_{\nu=1,2}Z^{(\nu)}_{FA}
\left ({U^{2}_{\nu{\bf k}}\over\omega-E_{\nu{\bf k}}}+
{V^{2}_{\nu{\bf k}}\over\omega+E_{\nu{\bf k}}}\right), \\
G_{T}({\bf k},\omega)&\approx&{1\over 2}\sum_{\nu=1,2}(-1)^{\nu+1}
Z^{(\nu)}_{FA}\left ({U^{2}_{\nu{\bf k}}\over\omega-E_{\nu{\bf k}}}
+{V^{2}_{\nu{\bf k}}\over\omega+E_{\nu{\bf k}}}\right ), \\
\Gamma^{\dagger}_{L}({\bf k},\omega)&=&{1\over 2}\sum_{\nu=1,2}
Z^{(\nu)}_{FA}{\bar{\Delta}_{hz}^{(\nu)}({\bf k})\over 2E_{\nu{\bf k
}}}\left ({1\over\omega-E_{\nu{\bf k}}}-{1\over\omega+E_{\nu{\bf k}}
}\right ), \\
\Gamma^{\dagger}_{T}({\bf k},\omega)&=&{1\over 2}\sum_{\nu=1,2}
(-1)^{\nu+1}Z^{(\nu)}_{FA}{\bar{\Delta}_{hz}^{(\nu)}({\bf k})\over
2E_{\nu{\bf k}}}\left ({1\over\omega-E_{\nu{\bf k}}}-{1\over\omega
+E_{\nu{\bf k}}}\right ),~~~~~
\end{eqnarray}
\end{subequations}
with the electron quasiparticle coherence factors,
\begin{subequations}
\begin{eqnarray}
U^{2}_{\nu{\bf k}}\approx V^{2}_{h\nu{\bf k}-{\bf k}_{AF}},\\
V^{2}_{\nu{\bf k}}\approx U^{2}_{h\nu{\bf k}-{\bf k}_{AF}},
\end{eqnarray}
\end{subequations}
where the electron coherent weights $Z^{(\nu)}_{FA}=Z^{(\nu)}_{hFA}
/2$, and the electron quasiparticle spectrum $E_{\nu{\bf k}}\approx
E_{h\nu{\bf k}-{\bf k_{AF}}}$. As a result, these electron normal
and anomalous Green's functions in Eq. (68) are typical bilayer BCS
formalism with the d-wave gap function \cite{bcs}. In this sense,
the peak-dip-hump structure in the bilayer cuprate superconductors
is unrelated to superconductivity.

\subsection{Electronic structure of the bilayer cuprate
superconductors in the normal state}

In this subsection, we discuss the electronic structure of the
bilayer cuprate superconductors in the normal state \cite{lan1}. As
the discussions of the electronic structure of the single layer case
in section 3, the dressed holon longitudinal and transverse parts of
the pairing order parameter of the bilayer cuprate superconductors
$\Delta_{hL}=0$ and $\Delta_{hT}=0$ (then the longitudinal and
transverse parts of the superconducting gap parameter $\Delta_{L}=0$
and $\Delta_{T}=0$) in the normal state. In this case, the
longitudinal and transverse parts of the full dressed holon normal
Green's function in Eqs. (53a) and (53b) are reduced as \cite{lan1},
\begin{subequations}
\begin{eqnarray}
g_{L}({\bf k},\omega)&=&{1\over 2}\sum_{\nu=1,2}{Z^{(\nu)}_{hFA}
\over\omega-\bar{\xi}_{\nu{\bf k}}}, \\
g_{T}({\bf k},\omega)&=&{1\over 2}\sum_{\nu=1,2}(-1)^{\nu+1}{
Z^{(\nu)}_{hFA }\over \omega-\bar{\xi}_{\nu{\bf k}}},
\end{eqnarray}
\end{subequations}
while the equations satisfied by the dressed holon quasiparticle
coherent weights $Z^{(1)}_{hFA}$ and $Z^{(2)}_{hFA}$ in Eqs. (63c)
and (63d) are reduced as,
\begin{subequations}
\begin{eqnarray}
{1\over Z_{hFA}^{(1)}}&=&1+{1\over 32N^{2}}\sum_{{\bf q,p}}
\sum_{\nu,\nu',\nu''}[1+(-1)^{\nu+\nu'+\nu''+1}]C_{\nu\nu''}({\bf
p+k}_{N})Z^{(\nu'')}_{hFA}{B_{\nu'{\bf p}}B_{\nu{\bf q}}\over
\omega_{\nu'{\bf p}} \omega_{\nu{\bf q}}}\nonumber \\
&\times&\left({R^{(1)}_{\nu\nu'\nu''}({\bf q,p})\over
[\omega_{\nu'{\bf p}}-\omega_{\nu{\bf q}}+\bar{\xi}_{\nu''{\bf
p-q+k}_{N}}]^{2}}+{R^{(2)}_{\nu\nu'\nu''}({\bf q,p})\over
[\omega_{\nu'{\bf p}}- \omega_{\nu{\bf q}}-\bar{\xi}_{\nu''{\bf
p-q+k}_{N}}]^{2}}\right . \nonumber \\
&+&\left . {R^{(3)}_{\nu\nu'\nu''}({\bf q,p})\over [\omega_{\nu'{\bf
p}}+\omega_{\nu{\bf q}}+\bar{\xi}_{\nu''{\bf p-q+k}_{N}}]^{2}}+
{R^{(4)}_{\nu\nu'\nu''}({\bf q,p})\over [\omega_{\nu'{\bf p}}+
\omega_{\nu{\bf q}}-\bar{\xi}_{\nu''{\bf p- q+k}_{N}}]^{2}} \right),
\\
{1\over Z_{hFA}^{(2)}}&=&1+{1\over 32N^{2}}\sum_{{\bf q,p}}
\sum_{\nu,\nu',\nu''}[1-(-1)^{\nu+\nu'+\nu''+1}] C_{\nu\nu''}({\bf
p+k}_{N})Z^{(\nu'')}_{hFA}{B_{\nu'{\bf p}}B_{\nu{\bf q}}\over
\omega_{\nu'{\bf p}} \omega_{\nu{\bf q}}}\nonumber \\
&\times&\left({R^{(1)}_{\nu\nu'\nu''}({\bf q,p})\over
[\omega_{\nu'{\bf p}}-\omega_{\nu{\bf q}}+\bar{\xi}_{\nu''{\bf
p-q+k}_{N}}]^{2}}+{R^{(2)}_{\nu\nu'\nu''}({\bf q,p})\over
[\omega_{\nu'{\bf p}}- \omega_{\nu{\bf q}}-\bar{\xi}_{\nu''{\bf
p-q+k}_{N}}]^{2}}\right . \nonumber \\
&+&\left . {R^{(3)}_{\nu\nu'\nu''}({\bf q,p})\over [\omega_{\nu'{\bf
p}}+\omega_{\nu{\bf q}}+\bar{\xi}_{\nu''{\bf p-q+k}_{N}}]^{2}}+
{R^{(4)}_{\nu\nu'\nu''}({\bf q,p})\over [\omega_{\nu'{\bf p}}+
\omega_{\nu{\bf q}}-\bar{\xi}_{\nu''{\bf p- q+k}_{N}}]^{2}} \right),
\end{eqnarray}
\end{subequations}
with
\begin{subequations}
\begin{eqnarray}
R^{(1)}_{\nu\nu'\nu''}({\bf q,p})&=&n_{F}(\bar{\xi}_{\nu''{\bf p-
q+k}_{N}})[n_{B}(\omega_{\nu'{\bf p}})-n_{B}(\omega_{\nu{\bf q}})]
\nonumber\\
&-&n_{B}(\omega_{\nu{\bf q}})n_{B}(-\omega_{\nu'{\bf p}}),~~~~~\\
R^{(2)}_{\nu\nu'\nu''}({\bf q,p})&=&n_{F}(\bar{\xi}_{\nu''{\bf p-
q+k}_{N}}) [n_{B}(\omega_{\nu{\bf q}})-n_{B}(\omega_{\nu'{\bf p}})]
\nonumber\\
&-&n_{B}(\omega_{\nu'{\bf p}})n_{B}(-\omega_{\nu{\bf q}}),~~~~~\\
R^{(3)}_{\nu\nu'\nu''}({\bf q,p})&=&[1-n_{F}(\bar{\xi}_{\nu''{\bf p
-q+k}_{N}})][1+n_{B}(\omega_{\nu{\bf q}})+n_{B}(\omega_{\nu'{\bf p
}})]\nonumber\\
&+&n_{B}(\omega_{\nu{\bf q}})n_{B}(\omega_{\nu'{\bf p}}),\\
R^{(4)}_{\nu\nu'\nu''}({\bf q,p})&=&n_{F}(\bar{\xi}_{\nu''{\bf p-
q+k}_{N}}) [1+n_{B}(\omega_{\nu{\bf q}})+n_{B}(\omega_{\nu'{\bf p}
})]\nonumber \\
&+& n_{B}(\omega_{\nu{\bf q}})n_{B}(\omega_{\nu'{\bf p}}).
\end{eqnarray}
\end{subequations}
As in the superconducting state, these two self-consistent equations
must be solved simultaneously with other self-consistent equations
in Eq. (65), then the longitudinal and transverse parts of the
electron Green's function in Eqs. (62a) and (62b) and the electron
spectral function in Eqs. (66a) and (66d) are reduced as,
\begin{subequations}
\begin{eqnarray}
G_{L}({\bf k},\omega)&=&{1\over 8N}\sum_{\bf p}\sum_{\mu\nu}
Z_{hFA}^{(\mu)}{B_{\nu{\bf p}}\over \omega_{\nu{\bf p}}} \left
({M^{(1)}_{\mu\nu}({\bf k,p})\over\omega+\bar{\xi}_{\mu{\bf p-k}}
-\omega_{\nu{\bf p}}}\right . \nonumber \\
&+&\left . {M^{(2)}_{\mu\nu}({\bf k,p})\over\omega+
\bar{\xi}_{\mu{\bf p-k}}+\omega_{\nu{\bf p}}}\right ),~~~~~\\
G_{T}({\bf k},\omega)&=&{1\over 8N}\sum_{\bf p}\sum_{\mu\nu}
(-1)^{\mu+\nu}Z_{hFA}^{(\mu)}{B_{\nu{\bf p}}\over\omega_{\nu{\bf p}
}}\left ({M^{(1)}_{\mu\nu}({\bf k,p})\over\omega+ \bar{\xi}_{\mu{\bf
p-k}}-\omega_{\nu{\bf p}}}\right. \nonumber\\
&+&\left. {M^{(2)}_{\mu\nu}({\bf k,p})\over\omega+
\bar{\xi}_{\mu{\bf p-k}}+ \omega_{\nu{\bf p}}}\right ),\\
A_{L}({\bf k},\omega)&=&\pi{1\over 4N}\sum_{\bf p}\sum_{\mu\nu}
Z_{hFA}^{(\mu)}{B_{\nu{\bf p}}\over\omega_{\nu{\bf p}}}
[M^{(1)}_{\mu\nu}({\bf k,p})\delta(\omega+\bar{\xi}_{\mu{\bf p-k}}
-\omega_{\nu{\bf p}}) \nonumber \\
&+&M^{(2)}_{\mu\nu}({\bf k,p})\delta(\omega+
\bar{\xi}_{\mu{\bf p-k}}+\omega_{\nu{\bf p}})], \\
A_{T}({\bf k},\omega)&=&\pi{1\over 4N}\sum_{\bf p}\sum_{\mu\nu}
(-1)^{\mu+\nu}Z_{hFA}^{(\mu)}{B_{\nu{\bf p}}\over\omega_{\nu{\bf p}
}}[M^{(1)}_{\mu\nu}({\bf k,p})\delta(\omega+\bar{\xi}_{\mu{\bf
p-k}}-\omega_{\nu{\bf p}})\nonumber \\
&+&M^{(2)}_{\mu\nu}({\bf k,p})\delta(\omega+ \bar{\xi}_{\mu{\bf
p-k}}+\omega_{\nu{\bf p}})].
\end{eqnarray}
\end{subequations}
where $M^{(1)}_{\mu\nu}({\bf k,p})=n_{F}(\bar{\xi}_{\mu{\bf p-k}})
+n_{B}(\omega_{\nu{\bf p}})$, $M^{(2)}_{\mu\nu}({\bf k,p})=1-n_{F}
(\bar{\xi}_{\mu{\bf p-k}})+n_{B}(\omega_{\nu{\bf p}})$, then the
bonding and antibonding electron spectral functions of the bilayer
cuprate superconductors in the normal state are obtained as,
$A^{+}({\bf k},\omega)=[A_{L}({\bf k},\omega)+A_{T}({\bf k},
\omega)]/2$ and $A^{-}({\bf k},\omega)= [A_{L}({\bf k},
\omega)-A_{T}({\bf k},\omega)]/2$, respectively.

We have performed a calculation for the electron spectral functions
of the bilayer cuprate superconductors in the normal state, and the
results of the bonding (solid line) and antibonding (dashed line)
electron spectral functions at (a) the $[\pi,0]$ point and (b)
$[\pi/2,\pi/2]$ point for $t/J=2.5$, $t'/t=0.15$, and
$t_{\perp}/t=0.3$ with $T=0.1J$ at $\delta=0.15$ are plotted in Fig.
12 in comparison with the corresponding experimental data of the
bilayer cuprate superconductor
Bi$_{2}$Sr$_{2}$CaCu$_{2}$O$_{8+\delta}$ at the $[\pi,0]$ point
\cite{fedorov} and $[\pi/2,\pi/2]$ point \cite{dlfeng2} in the
normal state, respectively. Apparently, (1) there is a double-peak
structure in the electron spectral function around the $[\pi,0]$
point, i.e., the bonding and antibonding quasiparticle peaks around
the $[\pi,0]$ point are located at the different positions, while
the bonding and antibonding peaks around the $[\pi/2,\pi/2]$ point
are located at the same position, which leads to that the bilayer
spilitting appears around the $[\pi,0]$ point, and is absent from
the vicinity of the $[\pi/2,\pi/2]$ point; (2) The position of the
antibonding peak at the $[\pi,0]$ point is more closer to the Fermi
energy than these for the bonding peak; (3) In analogy to the single
layer case \cite{guo1} discussion in section 3, both positions of
the quasiparticle peaks from the bonding and antibonding electron
spectral functions at the $[\pi,0]$ and $[\pi/2,\pi/2]$ points are
below the Fermi energy, but the positions of the peaks at the
$[\pi/2,\pi/2]$ point are more closer to the Fermi energy, which
indicates that the lowest energy states are located at the
$[\pi/2,\pi/2]$ point, in other words, the low energy spectral
weight with the majority contribution to the low-energy properties
of the bilayer cuprate superconductors in the normal state comes
from the $[\pi/2,\pi/2]$ point, in qualitative agreement with the
ARPES experimental data on the bilayer cuprate superconductors in
the normal state
\cite{shen1,fedorov,dlfeng2,dlfeng1,kordyuk,chuang}. In the above
calculations, we also find that the double-peak structure in the
electron spectral functions around the $[\pi,0]$ point is closely
related to the interlayer hopping form in Eq. (49). With decreasing
the values of $t_{\perp}$ and $J_{\perp}$, the distance between the
bonding and antibonding peaks in the electron spectral functions
decreases. When $t_{\perp}=0$ and $J_{\perp}=0$, we find that the
transverse part of the dressed holon Green's functions in Eq. (70b)
(then the transverse part of the electron Green's functions in Eq.
(73b) and transverse part of the electron spectral functions in Eq.
(73d)) is equal to the zero. In this case, the bonding electron
spectral function is exactly same as the antibonding electron
spectral function, then the electron spectral functions are reduced
to the single layer case \cite{guo1}.

\begin{figure}[t]
\begin{center}
\begin{minipage}[h]{100mm}
\epsfig{file=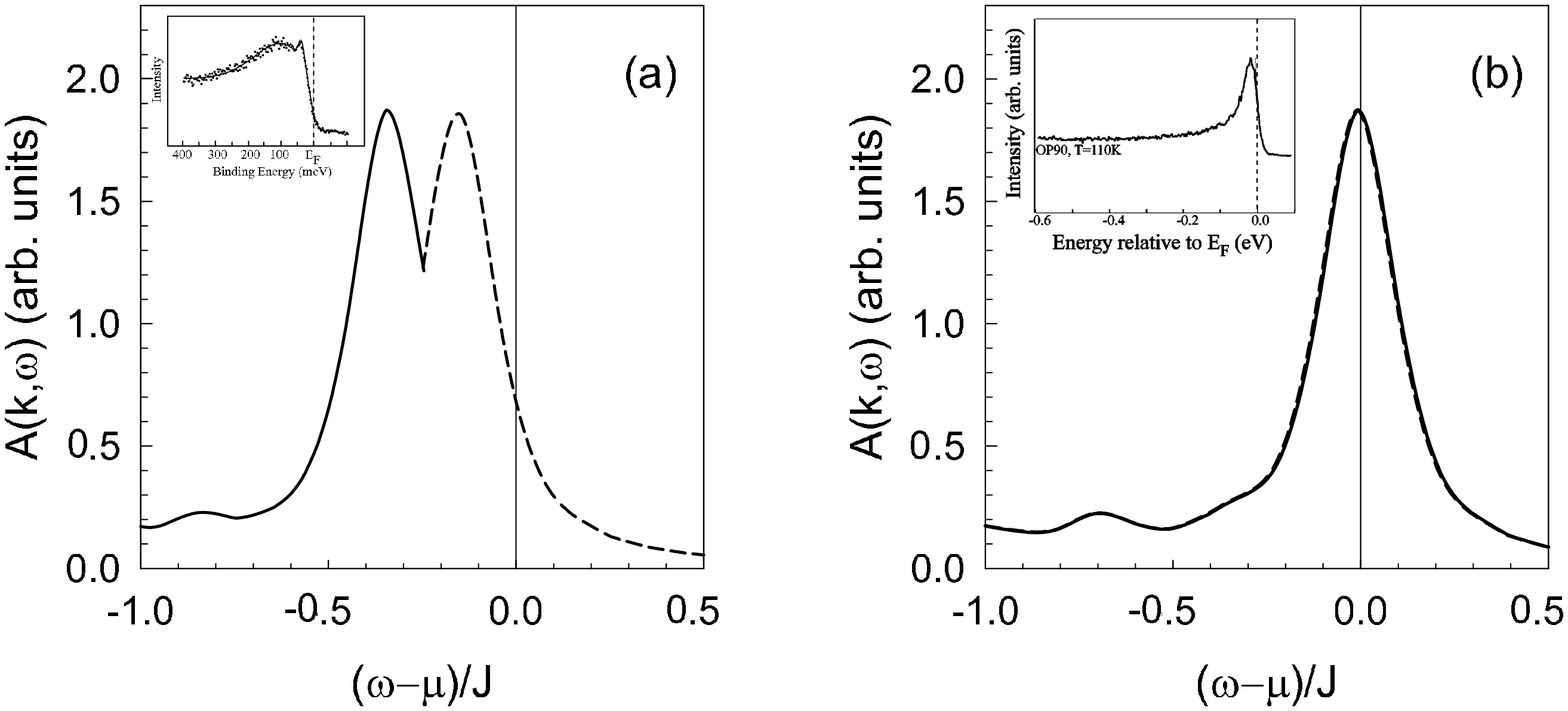, width=100mm}
\end{minipage}
\caption{The bonding (solid line) and antibonding (dashed line)
electron spectral functions of the bilayer cuprate superconductors
in the normal state as a function of energy at (a) the $[\pi,0]$
point and (b) $[\pi/2,\pi/2]$ point for $t/J=2.5$, $t'/t=0.15$, and
$t_{\perp}/t=0.3$ with $T=0.1J$ at $\delta=0.15$. Inset: the
corresponding experimental data of the bilayer cuprate
superconductor Bi$_{2}$Sr$_{2}$CaCu$_{2}$O$_{8+\delta}$ at the
$[\pi,0]$ point and $[\pi/2,\pi/2]$ point in the normal state taken
from Refs. \protect\cite{fedorov} and \protect\cite{dlfeng2},
respectively.}
\end{center}
\end{figure}

\begin{figure}[t]
\begin{center}
\begin{minipage}[h]{85mm}
\epsfig{file=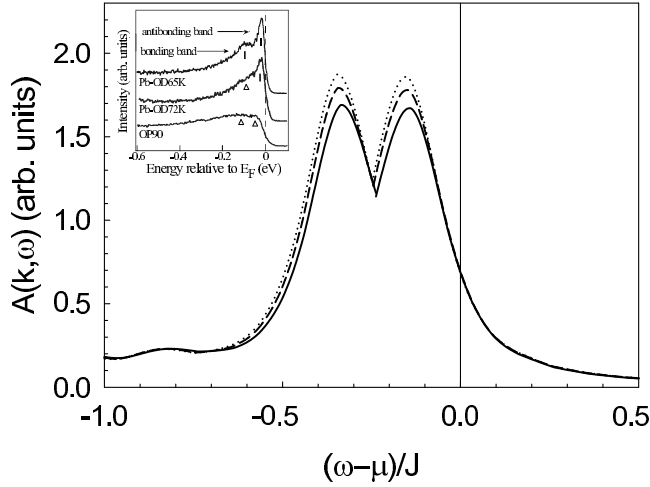, width=85mm}
\end{minipage}
\caption{The electron spectral functions of the bilayer cuprate
superconductors at $[\pi,0]$ point in the normal state for
$t/J=2.5$, $t'/t=0.15$, and $t_{\perp}/t=0.3$ with $T=0.1J$ at
$\delta=0.09$ (solid line), $\delta=0.12$ (dashed line), and
$\delta=0.15$ (dotted line). Inset: the corresponding experimental
data of the bilayer cuprate superconductor
Bi$_{2}$Sr$_{2}$CaCu$_{2}$O$_{8+\delta}$ in the normal state taken
from Ref. \protect\cite{dlfeng2}}
\end{center}
\end{figure}

For a better understanding of the physical properties of the
electron spectrum of the bilayer cuprate superconductors in the
normal state, we have studied the electron spectrum at different
doping concentrations, and the result of the electron spectral
functions at $[\pi,0]$ point for $t/J=2.5$, $t'/t=0.15$, and
$t_{\perp}/t=0.3$ with $T=0.1J$ at $\delta=0.09$ (solid line),
$\delta=0.12$ (dashed line), and $\delta=0.15$ (dotted line) are
plotted in Fig. 13 in comparison with the corresponding experimental
data of the bilayer cuprate superconductor
Bi$_{2}$Sr$_{2}$CaCu$_{2}$O$_{8+\delta}$ in the normal state
\cite{dlfeng2}, which indicates that with increasing the doping
concentration, both bonding and antibonding quasiparticle peaks
become sharper, and the spectral weights of these peaks increase in
intensity. Furthermore, we have also discussed the temperature
dependence of the electron spectrum of the bilayer cuprate
superconductors in the normal state, and the result of the electron
spectral functions at $[\pi,0]$ point for $t/J=2.5$, $t'/t=0.15$,
and $t_{\perp}/t=0.3$ at $\delta=0.15$ with $T=0.1J$ (solid line),
$T=0.05J$ (dashed line), and $T=0.01J$ (dotted line) are plotted in
Fig. 14 in comparison with the corresponding experimental data of
the bilayer cuprate superconductor
Bi$_{2}$Sr$_{2}$CaCu$_{2}$O$_{8+\delta}$ in the normal state
\cite{fedorov}, it is shown obviously that both bonding and
antibonding spectral weights are suppressed with increasing
temperatures. These results are also qualitatively consistent with
the ARPES experimental results on the bilayer cuprate
superconductors in the normal state
\cite{shen1,fedorov,dlfeng2,dlfeng1,kordyuk,chuang}.

\begin{figure}[t]
\begin{center}
\begin{minipage}[h]{85mm}
\epsfig{file=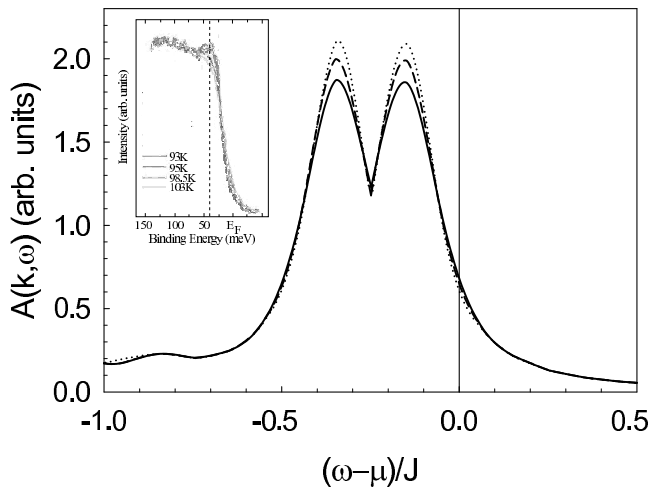, width=85mm}
\end{minipage}
\caption{The electron spectral functions of the bilayer cuprate
superconductors at $[\pi,0]$ point in the normal state for
$t/J=2.5$, $t'/t=0.15$, and $t_{\perp}/t=0.3$ at $\delta=0.15$ with
$T=0.1J$ (solid line), $T=0.05J$ (dashed line), and $T=0.01J$
(dotted line). Inset: the corresponding experimental data of the
bilayer cuprate superconductor
Bi$_{2}$Sr$_{2}$CaCu$_{2}$O$_{8+\delta}$ in the normal state taken
from Ref. \protect\cite{fedorov}.}
\end{center}
\end{figure}

\begin{figure}[t]
\begin{center}
\begin{minipage}[h]{100mm}
\epsfig{file=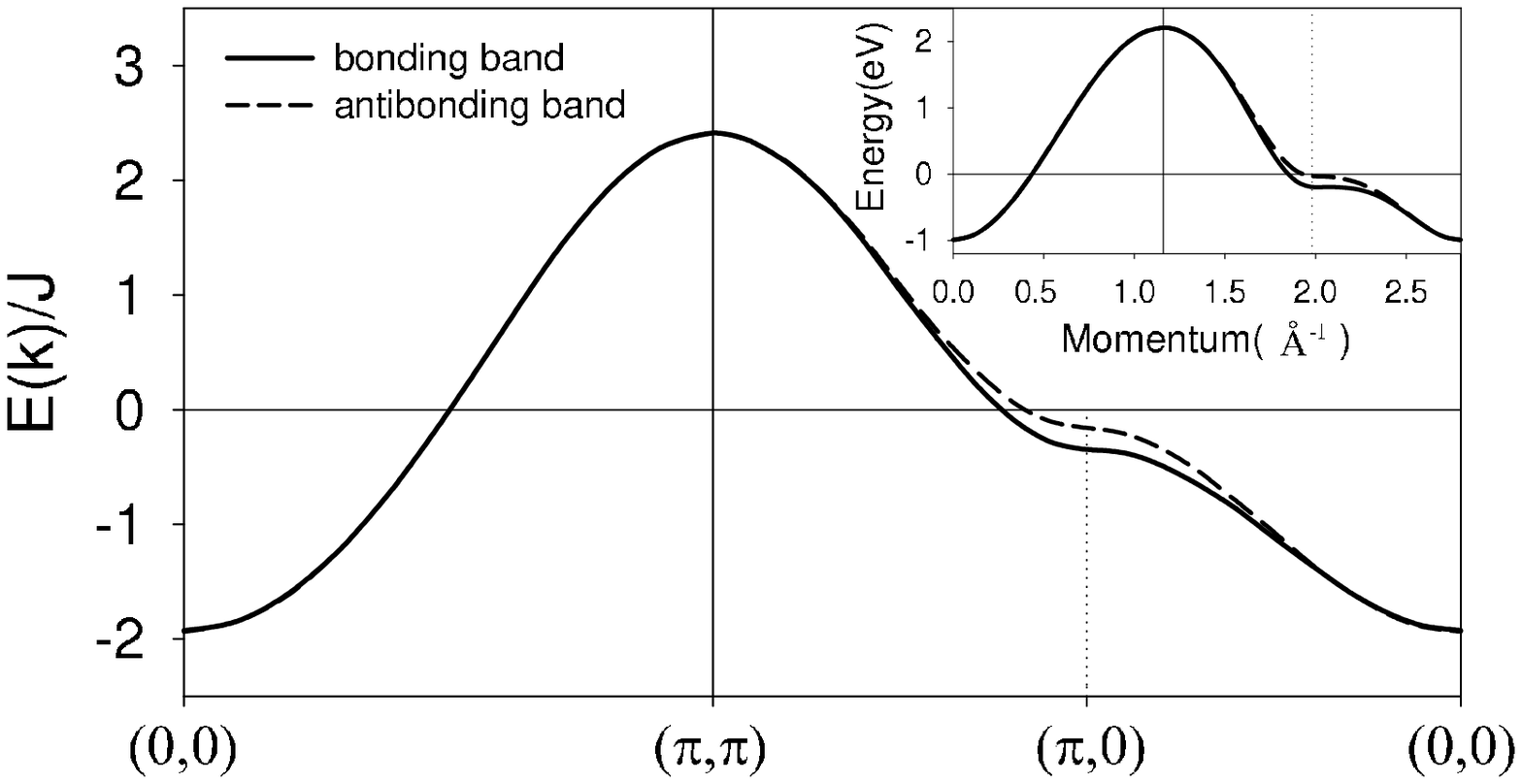, width=100mm}
\end{minipage}
\caption{The positions of the lowest energy quasiparticle peaks in
the bonding (solid line) and antibonding (dashed line) electron
spectra as a function of momentum with $T=0.1J$ at $\delta=0.15$ for
$t/J=2.5$, $t'/t=0.15$, and $t_{\perp}/t=0.3$. Inset: the
corresponding result from the tight binding fit to the experimental
data of the bilayer cuprate superconductor
Bi$_{2}$Sr$_{2}$CaCu$_{2}$O$_{8+\delta}$ in the normal state taken
from Ref. \protect\cite{kordyuk}.}
\end{center}
\end{figure}

For considering the quasiparticle dispersion of the bilayer cuprate
superconductors in the normal state, we have made a series of
calculations for both bonding and antibonding electron spectral
functions at different momenta, and find that the lowest energy
peaks in the normal state are well defined at all momenta. In
particular, the positions of both bonding and antibonding
quasiparticle peaks as a function of energy $\omega$ for momentum
${\bf k}$ in the vicinity of the $[\pi,0]$ point are almost not
changeable as in the single layer case, which leads to the unusual
quasiparticle dispersion around the $[\pi,0]$ point. To show this
broad feature clearly, we plot the positions of the lowest energy
quasiparticle peaks in the bonding and antibonding electron spectra
as a function of momentum along the high symmetry directions with
$T=0.1J$ at $\delta=0.15$ for $t/J=2.5$, $t'/t=0.15$, and
$t_{\perp}/t=0.3$ in Fig. 15. For comparison, the corresponding
result from the tight binding fit to the experimental data of the
bilayer cuprate superconductor
Bi$_{2}$Sr$_{2}$CaCu$_{2}$O$_{8+\delta}$ in the normal state
\cite{kordyuk} is also shown in Fig. 15 (inset). It is shown that in
analogy to the single layer case \cite{guo1}, both electron bonding
and antibonding quasiparticles around the $[\pi,0]$ point disperse
very weakly with momentum, and then the two main flat bands appear,
while the Fermi energy is only slightly above these flat bands.
Moreover, this bilayer energy band splitting reaches its maximum at
the $[\pi,0]$ point, in qualitative agreement with these obtained
from the ARPES experimental measurements on the bilayer cuprate
superconductors in the normal state
\cite{shen1,dlfeng1,kordyuk,chuang}.

In comparison with the single layer case \cite{guo2} discussed in
section 3, we therefore find that the essential physics of the
double-peak structure of the electron spectral function of the
bilayer cuprate superconductors around the $[\pi,0]$ point in the
normal state is dominated by the bilayer interaction. Since the full
electron Green's function in the bilayer cuprate superconductors in
the normal state is divided into the longitudinal and transverse
parts, respectively, due to the bilayer interaction, then these
longitudinal and transverse Green's functions (then the bonding and
antibonding electron spectral functions and corresponding
quasiparticle dispersions) are obtained beyond the mean-field
approximation by considering the dressed holon fluctuation due to
the spin pair bubble, therefore the nature of the bonding and
antibonding electron spectral functions of the bilayer cuprate
superconductors in the normal state is closely related to the strong
interaction between the dressed holons (then electron
quasiparticles) and collective magnetic excitations. In this case,
the single-particle hoppings in the bilayer $t$-$J$ model are
strongly renormalized by the magnetic interaction as in the single
layer case discussed in section 3. As a consequence, both bonding
and antibonding quasiparticle bandwidths are reduced to the order of
(a few) $J$, and then the energy scales of both bonding and
antibonding quasiparticle energy bands are controlled by the
magnetic interaction. These renormalizations for both bonding and
antibonding energy bands are then responsible for the unusual
bonding and antibonding electron quasiparticle spectra and
production of the two main flat bands around the $[\pi,0]$ point.

\section{Summary and discussions}

In the charge-spin separation fermion-spin theory \cite{feng1}, the
physical electron is decoupled completely as the gauge invariant
dressed holon and spin, with the dressed holon keeps track of the
charge degree of freedom together with some effects of the spin
configuration rearrangements due to the presence of the doped hole
itself, while the spin operator keeps track of the spin degree of
freedom. In particular, we have shown that the charge-spin
separation fermion-spin transformation in Eq. (9) is a natural
representation for the constrained electron under the decoupling
scheme. The advantage of this charge-spin separation fermion-spin
theory is that the electron local constraint for single occupancy is
satisfied in analytical calculations. Moreover, both dressed holon
and spin are gauge invariant, and in this sense, they are real and
can be interpreted as the physical excitations \cite{feng1}. Within
this charge-spin separation fermion-spin theory, we have developed a
kinetic energy driven superconducting mechanism \cite{feng2}, where
the dressed holons interact occurring directly through the kinetic
energy by exchanging the spin excitations, leading to a net
attractive force between the dressed holons, then the electron
Cooper pairs originating from the dressed holon pairing state are
due to the charge-spin recombination, and their condensation reveals
the superconducting ground-state. Furthermore, this superconducting
state is controlled by both superconducting gap parameter and
quasiparticle coherence, then the maximal superconducting transition
temperature occurs around the optimal doping, and decreases in both
underdoped and overdoped regimes \cite{feng3,feng4}.

In summary, we have given a brief review of the electronic structure
of the single layer and bilayer cuprate superconductors in both
superconducting and normal states \cite{feng4,guo1,guo2,lan1,lan2}.
Within the framework of the kinetic energy driven superconductivity,
we show the spectral weight in both superconducting and normal
states increases with increasing doping, and decreases with
increasing temperatures \cite{feng4,feng5,guo1,guo2,lan1,lan2}. In
the normal state, although the lowest energy states are located at
the $[\pi/2,\pi/2]$ point, the quasiparticle dispersion exhibits the
unusual flat band behavior around the $[\pi,0]$ point below the
Fermi energy. In corresponding to this unusual normal state flat
band behavior around the $[\pi,0]$ point, the superconducting
quasiparticles around the $[\pi,0]$ disperse very weakly with
momentum. In particular, we show that one of the universal features
is that the d-wave superconducting state in cuprate superconductors
is the conventional BCS like, so that the basic BCS formalism with
the d-wave gap function is still valid in quantitatively reproducing
of the doping dependence of the superconducting gap parameter, the
doping dependence of the superconducting transition temperature, and
the low energy electronic structure of cuprate superconductors
\cite{feng4,feng5,guo2,lan2}, although the pairing mechanism is
driven by the kinetic energy by exchanging spin excitations, and
other exotic properties are beyond the BCS formalism. Furthermore,
we have shown that the observed peak-dip-hump structure in the
bilayer cuprate superconductors in the superconducting state and
double-peak structure in the normal state around the $[\pi,0]$ point
is mainly caused by the bilayer splitting, with the superconducting
quasiparticle peak in the superconducting state (lower energy
quasiparticle peak in the normal state) being related to the
antibonding component, and the hump in the superconducting state
(higher energy quasiparticle peak in the normal state) being formed
by the bonding component. In this sense, the peak-dip-hump structure
in the bilayer cuprate superconductors in the superconducting state
is unrelated to superconductivity. Our these theoretical results
\cite{feng4,feng5,guo1,guo2,lan1,lan2} also show that the striking
behavior of the electronic structure in cuprate superconductors is
intriguingly related to the strong coupling between the electron
quasiparticles and collective magnetic excitations. Based on the
kinetic energy driven superconductivity, we \cite{cheng} have
discussed the electronic structure of the electron-doped cuprate
superconductors, and the results show explicitly that although the
electron-hole asymmetry is observed in the phase diagram, the
electronic structure of the electron-doped cuprates in the
superconducting state is similar to that in the hole-doped case.

\begin{center}
{\bf Acknowledgements}
\end{center}

This work was supported by the National Natural Science Foundation
of China under Grant Nos. 90403005 and 10774015, and the funds from
the Ministry of Science and Technology of China under Grant Nos.
2006CB601002 and 2006CB921300.



\begin{thebibliography}{0}

\bibitem [*] {add} The present address: Department of Physics,
Capital Normal University, Beijing 100037, China.

\bibitem{bednorz} J. G. Bednorz and K.A. M\"uller, Z. Phys. B
{\bf 64}, 189 (1986).

\bibitem{kastner} See, e.g., M. A.Kastner, R. J. Birgeneau, G.
Shirane, and Y. Endoh, Rev. Mod. Phys. {\bf 70}, 897 (1998), and
references therein.

\bibitem{shen1} See, e.g., Z. X. Shen, and D. S. Dessau, Phys. Rep.
{\bf 253}, 1 (1995), and referenes therein.

\bibitem{shen} See, e.g., A. Damascelli, Z. Hussain, and Z. X.
Shen, Rev. Mod. Phys. {\bf 75}, 473 (2003), and references therein.

\bibitem{campuzano} See, e.g., J. Campuzano, M. Norman, and M.
Randeira, in {\it Physics of Superconductors}, vol. II, edited by K.
Bennemann and J. Ketterson (Springer, Berlin Heidelberg New York,
2004), p. 167, and references therein.

\bibitem{anderson1} P. W. Anderson, Science {\bf 235}, 1196 (1987);
P. W. Anderson, in {\it Frontiers and Borderlines in Many Particle
Physics}, edited by R. A. Broglia and J. R. Schrieffer
(North-Holland, Amsterdam, 1987), p. 1.

\bibitem{anderson2} P. W. Anderson, Phys. Rev. Lett. {\bf 67}, 2092
(1991); Science {\bf 288}, 480 (2000); Physica C {\bf 341-348}, 9
(2000); cond-mat/0108522 (unpublished).

\bibitem{anderson3} P. W. Anderson, arXiv:0709.0656 (unpublished).

\bibitem {laughlin} R. B. Laughlin, Phys. Rev. Lett. {\bf 79}, 1726
(1997); J. Low. Tem. Phys. {\bf 99}, 443 (1995).

\bibitem{kim} Z. X. Shen, W. E. Spicer, D. M. King, D. S. Dessau,
and B. O. Wells, Science {\bf 267}, 343 (1995); C. Kim, P. J. White,
Z. X. Shen, T. Tohyama, Y. Shibata, S. Maekawa, B. O. Wells, Y. J.
Kim, R. J. Birgeneau, and M. A. Kastner, Phys. Rev. Lett. {\bf 80},
4245 (1998).

\bibitem{dessau} D. S. Dessau, Z. X. Shen, D. M. King, D. S.
Marshall, L. W. Lombardo, P. H. Dickinson, A. G. Loeser, J. DiCarlo,
C. H. Park, A. Kapitulnik, and W. E. Spicer, Phys. Rev. Lett. {\bf
71}, 2781 (1993).

\bibitem{wells} B. O. Wells, Z. X. Shen, A. Matsuura, D. M. King,
M. A. Kastner, M. Greven, and R. J. Birgeneau, Phys. Rev. Lett. {\bf
74}, 964 (1995); D. S. Marshall, D. S. Dessau, A. G. Loeser, C. H.
Park, A. Y. Matsuura, J. N. Eckstein, I. Bozovic, P. Fournier, A.
Kapitulnik, W. E. Spicer, and Z. X. Shen, Phys. Rev. Lett. {\bf 76},
4841 (1996).

\bibitem{ding} H. Ding, J. R. Engelbrecht, Z. Wang, J. C. Campuzano,
S. C. Wang, H. B. Yang, R. Rogan, T. Takahashi, K. Kadowaki, and D.
G. Hinks, Phys. Rev. Lett. {\bf 87}, 227001 (2001).

\bibitem{matsui} H. Matsui, T. Sato, T. Takahashi, S. C. Wang, H.
B. Yang, H. Ding, T. Fujii, T. Watanabe, and A. Matsuda, Phys. Rev.
Lett. {\bf 90}, 217002 (2003).

\bibitem{campuzano1} J. C. Campuzano, H. Ding, M. R. Norman, M.
Randeira, A. F. Bellman, T. Yokoya, T. Takahashi, H.
Katayama-Yoshida, T. Mochiku, and K. Kadowaki, Phys. Rev. B {\bf
53}, R14737 (1996).

\bibitem{yamada} K. Yamada, C.H. Lee, K. Kurahashi, J. Wada, S.
Wakimoto, S. Ueki, H. Kimura, Y. Endoh, S. Hosoya, G. Shirane, R. J.
Birgeneau, M. Greven, M.A. Kastner, and Y. J. Kim, Phys. Rev. B {\bf
57}, 6165 (1998).

\bibitem{dai} P. Dai, H. A. Mook, R. D. Hunt, and F. Do\~gan, Phys.
Rev. B {\bf 63}, 054525 (2001); P. Bourges, B. Keimer, S. Pailh\'es,
L. P. Regnault, Y. Sidis, and C. Ulrich, Physica C {\bf 424}, 45
(2005).

\bibitem{arai} M. Arai, T. Nishijima, Y. Endoh, T. Egami, S.
Tajima, K. Tomimoto, Y. Shiohara, M. Takahashi, A. Garret, and S. M.
Bennington, Phys. Rev. Lett. {\bf 83}, 608 (1999); S. M. Hayden, H.
A. Mook, P. Dai, T. G. Perring, and F. Do\~gan, Nature {\bf 429},
531 (2004); C. Stock, W. J. Buyers, R. A. Cowley, P. S. Clegg, R.
Coldea, C. D. Frost, R. Liang, D. Peets, D. Bonn, W. N. Hardy, and
R. J. Birgeneau, Phys. Rev. B {\bf 71}, 024522 (2005).

\bibitem{bcs} J. R. Schrieffer, {\it Theory of Superconductivity},
Benjamin, New York, 1964.

\bibitem{dessau1} D. S. Dessau, B. O. Wells, Z. X. Shen, W. E.
Spicer, A. J. Arko, R. S. List, D. B. Mitzi, and A. Kapitulnik,
Phys. Rev. Lett. {\bf 66}, 2160 (1991); Y. Hwu, L. Lozzi, M. Marsi,
S. La Rosa, M. Winokur, P. Davis, M. Onellion, H. Berger, F. Gozzo,
F. L\'evy, and G. Margaritondo, Phys. Rev. Lett. {\bf 67}, 2573
(1991).

\bibitem{randeria} Mohit Randeria, Hong Ding, J. C. Campuzano, A.
Bellman, G. Jennings, T. Yokoya, T. Takahashi, H. Katayama-Yoshida,
T. Mochiku, and K. Kadowaki, Phys. Rev. Lett. {\bf 74}, 4951 (1995);
H. Ding, T. Yokoya, J. C. Campuzano, T. Takahashi, M. Randeria, M.
R. Norman, T. Mochiku, K. Kadowaki, and J. Giapintzakis, Nature {\bf
382}, 51 (1996).

\bibitem{fedorov} A. V. Fedorov, T. Valla, P. D. Johnson, Q. Li, G.
D. Gu, and N. Koshizuka, Phys. Rev. Lett. {\bf 82}, 2179 (1999).

\bibitem{lu} D. H. Lu, D. L. Feng, N. P. Armitage, K. M. Shen, A.
Damascelli, C. Kim, F. Ronning, Z. X. Shen, D. A. Bonn, R. Liang, W.
N. Hardy, A. I. Rykov, and S. Tajima, Phys. Rev. Lett. {\bf 86},
4370 (2001).

\bibitem{sato} T. Sato, H. Matsui, S. Nishina, T. Takahashi, T.
Fujii, T. Watanabe, and A. Matsuda, Phys. Rev. Lett. {\bf 89},
067005 (2002); D. L. Feng, A. Damascelli, K. M. Shen, N. Motoyama,
D. H. Lu, H. Eisaki, K. Shimizu, J.-i. Shimoyama, K. Kishio, N.
Kaneko, M. Greven, G. D. Gu, X. J. Zhou, C. Kim, F. Ronning, N. P.
Armitage, and Z. X Shen, Phys. Rev. Lett. {\bf 88}, 107001 (2002).

\bibitem{dlfeng2} D. L. Feng, C. Kim, H. Eisaki, D. H. Lu, A.
Damascelli, K. M. Shen, F. Ronning, N. P. Armitage, N. Kaneko1, M.
Greven, J.-i. Shimoyama, K. Kishio, R. Yoshizaki, G. D. Gu, and Z.
X. Shen, Phys. Rev. B {\bf 65}, 220501(R) (2002); A. A. Kordyuk, S.
V. Borisenko, M. S. Golden, S. Legner, K. A. Nenkov, M. Knupfer, J.
Fink, H. Berger, L. Forr\'o, and R. Follath, Phys. Rev. B {\bf 66},
014502 (2002); Y.-D. Chuang, A. D. Gromko, A. V. Fedorov, Y. Aiura,
K. Oka, Yoichi Ando, D. S. Dessau, cond-mat/0107002 (unpublished).

\bibitem{ding9} K. Terashima, H. Matsui, D. Hashimoto, T. Sato, T.
Takahashi, H. Ding, T. Yamamoto and K. Kadowaki, Nature Phys. {\bf
2}, 27 (2006).

\bibitem{gros} C. Gros, Phys. Rev. B {\bf 38}, 931 (1988); T. K.
Lee and Shiping Feng, Phys. Rev. B {\bf 38}, 11809 (1988); S.
Sorella, G. B. Martins, F. Becca, C. Gazza, L. Capriotti, A. Parola,
and E. Dagotto, Phys. Rev. Lett. {\bf 88}, 117002 (2002); D.
Poilblanc and D. J. Scalapino, Phys. Rev. B {\bf 66}, 052513 (2002);
D. Eichenberger and D. Baeriswyl, arXiv:0708.2795 (unpublished); L.
Spanu, M. Lugas, F. Becca, and S. Sorella, arXiv:0709.2850
(unpublished).

\bibitem{dagotto} See, e.g., E. Dagotto, Rev. Mod. Phys. {\bf 66},
763 (1994); Th. Maier, M. Jarrell, T. Pruschke, and M. H. Hettler,
Rev. Mod. Phys. {\bf 77}, 1027 (2005).

\bibitem{feng1} Shiping Feng, Jihong Qin, and Tianxing Ma, J.
Phys. Condens. Matter {\bf 16}, 343 (2004); Shiping Feng, Tianxing
Ma, and Jihong Qin, Mod. Phys. Lett. B {\bf 17}, 361 (2003).

\bibitem{feng2} Shiping Feng, Phys. Rev. B {\bf 68}, 184501 (2003).

\bibitem{feng3} Shiping Feng, Tianxing Ma, and Huaiming Guo, Physica C
{\bf 436}, 14 (2006); Shiping Feng and Huaiming Guo, in {\ it
Proceedings of the International Conference on Materials and
Mechanisms of Superconductivity and High Temperature Superconductors
VIII}, Dresden, Germany, 2006 [Physica C {\bf 460-462}, 230 (2007)];
Tianxing Ma, Huaiming Guo, and Shiping Feng, Mod. Phys. Lett. B {\bf
18}, 895 (2004).

\bibitem{feng4} Shiping Feng and Tianxing Ma, Phys. Lett. A {\bf 350},
138 (2006).

\bibitem{feng5} Shiping Feng and Tianxing Ma, in {\it Superconductivity
Research Horizons}, edited by Eugene H. Peterson (Nova Science
Publishers, Nrw York, 2007) chapter 5, pp. 129-158, and references
therein.

\bibitem{guo1} Huaiming Guo and Shiping Feng, Phys. Lett. A
{\bf 355}, 473 (2006).

\bibitem{guo2} Huaiming Guo and Shiping Feng, Phys. Lett. A
{\bf 361}, 382 (2007).

\bibitem{lan1} Yu Lan, Jihong Qin, and Shiping Feng, Phys. Rev. B
{\bf 75}, 134513 (2007).

\bibitem{lan2} Yu Lan, Jihong Qin, and Shiping Feng, Phys. Rev. B
{\bf 76}, 014533 (2007).

\bibitem {haldane} F. D. M. Haldane, Phys. Rev. Lett. {\bf 45}, 1358
(1980); Phys. Lett. A {\bf 81}, 153 (1981); J. Solyom, Adv. Phys.
{\bf 28}, 201(1979).

\bibitem {ogata} H. Yokoyama and M. Ogata, Phys. Rev. Lett.
{\bf 67}, 3610 (1991); F.F. Assaad and D. W\" urtz, Phys. Rev. B
{\bf 44}, 2681 (1991).

\bibitem {kim2} C. Kim, A. Y. Matsuura, Z. X. Shen, N. Motoyama,
H. Eisaki, S. Uchida, T. Tohyama, and S. Maekawa, Phys. Rev. Lett.
{\bf 77}, 4054 (1996); B. J. Kim, H. Koh, E. Rotenberg, S.-J. Oh, H.
Eisaki, N. Motoyama, S. Uchida, T. Tohyama, S. Maekawa, Z.-X. Shen
and C. Kim, Nature Phys. {\bf 2}, 397 (2006).

\bibitem {maekawa} See, e.g., S. Maekawa and T. Tohyama, Rep.
Prog. Phys. {\bf 64}, 383 (2001), and references therein.

\bibitem {cooper1} Joseph Orenstein, G. A. Thomas, A. J. Millis, S.
L. Cooper, D. H. Rapkine, T. Timusk, L. F. Schneemeyer, and J. V.
Waszczak, Phys. Rev. B {\bf 42}, 6342 (1990); D.B. Tanner and T.
Timusk, in {\it Physical Properties of High Temperature
Superconductors} III, edited by D.M. Ginsberg (World Scientific,
Singapore, 1992), p. 363, and references therein.

\bibitem {uchida1} S. Uchida, Physca C {\bf 282-287}, 12 (1997),
and references therein.

\bibitem {uchida2} H. Takagi, B. Batlogg, H. L. Kao, J. Kwo, R. J.
Cava, J. J. Krajewski, and W. F. Peck, Jr., Phys. Rev. Lett. {\bf
69}, 2975 (1992).

\bibitem{ando1} Yoichi Ando, A. N. Lavrov, Seiki Komiya, Kouji
Segawa, and X. F. Sun, Phys. Rev. Lett. {\bf 87}, 017001 (2001);
Yoichi Ando, Kouji Segawa, A.N. Lavrov, Seiki Komiya, J. Low
Temperature Phys. {\bf 131}, 793 (2003).

\bibitem {hill} R. W. Hill, Cyril Proust, Louis Taillefer, P.
Fournier and R. L. Greene, Nature {\bf 414}, 711 (2001).

\bibitem {rice1} C. Gros, R. Joynt, and T.M. Rice, Phys. Rev. B
{\bf 36}, 381 (1987).

\bibitem {emery} L. Zhang, J. K. Jain, and V. J. Emery, Phys. Rev.
B {\bf 47}, 3368 (1993); Shiping Feng, J. B. Wu, Z. B. Su, and L.
Yu, Phys. Rev. B {\bf 47} 15192 (1993).

\bibitem {feng6} Shiping Feng, Z.B. Su, and L. Yu, Phys. Rev. B
{\bf 49}, 2368 (1994); Mod. Phys. Lett. B {\bf 7}, 1013 (1993).

\bibitem {ioffe} P.B. Wiegmann, Phys. Rev. Lett. {\bf 60}, 821
(1988).

\bibitem {dagotto2} George B. Martins, Robert Eder, and Elbio
Dagotto, Phys. Rev. B {\bf 60}, R3716 (1999); G. B. Martins, J. C.
Xavier, C. Gazza, M. Vojta, and E. Dagotto, Phys. Rev. B {\bf 63},
014414 (2000); G. B. Martins, C. Gazza, J. C. Xavier, A. Feiguin,
and E. Dagotto, Phys. Rev. Lett. {\bf 84}, 5844 (2000).

\bibitem {plakida} N. M. Plakida, Condens. Matter Phys. {\bf 5},
707 (2002).

\bibitem{tanaka} K. Tanaka, T. Yoshida, A. Fujimori, D. H. Lu, Z.
X. Shen, X. J. Zhou, H. Eisaki, Z. Hussain, S. Uchida, Y. Aiura, K.
Ono, T. Sugaya, T. Mizuno, and I. Terasaki, Phys. Rev. B {\bf 70},
092503 (2004).

\bibitem{tsuei} See, e.g., C.C. Tsuei and J.P. Kirtley, Rev. Mod.
Phys. {\bf 72}, 969 (2000).

\bibitem{shen2} Z. X. Shen, D. S. Dessau, B. O. Wells, D. M. King,
W. E. Spicer, A. J. Arko, D. Marshall, L. W. Lombardo, A.
Kapitulnik, P. Dickinson, S. Doniach, J. DiCarlo, A. G. Loeser, and
C. H. Park, Phys. Rev. Lett. {\bf 70}, 1553 (1993); H. Ding, M. R.
Norman, J. C. Campuzano, M. Randeria, A. F. Bellman, T. Yokoya, T.
Takahashi, T. Mochiku, and K. Kadowaki, Phys. Rev. B {\bf 54}, R9678
(1996).

\bibitem{bozovic} I. Bozovic, G. Logvenov, M. A. J. Verhoeven, P.
Gaputo, E. Goldobin, and T. H. Geballe, Nature {\bf 422}, 873
(2003).

\bibitem{feng7} Shiping Feng and Yun Song, Phys. Rev. B {\bf 55},
642 (1997).

\bibitem{kondo} J. Kondo and K. Yamaji, Prog. Theor. Phys.
{\bf 47}, 807 (1972).

\bibitem{feng8} Shiping Feng and Zhongbing Huang, Phys. Lett. A
{\bf 232}, 293 (1997); Feng Yuan, Jihong Qin, Shiping Feng, and W.
Y. Chen, Phys. Rev. B {\bf 67}, 134505 (2003); Shiping Feng, Feng
Yuan, and Weiqiang Yu, Eur. Phys. J. B {\bf 15}, 607(2000); Shiping
Feng, Feng Yuan, Weiqiang Yu, and Pengpeng Zhang, Phys. Rev. B {\bf
60}, 7565(1999).

\bibitem{graf} J. Graf, G.-H. Gweon, K. McElroy, S. Y. Zhou, C.
Jozwiak, E. Rotenberg, A. Bill, T. Sasagawa, H. Eisaki, S. Uchida,
H. Takagi, D.-H. Lee, and A. Lanzara, Phys. Rev. Lett. {\bf 98},
067004 (2007); B. P. Xie, K. Yang, D. W. Shen, J. F. Zhao, H. W. Ou,
J. Wei, S. Y. Gu, M. Arita, S. Qiao, H. Namatame, M. Taniguchi, N.
Kaneko, H. Eisaki, K. D. Tsuei, C. M. Cheng, I. Vobornik, J. Fujii,
G. Rossi, Z. Q. Yang, and D. L. Feng, Phys. Rev. Lett. {\bf 98},
147001 (2007); K. Byczuk, M. Kollar, K. Held, Y.-F. Yang, I. A.
Nekrasov, Th. Pruschke and D. Vollhardt, Nature Phys. {\bf 3}, 168
(2007).

\bibitem{eliashberg} G. M. Eliashberg, Sov. Phys. JETP {\bf 11},
696 (1960); D. J. Scalapino, J. R. Schrieffer, and J. W. Wilkins,
Phys. Rev. {\bf 148}, 263 (1966).

\bibitem{wen} S. Huefner, M. A. Hossain, A. Damascelli, G. A.
Sawatzky, arXiv:0706.4282 (unpublished).

\bibitem{tallon} See, e.g., J.L. Tallon, J.W. Loram, J.R. Cooper,
C. Panagopoulos, and C. Bernhard, Phys. Rev. B {\bf 68}, 180501 (R)
(2003).

\bibitem{campuzano2} J. C. Campuzano, H. Ding, M. R. Norman, H. M.
Fretwell, M. Randeira, A. Kaminski, J. Mesot, T. Takeuchi, T. Sato,
T. Yokoya, T. Takahashi, T. Mochiku, K. Kadowaki, P. Guptasarma, D.
G. Hinks, Z. Konstantinovic, Z. Z. Li, and H. Raffy, Phys. Rev.
Lett. {\bf 83}, 3709 (1999); M. R. Norman, H. Ding, J. C. Campuzano,
T. Takeuchi, M. Randeria, T. Yokoya, T. Takahashi, T. Mochiku, and
K. Kadowaki, Phys. Rev. Lett. {\bf 79}, 3506 (1997).

\bibitem{uemura} Y. J. Uemura, G. M. Luke, B. J. Sternlieb, J. H.
Brewer, J. F. Carolan, W. N. Hardy, R. Kadono, J. R. Kempton, R. F.
Kiefl, S. R. Kreitzman, P. Mulhern, T. M. Riseman, D. L. Williams,
B. X. Yang, S. Uchida, H. Takagi, J. Gopalakrishnan, A. W. Sleight,
M. A. Subramanian, C. L. Chien, M. Z. Cieplak, G. Xiao, V. Y. Lee,
B. W. Statt, C. E. Stronach, W. J. Kossler, and X. H. Yu, Phys. Rev.
Lett. {\bf 62}, 2317 (1989); Y. J. Uemura, L. P. Le, G. M. Luke, B.
J. Sternlieb, W. D. Wu, J. H. Brewer, T. M. Riseman, C. L. Seaman,
M. B. Maple, M. Ishikawa, D. G. Hinks, J. D. Jorgensen, G. Saito,
and H. Yamochi, Phys. Rev. Lett. {\bf 66}, 2665 (1991).

\bibitem{sato6} T. Sato, T. Kamiyama, Y. Naitoh, T. Takahashi, I.
Chong, T. Terashima, and M. Takano, Phys. Rev. B {\bf 63}, 132502
(2001); T. Sato, T. Kamiyama, T. Takahashi, J. Mesot, A. Kaminski,
J. C. Campuzano, H. M. Fretwell, T. Takeuchi, H. Ding, I. Chong, T.
Terashima, and M. Takano, Phys. Rev. B {\bf 64}, 054502 (2001).

\bibitem{zhou1} X. J. Zhou, T. Yoshida, S. A. Kellar, P. V.
Bogdanov, E. D. Lu, A. Lanzara, M. Nakamura, T. Noda, T. Kakeshita,
H. Eisaki, S. Uchida, A. Fujimori, Z. Hussain, and Z. X. Shen, Phys.
Rev. Lett. {\bf 86}, 5578 (2001); T. Yoshida, X. J. Zhou, T.
Sasagawa, W. L. Yang, P. V. Bogdanov, A. Lanzara, Z. Hussain, T.
Mizokawa, A. Fujimori, H. Eisaki, Z. X. Shen, T. Kakeshita, and S.
Uchida, Phys. Rev. Lett. {\bf 91}, 027001 (2003); X. J. Zhou, T.
Yoshida, D.-H. Lee, W. L. Yang, V. Brouet, F. Zhou, W. X. Ti, J. W.
Xiong, Z. X. Zhao, T. Sasagawa, T. Kakeshita, H. Eisaki, S. Uchida,
A. Fujimori, Z. Hussain, and Z. X. Shen, Phys. Rev. Lett. {\bf 92},
187001 (2004); K. Terashima, H. Matsui, T. Sato, T. Takahashi, M.
Kofu, and K. Hirota, Phys. Rev. Lett. {\bf 99}, 017003 (2007).

\bibitem{takeuchi6} T. Takeuchi, T. Yokya, S. Shin, K. Jinno, M.
Matsuura, T. Kondo, H. Ikuta, and U. Mizutani, J. Electron
Spectrosc. Relat. Phenom. {\bf 114-116}, 629 (2001).

\bibitem{landau} L. D. Landau, Sov. Phys. JETP {\bf 3}, 920 (1956);
Sov. Phys. JETP {\bf 5}, 101 (1957); Sov. Phys. JETP {\bf 8}, 70
(1959).

\bibitem{rice19} T. M. Rice, Physica C {\bf 282-287}, xix (1997).

\bibitem{guo3} Huaiming Guo and Shiping Feng, In {\it Proceedings
of the Sixth International Conference on New Theories, Discoveries
and Applications of Superconductors and Related Materials}, Sydney,
Australia, January 9-11, 2007 [Int. J. Mod. Phys. B {\bf 21}, 3108
(2007)].

\bibitem{lee} Cody P. Nave, Dmitri A. Ivanov, and Patrick A. Lee,
Phys. Rev. B {\bf 73}, 104502 (2006);  Samuel Bieri and Dmitri
Ivanov, Phys. Rev. B {\bf 75}, 035104 (2007); S. Yunoki, Phys. Rev.
B {\bf 74}, 180504(R) (2006);  Hong-Yu Yang, Fan Yang, Yong-Jin
Jiang, and Tao Li, cond-mat/0604488 (unpublished).

\bibitem{kondo6} Takeshi Kondo, Tsunehiro Takeuchi, Adam Kaminski,
Syunsuke Tsuda, and Shik Shin, Phys. Rev. Lett. {\bf 98}, 267004
(2007).

\bibitem{dlfeng1} D. L. Feng, N. P. Armitage, D. H. Lu, A. Damascelli,
J. P. Hu, P. Bogdanov, A. Lanzara, F. Ronning, K. M. Shen, H.
Eisaki, C. Kim, Z. X. Shen, J.-i. Shimoyama, and K. Kishio, Phys.
Rev. Lett. {\bf 86}, 5550 (2001).

\bibitem{kordyuk} A. A. Kordyuk, S. V. Borisenko, M. Knupfer, and
J. Fink, Phys. Rev. B {\bf 67}, 064504 (2003); A. A. Kordyuk and S.
V. Borisenko, Low Temp. Phys. {\bf 32}, 298 (2006).

\bibitem{chuang} Y.-D. Chuang, A. D. Gromko, A. Fedorov, Y. Aiura,
K. Oka, Y. Ando, H. Eisaki, S. I. Uchida, and D. S. Dessau, Phys.
Rev. Lett. {\bf 87}, 117002 (2001); P. V. Bogdanov, A. Lanzara, X.
J. Zhou, S. A. Kellar, D. L. Feng, E. D. Lu, H. Eisaki, J.-I.
Shimoyama, K. Kishio, Z. Hussain, and Z. X. Shen, Phys. Rev. B {\bf
64}, 180505(R) (2001).

\bibitem{borisenko} S. V. Borisenko, A. A. Kordyuk, T. K. Kim, S.
Legner, K. A. Nenkov, M. Knupfer, M. S. Golden, J. Fink, H. Berger,
and R. Follath, Phys. Rev. B {\bf 66}, 140509(R) (2002).

\bibitem{chakarvarty} O. K. Anderson, A. I. Liechtenstein, O. Jepsen,
and F. Paulsen, J. Phys. Chem. Solids {\bf 56}, 1573 (1995); A. I.
Liechtenstein, O. Gunnarsson, O. K. Anderson, and R. M. Martin,
Phys. Rev. B {\bf 54}, 12505 (1996); S. Chakarvarty, A. Sudbo, P. W.
Anderson, and S. Strong, Science {\bf 261}, 337 (1993).

\bibitem{cheng} Li Cheng, Huaiming Guo, and Shiping Feng, Phys.
Lett. A {\bf 366}, 137 (2007).

\end{thebibliography}
\end{document}